\documentclass[12pt]{article}
\pdfoutput=1
\usepackage{amsmath,amssymb,epsfig,amsfonts}
\usepackage{graphicx,subfigure}
\usepackage[usenames, dvipsnames]{color}
\usepackage[backref]{hyperref}
\usepackage{cite}
\usepackage{verbatim} 
\usepackage{slashed}
\usepackage{multirow} 
\usepackage{color}
\usepackage{ulem}

\makeatletter

\newcommand{\be}{\begin{equation}}
\newcommand{\ee}{\end{equation}}
\newcommand{\ba}{\begin{aligned}}
\newcommand{\ea}{\end{aligned}}
\newcommand{\su}{\mathfrak{su}}
\newcommand{\so}{\mathfrak{so}}

\newcommand{\dd}{\delta}

\newcommand{\ti}{\widetilde}
\newcommand{\wat}{\widehat}
\newcommand{\lp}{\left(}
\newcommand{\rp}{\right)}
\newcommand{\un}{\underline}

\newcommand{\gf}{\sqrt{|g_{4}|}}

\def\uA{{\underline{A}}}
\def\uB{{\underline{B}}}
\def\uC{{\underline{C}}}
\def\uD{{\underline{D}}}

\newcommand{\cN}{\mathcal{N}}
\newcommand{\cM}{\mathcal{M}}

\newcommand{\nn}{\nonumber}
\newcommand{\bea}{\begin{eqnarray}}
\newcommand{\eea}{\end{eqnarray}}

\renewcommand{\S}{{\rm S}}
\newcommand{\R}{{\mathbb R}}

\def\Tr{\mathop{\mathrm{Tr}}\nolimits}
\def\tr{\mathop{\mathrm{Tr}}\nolimits}

\def\half{{\frac{1}{2}}}

\def\p{\partial}

\def\unit{{1\kern-.65ex {\rm l}}}
\def\1{{1\kern-.65ex {\rm l}}}

\def\CD{{\cal D}}

\def\CS{{\cal S}}

\def\bbC{{\mathbb{C}}}

\def\bbR{{\mathbb{R}}}

\def\bbZ{{\mathbb{Z}}}

\newcount\hour \newcount\minute
\hour=\time \divide \hour by 60
\minute=\time
\def\now{%
\ifnum \hour<13
  \ifnum \hour=0 \advance \hour by 12 \number\hour:\else \number\hour:\fi%
     \ifnum \minute<10 0\fi%
     \number\minute%
\ A.M.%
\else \advance \hour by -12 \number\hour:%
  \ifnum \minute<10 0\fi%
  \number\minute%
  \ P.M.%
\fi%
}

\makeatother

\begin{document}

\baselineskip=18pt  
\numberwithin{equation}{section}  
\allowdisplaybreaks  


%
%


\thispagestyle{empty}

\vspace*{-2cm} 
\begin{flushright}
{\tt KCL-MTH-16-02}\\
\end{flushright}

\vspace*{1cm} 
\begin{center}
{\LARGE  M5-branes on $S^2 \times M_4$:\\
\bigskip
Nahm's Equations and 4d Topological Sigma-models
}

 \vspace*{1.8cm}
{Benjamin Assel, Sakura Sch\"afer-Nameki,  and Jin-Mann Wong}\\

 \vspace*{.5cm} 
{\it Department of Mathematics, King's College London, \\
  The Strand, London, WC2R 2LS,  UK}\\
  
 {\tt {gmail:$\,$ benjamin.assel, sakura.schafer.nameki, jinmannwong}}

\vspace*{0.8cm}
\end{center}
\vspace*{.5cm}

\noindent
We study the 6d $N=(0,2)$ superconformal field theory, which describes multiple M5-branes,  on the product space $S^2 \times M_4$, and suggest a correspondence between a 2d $N=(0,2)$ half-twisted gauge theory on $S^2$ and a topological sigma-model on the four-manifold $M_4$. To set up this correspondence, we determine in this paper the dimensional reduction of the 6d $N=(0,2)$ theory on a two-sphere and derive that the four-dimensional theory is a sigma-model into the moduli space of solutions to Nahm's equations, or equivalently the moduli space of $k$-centered $SU(2)$ monopoles, where $k$ is the number of M5-branes. 
We proceed in three steps: we reduce the 6d abelian theory to a 5d Super-Yang-Mills theory on $I \times M_4$, with $I$ an interval, then non-abelianize the 5d theory and finally reduce this to 4d. 
In the special case, when $M_4$ is a Hyper-K\"ahler manifold, we show that the dimensional reduction gives rise to a topological sigma-model based on tri-holomorphic maps. 
Deriving the theory on a general $M_4$ requires knowledge of the metric of the target space. For $k=2$ the  target space is the Atiyah-Hitchin manifold and we twist the theory to obtain a topological sigma-model, which has both scalar fields and self-dual two-forms.

\newpage

\tableofcontents


\section{Introduction}

The six-dimensional $N=(0,2)$ superconformal theory (SCFT) with an ADE type gauge group is believed to describe the theory on multiple M5-branes. The equations of motion  in six dimensions are known only for the abelian theory \cite{Howe:1996yn, Howe:1997fb}, and a Lagrangian formulation of this theory is believed to not exist. However, in the last few years, much progress has been made in uncovering properties of this elusive theory by considering compactifications to lower dimensions. Compactification of the 6d theory on a product $S^d \times M_{6-d}$ has resulted in correspondences between supersymmetric gauge theories on $d$-dimensional spheres $S^d$ and conformal/topological field theories on a $6-d$ dimensional manifold $M_{6-d}$. 
The goal of this paper is to consider the compactification of the 6d theory on a four-manifold $M_4$ times a two-sphere $S^2$ and to determine the topological theory on $M_4$. The particular background that we consider is a half-topological twist along the $S^2$, together with a Vafa-Witten-like twist on $M_4$, and we will find that the theory on $M_4$ is a twisted version of a sigma-model into the  moduli space of $SU(2)$ monopoles with $k$ centers, where $k$ is the number of M5-branes, or equivalently, the moduli space of Nahm's equations \cite{Nahm:1979yw} with certain singular boundary conditions. This suggests the existence of a correspondence between this topological sigma-model on $M_4$ and a two-dimensional $(0,2)$ theory, with a half-twist. This fits into the correspondences studied in the last years, which we shall now briefly summarize.

For $d=4$, the Alday-Gaiotto-Tachikawa (AGT) correspondence \cite{Alday:2009aq} connects 4d $N=2$ supersymmetric gauge theories on $S^4$ with Liouville or Toda theories on Riemann surfaces $M_2$. Correlation functions in Toda theories are equal to the partition function of an $N=2$ supersymmetric gauge theory, which depends on the Riemann surface $M_2$. Such 4d $N=2$ gauge theories obtained by dimensional reduction of the 6d $N=(0,2)$ theories were first studied by Gaiotto in \cite{Gaiotto:2009we}, generalizing the Seiberg-Witten construction \cite{Seiberg:1994aj}.
For $d=3$, a correspondence between 3d supersymmetric gauge theories, labeled by three-manifolds $M_3$, and complex Chern-Simons theory on $M_3$ was proposed in \cite{Dimofte:2011ju, Terashima:2011qi}, also refered to as the 3d-3d correspondence. This correspondence has a direct connection to the AGT correspondence by considering three-manifolds, which are a Riemann surface $M_2$ times an interval $I$, $M_3= M_2 \times_\varphi I$, whose endpoints are identified modulo the action of an element $\varphi$ of the mapping class group of $M_2$. On the dual gauge theory side, the mapping class group action translates into a generalized S-duality, and the three-dimensional gauge theories, dual to complex Chern-Simons theory are obtained on duality defects in the 4d $N=2$ Gaiotto theory. The 3d-3d correspondence was ultimately derived from a direct dimensional reduction of the 6d $(0,2)$ theory on a three-sphere via 5d by Cordova and Jafferis \cite{Cordova:2013bea, Cordova:2013cea}.

Other dimensional reductions concern the case of $T^d \times M_{6-d}$. The circle-reduction is known to give rise to $N=2$ 5d Super-Yang-Mills (SYM) \cite{Seiberg:1997ax}. The case of $d=2$ gives rise to $N=4$ SYM with the Vafa-Witten twist \cite{Vafa:1994tf} along $M_4$ \cite{Gadde:2013sca}, which yields a duality between a 2d $N=(0,2)$ gauge theory on $T^2$ and the Vafa-Witten theory on $M_4$. Some results on twisted M5-branes have appeared in \cite{Gran:2014lia}.

\begin{figure}
\centering
\includegraphics[width=11cm]{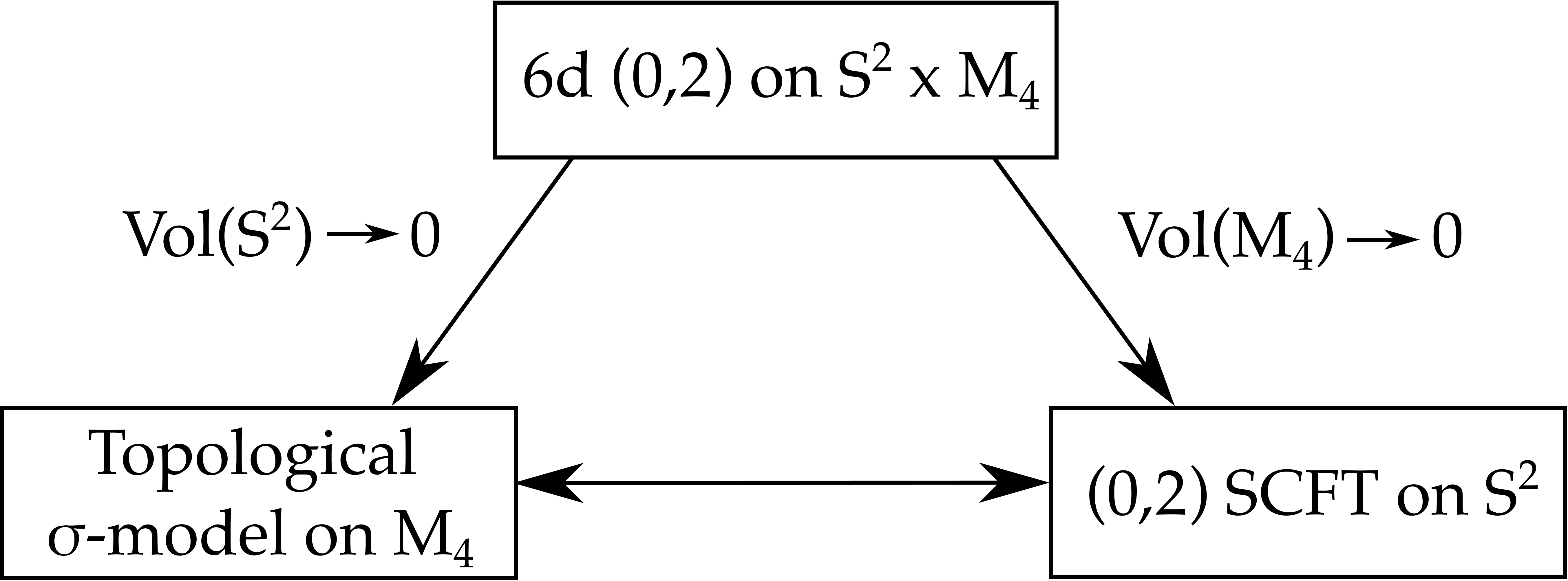}
\caption{4d-2d correspondence between the reduction of the 6d $(0,2)$ theory on $M_4$ to a 2d $(0,2)$ SCFT on $S^2$, and 
the `dual' 4d topological sigma-model from $M_4$ into the Nahm or monopole moduli space, which is obtained in this paper by reducing the 6d theory on a two-sphere. 
\label{fig:4d2dCorrespondence}}
\end{figure}

Both the AGT and 3d-3d correspondences uncovered very deep and surprising relations between supersymmetric gauge theories and two/three-manifolds, their geometry and moduli spaces.  
In view of this a very natural question is to ask, whether we can obtain insights into four-manifolds, as well as the dual two-dimensional gauge theories obtained by dimensional reduction  of the 6d $(0,2)$ theory. Here, unlike the AGT case, the theory on the four-manifold is a topological theory, and the gauge theory lives in the remaining two dimensions and has (half-twisted) $N=(0,2)$ supersymmetry. A schematic depiction of this is given in figure 
\ref{fig:4d2dCorrespondence}.  More precisely, we propose a correspondence between a 4d topological sigma-model and a 2d half-twisted $N=(0,2)$ gauge theory. In particular we expect that topological observables in the 4d theory can be mapped to the partition function and other supersymmetric observables of the 2d theory. Note that the $S^2$ partition function defined with the topological half-twist \cite{Witten:1993yc} is ambiguous as explained in \cite{Gomis:2015yaa}. However the analysis of counter-terms (and therefore ambiguities) must be revisited in the context of the embedding in 6d conformal supergravity, which is our set-up. In particular, the 2d counterterms should originate from 6d counter-terms. 
Recent results on localization in 2d $(0,2)$ theories have appeared in \cite{Closset:2015ohf}, albeit only for theories that have $(2,2)$ loci. The theories obtained from the reduction in this paper do not necessarily have such a $(2,2)$ locus. 

From a brane picture, the theory we consider can be obtained by compactifying $k$ M5-branes on a co-associated cycle in $G_2$ \cite{Bershadsky:1995qy, Gauntlett:2000ng}. The two-dimensional theory that is transverse to the co-associative cycle has $(0,2)$ supersymmetry, and we consider this on a two-sphere, with an additional topological half-twist.

The first question in view of this proposal is to determine what the topological theory on $M_4$ is. 
There are various ways to approach this question. The simplest case is the abelian theory, which on $S^2\times \mathbb{R}^{1,3}$ gives rise to a 4d free $N=2$ hyper-multiplet \cite{Maldacena:2000mw}, which we shall view as a sigma-model into the one-monopole moduli space. On a general four-manifold $M_4$, we will show that in the topologically twisted reduction, the abelian theory integrates indeed to a ``twisted version" of a hyper-multiplet, where the fields are a compact scalar and self-dual two-form on $M_4$. 

For the general, non-abelian case, this 4d-2d correspondence can in principle be connected to the 3d-3d correspondence by considering the special case of $M_4= M_3 \times_\varphi I$, where $I$ is an interval, similar to the derivation of the 3d-3d correspondence from AGT. In this paper we will refrain from considering this approach, and study instead the reduction via 5d SYM, in the same spirit as \cite{Cordova:2013bea, Cordova:2013cea}.

\begin{figure}
\centering
\includegraphics[width=8cm]{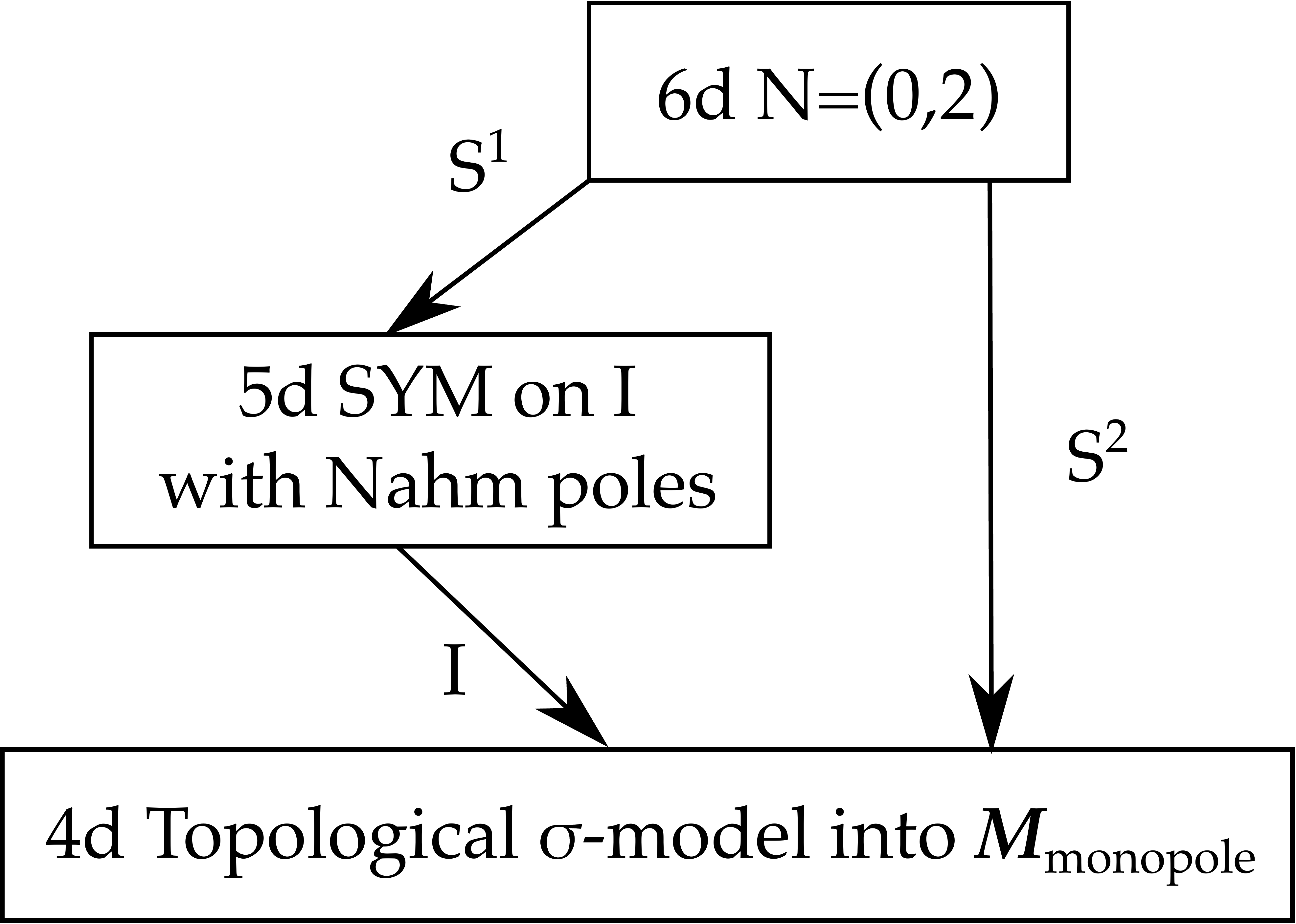}
\caption{
The dimensional reduction of the 6d $N=(0,2)$ theory on an $S^2$, viewed as a circle-fibration along an interval $I$,   is determined by dimensional reduction via 5d SYM. The scalars of the  5d theory satisfy the Nahm equations, with Nahm pole boundary conditions at the endpoints of the interval. The 4d theory is a topological sigma-model into the moduli space of solutions to these Nahm equations, or equivalently the moduli space of monopoles. 
\label{fig:654Redux}}
\end{figure}

We first consider the dimensional reduction on flat $M_4$, and then topologically twist the resulting 4d $N=2$ theory. We restrict to the $U(k)$ gauge  groups, but in principle the analysis holds also for the $D$ and $E$ type. 
To determine the flat space reduction, we view the $S^2$ in terms of a circle-fibration over an interval, where the circle-fiber shrinks to zero-radius at the two endpoints. We determine the 5d supergravity background, which corresponds to the dimensional reduction of the 6d theory on $S^2$. The resulting theory is 5d SYM on an interval, where the scalars satisfy Nahm pole boundary conditions \cite{Witten:2011zz, Gaiotto:2011xs}. Further dimensional reduction to 4d requires to consider scalars, that satisfy Nahm's equations. The resulting theory is a 4d sigma-model into the moduli space of solutions of Nahm's equations, which is isomorphic to the moduli space of $k$-centered monopoles \cite{Donaldson:1985id} and has a natural Hyper-K\"ahler structure.  Much of the geometry of the moduli space is known, in particular for one- or two-monopoles \cite{Atiyah:1988jp}, and a more algebraic formulation in terms of Slodowy-slices exists following \cite{Kronheimer2004, MR2309936, Bielawski01041997}. The latter description is particularly amenable for the characterization of $N=2$ Gaiotto theories with finite area for the Riemann surface as studied in \cite{Gaiotto:2011xs}. Figure \ref{fig:654Redux} summarizes our dimensional reduction procedure. 

The 4d $N=2$ supersymmetric sigma-model for flat $M_4$ falls into the class of models obtained in \cite{AlvarezGaume:1981hm,Bagger:1983tt}. We find that the coupling constant of the 4d sigma-model is given in terms of the area of the two-sphere. To define this sigma-model on a general four-manifold requires 
topologically twisting the theory with the R-symmetry of the 4d theory. One of the complications is that the $SU(2)$ R-symmetry of the 4d theory gets identified with an $SU(2)$ isometry of the Hyper-K\"ahler target. The twisting requires thus a precise knowledge of how the coordinates of the monopole moduli space transform under the $SU(2)$ symmetry.  This is known only in the case of one- and two-monopoles, where a metric  has been determined explicitly\cite{Atiyah:1988jp}. In these cases, we shall describe in section \ref{sec:TwoForms} the topological sigma-models, which have both scalars and self-dual two-form fields on $M_4$. The sigma-model into the one-monopole moduli space $S^1 \times \bbR^3$, corresponding to the reduction of the abelian theory to a free 4d hypermultiplet, gives rise upon twisting to a (free) theory on $M_4$ with a compact scalar and a self-dual two-form, and belongs to the class of 4d A-model of \cite{Kapustin:2010ag}. The sigma-model into the two-monopole moduli space, which is closely related to the Atiyah-Hitchin manifold, gives rise to an exotic sigma-model of scalars and self-dual two-forms obeying constraints. 
Sigma-models in 4d are non-renormalizable and infrared free, however, the observables of the topologically twisted theory are independent of the RG flow and can in principle be computed in the weak coupling regime. 

In the case of $M_4$ a Hyper-K\"ahler manifold, the holonomy is reduced and the twisting does not require knowledge of the R-symmetry transformations of the coordinate fields. This is discussed in section \ref{subsec:TopoTwist}, and the topological sigma-model  that we find upon twisting is the one studied in \cite{Anselmi:1993wm} by Anselmi and Fr\`e for almost quaternionic target spaces.

In this paper we focus on the reduction of the 6d $(0,2)$ theory on a two-sphere, however, as we emphasize in section \ref{sec:5dTheory}, the reduction would proceed in the same way with the addition of two arbitrary `punctures' on the two-sphere, characterizing BPS defects of the 6d non-abelian theory. In the intermediate 5d theory, it would result in different Nahm-pole boundary conditions for scalar fields at the two ends of the interval and the final flat space four-dimensional theory would be a sigma-model into the moduli space of solutions of Nahm's equations with these modified Nahm-pole boundary conditions.

We should also remark upon the connection of our results to the paper by Gadde, Gukov and Putrov \cite{Gadde:2013sca}, who consider the torus-reduction of the M5-brane theory. The topological twist along $M_4$ is the same in their setup as in our construction. Thus, the dictionary to the data of the 2d theory as developed in \cite{Gadde:2013sca}, such as its dependence on the topological/geometric data of $M_4$, should hold in our case as well. For instance, the rank of the 2d gauge group is determined by $b^2(M_4)$. 
The key difference is however, that we consider this 2d theory on $S^2$, and topologically twist the chiral supersymmetry. Interestingly, the reduction of the 6d theory on either $T^2$ or $S^2$ with half-twist gives rather distinct 4d topological theories: in the former, the 4d $N=4$ SYM theory with Vafa-Witten twist, in the latter, we find a four-dimensional topological sigma-model into the monopole moduli space, which for general $M_4$ has both scalars as well as self-dual two-forms. The appearance of self-dual two-forms is indeed not surprising in this context, as the topological twist along $M_4$ is precisely realized in terms of M5-branes wrapping a co-associative cycle in $G_2$, which locally is given in terms of the bundle of self-dual two-forms $\Omega^{2+}(M_4)$ \cite{Joyce}.

The plan of the paper is as follows. We begin in section \ref{sec:ansatz} by setting up the various topological twists of the 6d $N=(0,2)$ theory on $S^2 \times M_4$, and provide the supergravity background and Killing spinors, for the $S^2$ reduction with the half-twist.  In section \ref{sec:5dTheory} we dimensionally reduce the 6d theory to 5d SYM on an interval times $\mathbb{R}^4$, with Nahm pole boundary conditions for the scalar fields. In particular we study this with a generic squashed metric on $S^2$ and in a special `cylinder' limit. The reduction to 4d is then performed in section \ref{sec:SigmaModNahm}, where we show that the fields have to take values in the moduli space of Nahm's equations, and determine the $N=2$ supersymmetric sigma-model on $\mathbb{R}^4$. The action can be found in (\ref{4dUntwisted}), as well as in the form of the models of \cite{AlvarezGaume:1981hm,Bagger:1983tt} in (\ref{Final4dUntwisted}). In sections \ref{sec:4dHK} and \ref{sec:TwoForms} we study the associated topological sigma-models:
in section \ref{sec:4dHK} we consider the case of $M_4$ a Hyper-K\"ahler manifold, and show that this gives rise to the topological sigma-model in \cite{Anselmi:1993wm}. The action can be found in (\ref{4dHKSigmaModel}). 
We furthermore connect this to the dimensional reduction of the topologically twisted 5d SYM theory and show that both approaches yield the same 4d sigma-model in appendix \ref{sec:TopSigM4}. 
In section \ref{sec:TwoForms}, we let $M_4$ be a general four-manifold, but specialize to the case of one- or two- monopole moduli spaces, and use the explicit metrics to determine the topological field theory. In this case, the bosonic fields are scalars and self-dual two-forms on $M_4$. The action for $k=1$ is (\ref{4dAbelian}) and for $k=2$ 
we obtain (\ref{S2monopM4}). 
We close with some open questions in section \ref{sec:Conc}, and provide details on our conventions and computational intricacies in the appendices.


\section{Topological Twists and Supergravity Backgrounds}
\label{sec:ansatz}

This section serves two purposes: firstly, to explain the possible twists of the 6d $N=(0,2)$ theory on a two-sphere $S^2$, and secondly, to determine the supergravity background associated to the topological half-twist on $S^2$.

\subsection{Twists of the M5-brane on $M_4$}
\label{ssec:TwistsM4}

We consider the compactification of the M5-brane theory, i.e. the six-dimensional $N=(0,2)$ theory, on $M_4 \times S^2$, where $M_4$ is a four-dimensional manifold. More generally, we can consider the twists for reductions on general Riemann surfaces $\Sigma$ instead of $S^2$. 
We will determine the 4d theory that is obtained upon dimensional reduction on the $S^2$, and consider this theory on a general four-manifold $M_4$. Supersymmetry of this theory requires that certain background fields are switched on, which correspond to twisting the theory -- both along $M_4$ as well as along $S^2$. 
The twisting procedure requires to identify part of the Lorentz algebra of the flat space theory with a subalgebra of the R-symmetry.
The R-symmetry and Lorentz algebra of the M5-brane theory on $\mathbb{R}^{6}$ are\footnote{In the dimensional reduction via 5d SYM, we will in fact consider the Lorentzian theory to derive the theory on $\mathbb{R}^{1,3}$. As we have in mind a compactification on a compact four-manifold $M_4$, we will discuss here the Euclidean version.}
\be
\mathfrak{sp}(4)_R \oplus \mathfrak{so}(6)_L \,. 
\ee
The supercharges transform in the $({\bf 4}, \bar{\bf 4})$ spinor representation (the same representation as the fermions in the theory, see appendix \ref{app:conventions}). 
The product structure of the space-time implies that we decompose the Lorentz algebra as
\be\label{LTwist1}
\mathfrak{so}(6)_L \quad \rightarrow \quad \mathfrak{so}(4)_L \oplus \mathfrak{so}(2)_L \cong \mathfrak{su}(2)_\ell \oplus \mathfrak{su}(2)_r \oplus \mathfrak{so}(2)_L \,.
\ee
We can consider the following twists of the theory along $M_4$. Either we identify an $\mathfrak{su}(2)$ subalgebra of both Lorentz and R-symmetry, or we twist with the full $\mathfrak{so}(4)$.  

On $M_4$ there are two {$\su(2)$} twists that we can consider.  
In the first instance consider the decomposition of the R-symmetry as 
\be\label{RTwist1}
\mathfrak{sp}(4)_R \quad \rightarrow \quad \mathfrak{su}(2)_R \oplus \mathfrak{so}(2)_R
\ee
and the $\mathfrak{su}(2)_\ell$ is twisted  by $\mathfrak{su}(2)_R$. That is we replace $\su(2)_\ell$ by the diagonal $\su(2)_{\rm twist} \subset \su(2)_\ell \oplus \su(2)_R$ and define the twisted $\mathfrak{su}(2)$ generators by 
\be
T^a_{\rm twist} = \frac 12 \left(T^a_{\ell} + T^a_{R}\right) \,,
\ee
so that the twisted theory has the following symmetries
\be
\hbox{Twist 1}: \qquad \mathfrak{sp}(4)_R \oplus \mathfrak{so}(6)_L \quad \rightarrow \quad \mathfrak{su}(2)_{\rm twist} \oplus \mathfrak{su}(2)_r \oplus \mathfrak{so}(2)_R \oplus \mathfrak{so}(2)_L \,.
\ee
This twist is reminiscent of the  Vafa-Witten twist of 4d $N=4$ SYM \cite{Vafa:1994tf}. 
The supercharges decompose under (\ref{LTwist1}) and (\ref{RTwist1}) as 
\be\label{4Decomp}
\ba
 \mathfrak{sp}(4)_R \oplus \mathfrak{so}(6)_L \quad &\rightarrow \quad \mathfrak{su}(2)_{R} \oplus \mathfrak{so}(2)_R \oplus \mathfrak{su}(2)_\ell \oplus \mathfrak{su}(2)_r \oplus \mathfrak{so}(2)_L  \cr 
 ({\bf 4}, \overline{\bf 4}) \quad &\rightarrow \quad 
 (\bf{2}_{+1}\oplus \bf{2} _{-1}, (\bf{2,1})_{-1}\oplus(\bf{1,2})_{1}) \,,
\ea
\ee
which after the twist becomes
\be
\ba
 \mathfrak{sp}(4)_R \oplus \mathfrak{so}(6)_L \quad &\rightarrow \quad \mathfrak{su}(2)_{\rm twist} \oplus \mathfrak{su}(2)_r \oplus \mathfrak{so}(2)_R \oplus \mathfrak{so}(2)_L  \cr 
 ({\bf 4}, \bar{\bf 4}) \quad &\rightarrow \quad 
 ({\bf 1 } \oplus {\bf 3},{\bf 1} )_{+-} \oplus 
 ({\bf 1} \oplus {\bf 3},{\bf 1})_{--} \oplus 
 ({\bf 2} , {\bf 2})_{++}\oplus 
 ({\bf 2}, {\bf 2})_{-+} \,.
\ea
\ee
This yields two scalar supercharges on $M_4$, which are of the same negative 2d chirality under $\mathfrak{so}(2)_L$
\be
{  ({\bf 1 },{\bf 1} )_{+-} \oplus 
 ({\bf 1},{\bf 1})_{--}  } \,.
\ee
Upon reduction on $M_4$, this twist leads to a 2d theory with $N=(0,2)$ supersymmetry. In this paper we are not concerned with the reduction on $M_4$, but focus on the reverse, namely the theory on $M_4$. 
{This twist is compatible with a further twist along $S^2$ or more generally an arbitrary Riemann surface $\Sigma$, which identifies $\mathfrak{so}(2)_L$ with the remaining R-symmetry $\mathfrak{so}(2)_R$. This is the setup that we will study in this paper.}
 In the following we will first perform the reduction (and topological twisting) along the $S^2$, and then further twist the resulting four-dimensional theory on $M_4$. 

Finally, let us briefly discuss  alternative twists. 
We can use a different $\su(2)$ R-symmetry factor to twist the theory along $M_4$, namely we can use $\su(2)_1 \subset \su(2)_1 \oplus \su(2)_2 \simeq \so(4)_R \subset \mathfrak{sp}(4)_R$ decomposed as
\be\label{RTwist2}
\mathfrak{sp}(4)_R \quad \rightarrow \quad \mathfrak{su}(2)_1 \oplus \mathfrak{su}(2)_2 \,.
\ee
This twist leads upon reduction on $M_4$ to a 2d theory with $N=(0,1)$ supersymmetry.
\be
\ba
\hbox{Twist 2}: \qquad \mathfrak{sp}(4)_R \oplus \mathfrak{so}(6)_L 
&\quad \rightarrow \quad 
\mathfrak{su}(2)_{\rm twist} \oplus \su(2)_2 \oplus \su(2)_r \oplus \mathfrak{so}(2)_L \cr 
({\bf 4}, \bar{\bf  4}) 
&\quad \rightarrow \quad ({\bf 3} \oplus {\bf 1}, {\bf 1}, {\bf 1})_- \oplus ({\bf 2}, {\bf 1}, {\bf 2})_+ \oplus 
({\bf 2}, {\bf 2}, {\bf 1})_- \oplus ({\bf 1}, {\bf 2}, {\bf 2})_+ \,.
\ea
\ee
We can in fact further twist the $\su(2)_2$ with the remaining $\su(2)_r$ Lorentz symmetry on $M_4$. This corresponds to a total twist of the full $\so(4)_R$ with $\so(4)_L$ and is analogous to the geometric Langlands (or Marcus) twist of 4d $N=4$ SYM theory on $M_4$ \cite{Marcus:1995mq, Kapustin:2006pk}
\be\ba
\hbox{Twist 3}: \qquad \mathfrak{sp}(4)_R \oplus \mathfrak{so}(6)_L  &\quad \rightarrow \quad 
\mathfrak{so}(4)_{\rm twist} \oplus \mathfrak{so}(2)_L \cr 
({\bf 4}, \bar{\bf 4}) &\quad \rightarrow \quad ({\bf 3} \oplus {\bf 1}, {\bf 1})_- \oplus ({\bf 2}, {\bf 2})_+ \oplus ({\bf 2}, {\bf 2})_{-} \oplus ({\bf 1}, {\bf 1} \oplus {\bf 3})_+ \,,
\ea\ee
which has two scalar supercharges of opposite 2d chiralities
\be
({\bf 1}, {\bf 1})_+\ \oplus \  ({\bf 1}, {\bf 1})_-\,,
\ee
so that this twist leads upon reduction on $M_4$ to a 2d theory with $N=(1,1)$ supersymmetry. It is not compatible with a further topological twist  on $S^2$. 
Interestingly it was found in \cite{Bak:2015taa} that supersymmetry can be preserved by turning on suitable background supergravity fields on $M_4$. We will not study this background in this paper, but will return to this in the future.

We will now consider the setup of twist 1 and carry out the reduction of the 6d $N=(0,2)$ theory on $S^2 \times M_4$. As explained in the introduction  our strategy is to find the 6d supergravity background corresponding to the twisted theory along $S^2$, taking $M_4 =\mathbb{R}^4$ to begin with, and carry out the reduction to 4d, where we will finally twist the theory along an arbitrary $M_4$.


%



\subsection{Twisting on $S^2$}
\label{S2Twist}

For our analysis we first {consider the theory on $S^2 \times \mathbb{R}^4$ and the twist along $S^2$.}
 The Lorentz and R-symmetry groups reduce again as in (\ref{LTwist1}) and (\ref{RTwist1}). {The twist is implemented  by identifying $\so(2)_R$ with $\so(2)_L$ and we denote it $\so(2)_{\rm twist} \simeq \mathfrak{u}(1)_{\rm twist}$, whose generators are given by
 \be
U_{\rm twist} = U_{L}  + U_{R} \,.
\ee 
As we have seen this is compatible with the twist 1, discussed in the last subsection. 
\be
\hbox{$S^2$ Twist}: \qquad 
\mathfrak{so}(6)_L \oplus \mathfrak{sp}(4)_R \quad \rightarrow \quad \mathfrak{g}_{res}\cong \mathfrak{su}(2)_\ell \oplus \mathfrak{su}(2)_r \oplus \mathfrak{su}(2)_R\oplus \mathfrak{u}(1)_{\rm twist}  \,.
\ee
The residual symmetry group and decomposition of the supercharges and fermions is then
\be\label{spinordec}
\ba
\mathfrak{so}(6)_L \oplus \mathfrak{sp}(4)_R \quad &\rightarrow \quad \mathfrak{g}_{res}\cong \mathfrak{su}(2)_\ell \oplus \mathfrak{su}(2)_r \oplus \mathfrak{su}(2)_R\oplus \mathfrak{u}(1)_{\rm twist} \cr 
({\overline{\bf{4}}, {\bf{4}} })\quad & \rightarrow \quad 
({\bf 2}, {\bf 1},{\bf 2})_{0}\oplus
({\bf 2},{\bf 1},{\bf 2})_{-2}\oplus 
({\bf1},{\bf 2},{\bf 2})_{2}\oplus
({\bf 1},{\bf 2},{\bf 2})_{0} \,.
\ea
\ee
There are eight supercharges transforming as singlets on $S^2$ and transforming as Weyl spinors of opposite chirality on $M_4$ and doublets under the remaining R-symmetry.
The  fields of the 6d $(0,2)$ theory decompose as follows
\be
\ba
\mathfrak{so}(6)_L \oplus \mathfrak{sp}(4)_R \quad &\rightarrow \quad  \mathfrak{su}(2)_\ell \oplus \mathfrak{su}(2)_r \oplus \mathfrak{su}(2)_R\oplus \mathfrak{u}(1)_{L} \oplus \mathfrak{u}(1)_R \cr 
\Phi^{\wat{m}\wat{n}}= ({\bf 1}, {\bf 5}) \quad &\rightarrow \quad  
({\bf 1}, {\bf 1}, {\bf 1})_{0,2} \oplus  ({\bf 1}, {\bf 1}, {\bf 1})_{0,-2} \oplus  ({\bf 1}, {\bf 1}, {\bf 3})_{0,0} \cr 
\rho_{\underline{m}}^{\wat{m}} = (\overline{\bf 4}, {\bf 4}) \quad & \rightarrow \quad 
 {({\bf 1}, {\bf 2}, {\bf 2})_{+1, -1} \oplus  ({\bf 1}, {\bf 2}, {\bf 2})_{+1,+1} \oplus  ({\bf 2}, {\bf 1}, {\bf 2})_{-1,-1} \oplus  ({\bf 2}, {\bf 1}, {\bf 2})_{-1,+1}  }
 \cr 
 \mathcal{B}_{\uA\uB}= ({\bf 15}, {\bf 1}) \quad & \rightarrow \quad 
  ({\bf 1}, {\bf 1}, {\bf 1})_{0,0} \oplus  ({\bf 3}, {\bf 1}, {\bf 1})_{0,0} \oplus  ({\bf 1}, {\bf 3}, {\bf 1})_{0,0} \oplus  
  ({\bf 2}, {\bf 2}, {\bf 1})_{2,0} \oplus  ({\bf 2}, {\bf 2}, {\bf 1})_{-2,0} \,.
\ea
\ee
Note from the point of view of the 4d $N=2$ superalgebra, some of these fields transform in hyper-multiplets, however with a non-standard transformation under the R-symmetry, under which some of the scalars form a triplet. 
The standard transformation of the hyper-multiplet can be obtained using an additional $SU(2)$ symmetry \cite{MR2450736}. 
However, in the present situation, we have to use the R-symmetry as given in the above decomposition. 
Twisting with the $\mathfrak{su}(2)_\ell$ Lorentz with the remaining $\mathfrak{su}(2)_R$, i.e. 
\be
\mathfrak{su}(2)_{\rm twist} \cong \textrm{diag}( \mathfrak{su}(2)_\ell \oplus \mathfrak{su}(2)_R ) 
\ee
 the resulting topological theory has the following matter content
\be
\ba
\mathfrak{so}(6)_L \oplus \mathfrak{sp}(4)_R \quad &\rightarrow \quad \tilde{\mathfrak{g}}\cong  \mathfrak{su}(2)_{\rm twist} \oplus \mathfrak{su}(2)_r \oplus \mathfrak{u}(1)_{\rm twist} \cr 
\Phi^{\wat{m}\wat{n}}= ({\bf 1}, {\bf 5}) \quad &\rightarrow \quad  
({\bf 1}, {\bf 1})_{2} \oplus  ({\bf 1}, {\bf 1})_{-2} \oplus  ({\bf 3},  {\bf 1})_{0} \cr 
\rho_{\underline{m}}^{\wat{m}}= 
(\overline{\bf 4}, {\bf 4}) \quad & \rightarrow \quad 
  ({\bf 2},  {\bf 2})_{0} \oplus  ({\bf 2}, {\bf 2})_{2}   \oplus ( {\bf 1} \oplus  {\bf 3}, {\bf 1})_{-2} \oplus  ({\bf 1} \oplus {\bf 3}, {\bf 1})_{0} 
 \cr 
{\mathcal{B}}_{\uA\uB}= ({\bf 15}, {\bf 1}) \quad & \rightarrow \quad 
  ({\bf 1}, {\bf 1})_{0} \oplus  ({\bf 3}, {\bf 1})_{0} \oplus  ({\bf 1}, {\bf 3})_{0} \oplus  
  ({\bf 2}, {\bf 2})_{2} \oplus  ({\bf 2}, {\bf 2})_{-2} \,.
\ea
\ee
In the following it will be clear that the 6d scalars $\Phi$ give rise to scalars and a self-dual two-form on $M_4$. The fermions give rise to either vectors, or scalars and self-dual two-forms as well.  The fields appearing in the decomposition of the two-form $\mathcal{B}$ are not all independent due to the constraint of self-duality of $H=d\mathcal{B}$. They will give rise to a  vector field and a scalar on $M_4$. This matter content will be visible in the intermediate 5d description that we reach later in section \ref{sec:5dTheory}, however, after reducing the theory to 4d and integrating out massive fields, the matter content of the final 4d theories will be different.


\subsection{Supergravity Background Fields}
\label{sec:Ansaetze}

Before describing the details of the reduction, we should summarize our strategy. 
Our goal is to determine the dimensional reduction of the 6d $(0,2)$ theory with non-abelian $\mathfrak{u}(k)$ gauge algebra.  
For the abelian theory, the dimensional reduction is possible, using the equations of motions in 6d \cite{Howe:1996yn, Howe:1997fb}. However, for the non-abelian case, due to absence of a 6d formulation of the theory, we have to follow an alternative strategy. 
Our strategy is much alike to the derivation of complex Chern-Simons theory as the dimensional reduction on an $S^3$ in \cite{Cordova:2013cea}. First note, that the 6d theory on $S^1$ gives rise to  5d $N=2$ SYM theory. 
More generally, the  dimensional reduction of the 6d theory on a circle-fibration gives rise to a 5d SYM theory in a supergravity background \cite{Cordova:2013bea} (for earlier references see \cite{Kugo:2000hn, Kugo:2000af}). This theory has a non-abelian extension, consistent with gauge invariance and supersymmetry, which is then conjectured to be the dimensional reduction of the non-abelian 6d theory. 

More precisely, this approach requires first to determine the background of the 6d abelian theory as described in terms of the $N=(0,2)$ conformal supergravity theory \cite{Bergshoeff:1985mz, Bergshoeff:1999db}. The 5d background is determined by reduction on the circle fiber, and is then non-abelianized. We can  further reduce the theory along the remaining compact directions to determine the theory in 4d. 
For $S^3$, there is the Hopf-fibration, used in \cite{Cordova:2013cea} to derive the Chern-Simons theory in this two-step reduction process. In the present case of the two-sphere, we will fiber the $S^1$ over an interval $I$, and necessarily, the fibers will have to become singular at the end-points.

In the following we will prepare the analysis of the supergravity background.
By requiring invariance under the residual group of symmetries $\mathfrak{g}_{res}$ preserved by the topological twist on $S^2$, we derive ans\"atze for the background fields in 6d $N=(0,2)$ off-shell conformal supergravity fields. In the next section we will consider the Killing spinor equations and fix the background fields completely.

To begin with, the 6d metric on $S^2 \times \mathbb{R}^4$ is given by
\be
ds^2 \ = \ ds^2_{\bbR^4} + r^2  d\theta^2 + \ell (\theta)^2 \, d\phi^2  \, ,
\ee
with $\ell (\theta) = r \sin(\theta)$ for the round two-sphere and $\theta \in I = [0, \pi]$.  
More generally, $\ell(\theta)$ can be a function, which is smooth and interpolates between 
\be\label{ellAsump}
{\ell (\theta)\over r} \sim   \theta \,, \quad  \hbox{for} \quad \theta\rightarrow 0 \,,\qquad 
{\ell (\theta) \over r }\sim \pi -\theta \,, \quad  \hbox{for} \quad \theta\rightarrow \pi \,.  
\ee
We choose the frame 
\be
e^{A} = dx^{A}\,,\qquad 
e^{5} = r\, d\theta\,,\qquad 
e^{6} = \ell(\theta) \, d\phi\,.
\ee 
The corresponding non-vanishing components of the spin connection are 
\be
\omega^{56} = - \omega^{65} = - {\ell'(\theta) \over r}\, d\phi \,.
\ee
In the following the index conventions are such that all hatted indices refer to the  R-symmetry, all unhatted ones are Lorentz indices. The background fields for the {off-shell gravity} multiplet are summarized in table \ref{tab:6dgravfields}. Underlined Roman capital letters are flat 6d coordinates, underlined Greek are curved space indices in 6d, and middle Roman alphabet underlined indices are 6d spinors. All our conventions are summarized in appendix \ref{app:conventions}.

\begin{table}
\begin{center}
\begin{tabular}{|c|c|c|c|}
 \hline
\hbox{Label} & \hbox{Field} & $\mathfrak{sp}(4)_R$ & \hbox{Properties} \\ [2pt] \hline
$e^{\underline{A}}_{\underline{\mu}}$ & Frame & ${\bf 1}$ &  \\ [5pt]
$V_{\underline{A}}^{\widehat{B}\widehat{C}}$ & R-symmetry gauge field & {\bf 10} &  
$V_{\underline{A}}^{\widehat{B} \widehat{C}} = -  V_{\underline{A}}^{\widehat{C} \widehat{B}}$  \\ [5pt] 
$T^{\widehat{A}}_{[\underline{B} \underline{C} \underline{D}]}$ & Auxiliary 3-form  &{\bf 5}&  $T^{\widehat{A}} = - \star T^{\widehat{A}}$ \\ [5pt] 
$D_{(\widehat{A}\widehat{B})}$ &Auxiliary scalar& {\bf 14}& $D_{\widehat{A}\widehat{B}} = D_{\widehat{B} \widehat{A}} $, $D^{\widehat{A}}_{\widehat{A}} = 0$ 
\\[5pt] 
$b_{\underline{A}}$ & Dilatation gauge field & {\bf 1}&  
\\[5pt] \hline
\end{tabular}
\caption{The bosonic background fields for the 6d $(0,2)$ conformal supergravity.\label{tab:6dgravfields}}
\end{center}
\end{table}

Before making the ans\"atze for the background fields, we note the following decompositions of representations that these  background fields transform under, first for the Lorentz symmetry, 
\be
\ba
\mathfrak{so}(6)_L& \quad \rightarrow \quad  \mathfrak{su}(2)_\ell \oplus \mathfrak{su}(2)_r \oplus \mathfrak{u}(1)_L \cr
\uA:		\qquad	{\bf 6}&\quad	\rightarrow \quad ({\bf 2},{\bf 2})_{{\bf 0 }} \oplus({\bf 1},{\bf 1})_{{\bf 2 }}\oplus({\bf 1},{\bf 1})_{{\bf - 2 }}\cr
[\uB \uC \uD]^{(+)}:	\quad	{\bf 10}&\quad	\rightarrow \quad ({\bf 2},{\bf 2})_{{\bf 0 }} \oplus ({\bf 3},{\bf 1})_{{\bf 2 }} \oplus ({\bf 1},{\bf 3})_{{\bf -2 }} \cr
[ \uB \uC ]:	\quad	{\bf 15}&\quad	\rightarrow \quad ({\bf 2},{\bf 2})_{{\bf  2 }} \oplus({\bf 2},{\bf 2})_{{\bf - 2 }} \oplus ({\bf 3}, {\bf 1})_{{\bf 0 }} \oplus ({\bf 1}, {\bf 3})_{{\bf 0 }} \oplus ({\bf 1}, {\bf 1})_{{\bf 0 }}
\ea\ee
and also for the R-symmetry
\be
\ba
\mathfrak{so}(5)_R& \quad \rightarrow \quad \mathfrak{su}(2)_R \oplus \mathfrak{u}(1)_R \cr
\wat{A}:\qquad 					{\bf 5}& \quad \rightarrow \quad {\bf 3}_{{\bf 0 }}  \oplus {\bf 1}_{{\bf 2 }} \oplus {\bf 1}_{{\bf - 2 }} \cr
[\wat{B}\wat{C}]:\quad 			{\bf 10}&\quad	\rightarrow \quad {\bf 3}_{{\bf 0 }} \oplus {\bf 3}_{{\bf 2 }} \oplus {\bf 3}_{{\bf - 2 }} \oplus {\bf 1}_{{\bf 0 }} \cr
(\wat{B}\wat{C}):\quad 			
{\bf 14} &\quad \rightarrow \quad {\bf 5}_{{\bf 0 }} \oplus {\bf 3}_{{\bf 2 }}\oplus {\bf 3}_{{\bf - 2 }} \oplus {\bf 1}_{{\bf 2 }} \oplus {\bf 1}_{{\bf - 2 }} \oplus {\bf 1}_{{\bf 0 }}  \,.
\ea
\ee
The bosonic supergravity fields of 6d off-shell conformal maximal supergravity were determined in \cite{Bergshoeff:1985mz, Bergshoeff:1999db,  Riccioni:1997np, Kugo:2000hn, Cordova:2013bea}.
They are the frame $e^\uA_{\underline{\mu}}$ and
\be
\ba
T_{ [\uB\uC\uD]\wat{A}}  \,,\qquad 
V_{\uA \, [\wat{B}\wat{C}]} \rightarrow (dV)_{[\uA\uB] \, [\wat{C}\wat{D}]}  \,,\qquad
D_{(\wat{A}\wat{B})} \,,\qquad
b_{\uA} \rightarrow (db)_{[\uA\uB] } \, ,
\ea
\ee
where $dV$ and $db$ denote the field strength of the R-symmetry and dilatation gauge fields, respectively.
Furthermore $T_{ [\uB\uC\uD]\wat{A}}$ is anti-self-dual\footnote{In Euclidean signature, $T_{[\uB\uC\uD]\wat{A}}$ can be complexified and taken to satify $T = i \star T$.}  and $D_{(\wat{A}\wat{B})} $ is traceless
\be
T_{ [\uB\uC\uD]\wat{A}}  = T_{[\uB\uC\uD]^{(+)} \wat{A}} \,,\qquad \delta^{\wat{A}\wat{B}} D_{\wat{A}\wat{B}}= 0\,.
\ee
We shall now decompose these in turn under the residual symmetry group 
$\mathfrak{g}_{res} \cong \mathfrak{su}(2)_\ell \oplus \mathfrak{su}(2)_r \oplus \mathfrak{su}(2)_R\oplus \mathfrak{u}(1)_{\rm twist}$ and determine the components that transform trivially, and thus can take non-trivial background values. 
\begin{enumerate}
\item {$T_{ [\uB\uC\uD]\wat{A}}$:}
The decomposition under $\mathfrak{g}_{res}$ is given by
\be
\ba
({\bf 10}, {\bf 5}) \rightarrow 
&   ({\bf 2},{\bf 2},{\bf 3})_{({\bf 2})} \oplus ({\bf 3},{\bf 1},{\bf 3})_{({\bf 2})}  \oplus ({\bf 1},{\bf 3},{\bf 3})_{({\bf -2})} \oplus ({\bf 2},{\bf 2},{\bf 1})_{({\bf \pm 2})} \oplus ({\bf 3},{\bf 1},{\bf 1})_{({\bf 4})} \cr 
&\quad  \oplus ({\bf 3},{\bf 1},{\bf 1})_{({\bf 0})}
 \oplus ({\bf 1},{\bf 3},{\bf 1})_{({\bf 0})} \oplus({\bf 1},{\bf 3},{\bf 1})_{({\bf -4})}   \,.
\ea
\ee
This tensor product does not contain any singlet under $\mathfrak{g}_{res} $, so the backgrounds we consider have $T_{ [\uB\uC\uD]\wat{A}}= 0$.

\item {$V_{\uA [\widehat{B}\widehat{C}]}$}:
We are looking for components of the field strength  $(dV)_{[\uA\uB] \, [\widehat{C}\widehat{D}]}$ invariant under $ \mathfrak{g}_{res}$. The decomposition of $(dV)_{[\uA\uB] \, [\widehat{C}\widehat{D}]}$ is:
\be
\ba
({\bf 15} , {\bf 10}) \rightarrow   &  ({\bf 2},{\bf 2},{\bf 3})_{({\bf \pm 2})} \oplus ({\bf 3},{\bf 1},{\bf 3})_{({\bf 0})}  \oplus ({\bf 1},{\bf 3},{\bf 3})_{({\bf 0})} \oplus ({\bf 1},{\bf 1},{\bf 3})_{({\bf 0})} \oplus ({\bf 2},{\bf 2},{\bf 3})_{({\bf \pm 4})} \\ & \oplus 2\times({\bf 2},{\bf 2},{\bf 3})_{({\bf 0})}  \oplus({\bf 3},{\bf 1},{\bf 3})_{({\bf \pm 2})} \oplus ({\bf 1},{\bf 3},{\bf 3})_{({\bf \pm 2})} \oplus({\bf 1},{\bf 1},{\bf 3})_{({\bf \pm 2})}\\ & \oplus ({\bf 2},{\bf 2},{\bf 1})_{({\bf \pm 2})} \oplus ({\bf 3},{\bf 1},{\bf 1})_{({\bf 0})} \oplus ({\bf 1},{\bf 3},{\bf 1})_{({\bf 0})} 
\oplus ({\bf 1},{\bf 1},{\bf 1})_{({\bf 0})}   \,.
\ea
\ee
We see that we have a singlet that corresponds to turning on a flux on the $S^2$ and an ansatz for $V$ is given by
\be
V_{\phi \, \wat{x}\wat{y}}  = \half \, v(\theta)  \, \epsilon_{\wat{x}\wat{y}}  \,,
\ee
where $\wat x, \wat y$ run over the components $\wat B, \wat C = 4,5$, and the other components of $V$ vanish.

\item {$b_{\uA}$}: 
The field strength $(db)_{[\uA\uB]}$ decomposes under $\mathfrak{g}_{res}$ as
\be
({\bf 15}, {\bf 1}) \rightarrow   ({\bf 2},{\bf 2},{\bf 1})_{({\bf \pm 2})}\oplus ({\bf 3},{\bf 1},{\bf 1})_{({\bf 0})}\oplus ({\bf 1},{\bf 3},{\bf 1})_{({\bf 0})}\oplus ({\bf 1},{\bf 1},{\bf 1})_{({\bf 0})} \,.
\ee
There is a singlet, which corresponds to turning on a field strength on the $S^2$. In the following we will not consider this possibility. Note that any other choice can always be obtained by a conformal transformation with $K$, which  shifts $b_\uA$ \cite{Bergshoeff:1999db}.
In the following we thus set 
\be
b_\uA =0 \,.
\ee

\item {$D_{(\wat{A} \wat{B})}$}:
The decomposition under $\mathfrak{g}_{res} $ is given by
\be
\ba
({\bf 1},{\bf 14}) \rightarrow    ({\bf 1},{\bf 1},{\bf 5})_{({\bf 0})}\oplus ({\bf 1},{\bf 1},{\bf 3})_{({\bf \pm 2})}\oplus ({\bf 1},{\bf 1},{\bf 1})_{({\bf \pm 2})}\oplus ({\bf 1},{\bf 1},{\bf 1})_{({\bf 0})} \,.
\ea 
\ee
There is one singlet corresponding to the ansatz
\be
D_{\wat{a} \wat{b}} = d \, \delta_{\wat{a} \wat{b}} \quad , \quad D_{\wat{x} \wat{y}} =  - \frac 3 2 \, d \, \delta_{\wat{x}\wat{y}}  \, ,
\label{Dansatz}
\ee
with other components vanishing. The relative coefficients are fixed by the tracelessness condition on $D_{(\wat A \wat B)}$.
\end{enumerate}


\subsection{Killing spinors}

With the ans\"atze for the supergravity background fields we can now determine the conditions on the coefficients $v$ and $d$, to preserve supersymmetry. The background of the 6d supergravity is summarized 
in section \ref{sec:Ansaetze} and the Killing spinor equations \eqref{KSE1} and  \eqref{ChiEqn} 
are solved in appendix \ref{app:Kill}. In summary the background with $T_{ [\uB\uC\uD]\wat{A}}=b_\uA=0$ preserves half the supersymmetries if 
\be
\ba
v(\theta) &= - {\ell'(\theta) \over r}\cr 
d(\theta) &= {3\over 2} {\ell''(\theta) \over r^2 \ell (\theta)} \,,
\ea
\ee
where for the round two-sphere $\ell (\theta) =r  \sin(\theta)$, and the Killing spinor $\epsilon$ is constant and satisfies the following constraint
\be\label{ProjectionCond}
 (\Gamma^{\wat 4 \wat 5})^{\wat m}{}_{\wat n} \epsilon^{\wat n}  -\Gamma^{56} \epsilon^{\wat m}  =0 \,.
\ee
 The value of the R-symmetry gauge field $V^{56} = - {\ell'(\theta) \over r} d\phi = \omega^{56} $ and the fact that the preserved supersymmetries are generated by constant spinors indicates that this supergravity background realizes the topological twist on $S^2$, as expected.

Finally, recall that we chose a gauge for which $b_{\uA}=0$. Note that the background field $b_{\uA}$ can be fixed to an arbitrary other value by a special conformal transformation (see \cite{Bergshoeff:1999db}). The special conformal transformation does not act on the other background fields (they transform as scalars under these transformations), nor on the spinor $\epsilon^{\wat m}$, however it changes the spinor $\eta^{\wat m}$ parametrizing conformal supersymmetry transformations. Indeed one can show that the Killing spinor equations \eqref{KSE1} and \eqref{ChiEqn} are solved for an arbitrary $b_{\uA}$ by the same solution $\epsilon^{\wat m}$ together with
\be
\eta^{\wat m} \ = \ - \frac 12 b_\uA \Gamma^\uA \epsilon^{\wat m} \, .
\ee
In this way one can recover the gauge choice $b_{\underline{\mu}} = \alpha^{-1} \p_{\underline{\mu}} \alpha$ (with $\alpha =1/\ell$ in our conventions) of \cite{Cordova:2013bea}, although we will keep our more convenient choice $b_{\underline{\mu}} = 0$. For our gauge choice, the dimensional reduction to 5d is rederived in appendix \ref{app:6dto5dGen}.


\section{From 6d $(0,2)$ on $S^2$ to 5d SYM}
\label{sec:5dTheory}

We now proceed with the dimensional reduction of the six dimensional $N=(0,2)$ theory on $S^1$ to obtain 5d maximally supersymmetric Yang-Mills theory, as in \cite{Kugo:2000hn,Cordova:2013bea}. The main distinction in our case arises in subtle boundary conditions, which will have to be imposed on the fields along the 5d interval. 
All our conventions are summarized in appendix \ref{app:conventions}.

We should remark on an important point in the signature conventions: the reduction to the 5d SYM theory is accomplished in Lorentzian signature, $\bbR^4 \to \bbR^{1,3}$, where fields admit 6d reality conditions, however it would go through in Euclidean signature upon complexifying the fields in 6d and then imposing reality conditions in 5d. This amounts to Wick-rotating the Lorentzian 5d theory.  In later sections, when we study the 5d theory on a generic $M_4$, we adopt the Euclidean signature, which is compatible with the twist on $M_4$.


\subsection{The 6d $(0,2)$ Theory}

The abelian 6d $N=(0,2)$ theory contains a tensor multiplet, which is comprised of a two-form $\mathcal{B}$ with field strength $H= d\mathcal{B}$, five scalars $\Phi_{\wat{m}\wat{n}}$, and four Weyl spinors $\rho_{\underline{m}}^{\wat{m}}$ of negative chirality, which are  symplectic Majorana. The scalars satisfy $\Phi_{\wat{m}\wat{n}} = - \Phi_{\wat{n} \wat{m}}$ and $\Omega^{\wat{m}\wat{n}}\Phi_{\wat{m}\wat{n}} =0$. The equations of motion are (we will use the conventions of \cite{Bergshoeff:1999db})
\be
\ba
H^-_{\underline{\mu\nu\sigma}}-\frac{1}{2}\Phi_{\wat m\wat n} T^{\wat m\wat n}_{\underline{\mu \nu \sigma}}&=0\\
D^2\Phi_{\wat m\wat n}-\frac{1}{15}D^{\wat r\wat s}_{\wat m\wat n}\Phi_{\wat r\wat s}+\frac{1}{3}H^+_{\underline{\mu \nu\sigma}}T^{\underline{\mu \nu\sigma}}_{\wat m \wat n}&=0\\
\slashed{D}\rho^{\wat m}-\frac{1}{12}T^{\wat m\wat n}_{\underline{\mu \nu\sigma}}\Gamma^{\underline{\mu \nu\sigma}}\rho_{\wat n}&=0 \,.
\ea
\ee
Here $H^{\pm} = 1/2 (H \pm \star H)$ and the R-symmetry indices of the background fields have been transformed
from $\wat{A} \rightarrow \wat{m} \wat{n}$  using the Gamma-matrices as in (\ref{sp4indices}). 
The covariant derivatives are defined as follows
\be\ba
D_{\underline{\mu}} \rho^{\wat{m}} &= \left( \partial_{\underline{\mu}} - {5\over 2} b_{\underline{\mu}} + {1\over 4} \omega_{\underline{\mu}}^{\uA \uB}\Gamma_{\uA\uB}\right) \rho^{\wat{m}} - {1\over 2} V_{\underline{\mu} \wat{n}}^{\wat{m}} \rho^{\wat{n}} \cr 
D_{\underline{\mu}} \Phi^{\wat{m} \wat{n}} & = (\partial_{\underline{\mu}} - 2 b_{\underline{\mu}}) \Phi^{\wat{m}\wat{n}} + V_{\underline{\mu}\wat{r}}^{[\wat{m}} \Phi^{\wat{n}] \wat{r}} \cr 
D^2 \Phi^{\wat{m}\wat{n}} & =\left( \partial^{\uA} - 3 b^{\uA} + \omega_{\uB}^{\uB \uA} \right) D_{\uA} \Phi^{\wat{m}\wat{n}} + 
V^{\uA[\wat{m}}_{\wat{r}} D_{\uA} \Phi^{\wat{n}] \wat{r}} - {R_{6d}\over 5} \Phi^{\wat{m}\wat{n}} \,.
\ea
\ee
Here $R_{6d}$ is the 6d Ricci scalar. 
These equations are invariant under the following supersymmetry transformations 
\be
\ba
\delta \mathcal{B}_{\underline{\mu \nu}} & = - \epsilon^{\wat{m}} \Gamma_{\underline{\mu\nu}} \rho_{\wat{m}}\cr 
\delta \Phi^{\wat{m}\wat{n}} & =  - 4 \epsilon^{[\wat{m}} \rho^{\wat{n}]} - 
\Omega^{\wat{m}\wat{n}} \epsilon^{\wat{r}} \, \rho_{\wat{r}} \cr 
\delta \rho^{\wat{m}}& ={1\over 48} H^+_{\underline{\mu \nu\sigma}} \Gamma^{\underline{\mu \nu\sigma}}\epsilon^{\wat{m}} + {1\over 4}\slashed{D} \Phi^{\wat{m}\wat{n}} \epsilon_{\wat{n}} - \Phi^{\wat m \wat n} \eta_{\wat n} \,.
\ea
\ee
The dimensional reduction of these equations yields abelian 5d SYM in a general supergravity background. We will 
perform this reduction in a gauge choice where $b_{\uA}=0$, which is for instance different from
the choice in \cite{Cordova:2013bea}.
The details of this general reduction are given in appendix \ref{app:6dto5dGen}.
The 6d supergravity fields decompose as follows
\be 
\ba
e^{\underline{\mu}}_{\underline{A}}\quad \rightarrow\quad \begin{pmatrix}
e^{\mu'}_{A'}  & &e^\phi_{A'}\equiv C_{A'}\\
e^{\mu'}_6\equiv 0 & &e_6^\phi\equiv \alpha
\end{pmatrix} 
\ea
\qquad \qquad 
\ba
H &\quad \rightarrow \quad F= dA \cr
\rho^{ \underline{m} \wat m}&\quad \rightarrow\quad 
\begin{pmatrix}0\\
 i\rho^{m' \wat m}
\end{pmatrix} 
\cr
\Phi^{\wat m \wat n} &\quad \rightarrow \quad \Phi^{\wat m \wat n}\,,
\ea
\ee
where we used again the index conventions in appendix \ref{app:conventions}.  
The action of abelian 5d SYM theory in a general background is 
\be
S_{5d} = S_F + S_{\rm scalar} + S_{\rho}\,,
\label{Action5d}
\ee
where 
\be
\ba
S_F &= -\int [\alpha\tilde{F}\wedge \star_{5d}\tilde{F} + C \wedge F \wedge F] \cr
S_{\rm scalar} &= -\int d^5x \,\,\sqrt{|g|}\,\alpha^{-1}\left( D_{A'}\Phi^{\wat m\wat n}D^{A'}\Phi_{\wat m\wat n}+ 4 \Phi^{\wat m \wat n}F_{A'B'}T^{A'B'}_{\wat m \wat n}-\Phi_{\wat m\wat n}(M_{\Phi})^{\wat m\wat n}_{\wat r\wat s}\Phi^{\wat r\wat s}\right) \cr
S_\rho &= -\int d^5x \sqrt{|g|} \,\alpha^{-1}\rho_{m' \wat m}\left(i \slashed{D}^{m'}_{n'}\rho^{n' \wat m}+(M_\rho)^{m' \wat m }_{n'\wat n} \rho^{n'\wat n}\right)\,,
\ea
\ee
with all mass matrices defined in appendix \ref{app:6dto5dGen} and $\tilde{F}$ is defined as 
\be
\tilde{F} = F-\frac{1}{\alpha}\Phi_{\wat m \wat n} T^{\wat m \wat n} \,.
\ee


\subsection{5d SYM in the Supergravity Background}
\label{subsec:5dGenf}

We can now specialize to the 6d background $\mathbb{R}^4\times S^2$, including the 
background supergravity fields of section \ref{sec:ansatz} and determine the 5d SYM theory in the background, which corresponds to the 6d $(0,2)$ theory on $S^2$, by performing the dimensional reduction along the circle fiber. 
As shown in section \ref{sec:Ansaetze}, the only background fields for the 5d SYM theory, which are compatible with the residual symmetry group, are $D^{\wat m \wat n}_{\wat r \wat s}$ and $V_\phi^{\wat m \wat n}  \equiv S^{\wat m \wat n}$. 
With these background fields, and the action of the 5d SYM theory in a general background, that we derived in appendix \ref{app:6dto5dGen} in the gauge $b_{\uA}=0$, we can now determine the non-abelian 5d action in our background.

For our background the metric, graviphoton, $C_{A'}$, and the dilaton, $\alpha$, are given by
\be \label{eqn:5dback}
ds^2_5 = ds_{\mathbb{R}^4}^2 + r^2 d\theta^2 \,, \quad 
C_{A'} = 0\,, \quad \alpha = \frac{1}{\ell(\theta)} \,,\quad  0\leq\theta\leq \pi \,,
\ee
which means that $G = dC = 0$. 
Imposing these conditions and turning on only the background fields $D^{\wat m \wat n}_{rs}$ and $S^{\wat m \wat n}$ the full action is given by\footnote{The ratios of numerical prefactors are determined by supersymmetry. Note that our convention for the scalar fields and gauge fields is that they are anti-hermitian.} 
\begin{equation}
S= S_F +   S_{\rm scalar}+  S_\rho + S_{\rm int} \,,
\end{equation}
where 
\begin{equation} \ba \label{eqn:5daction}
S_{F}&=-{1\over 4}\int {1\over \ell(\theta)}\, \, \text{Tr}( F\wedge \star_{5d}F) \\
S_{\rm scalar}&={1\over 16}\int d^5x \,\,\sqrt{|g|}\, \ell(\theta)  \, \text{Tr}\left( \Phi^{\wat m\wat n}D^2\Phi_{\wat m\wat n}+\Phi^{\wat m\wat n}(M_{\Phi})_{\wat m\wat n}^{\wat r\wat s}\Phi_{\wat r\wat s}\right) \\
S_{\rho}&=-\int d^5x \,\, \sqrt{|g|}\, \ell(\theta)  \,\text{Tr} \left( i\rho_{m' \wat m} \slashed{D}^{m'}_{n'}\rho^{n' \wat m}+ \rho_{m' \wat m} (M_\rho)^{ \wat m \wat n}{}^{m'}_{n' } \rho^{n'}_{\wat n}\right) \,.
\ea
\end{equation}
Here, we non-abelianized the theory, and the covariant derivatives and mass matrices
\be
\ba
D_{\mu'} \Phi^{\wat m\wat n}
&=\partial_{\mu'}+{[A_{\mu'},\Phi^{\wat m \wat n}]}\\
D^2\Phi^{\wat m\wat n}
&=\partial^{\mu'}D_{\mu'}
\Phi^{\wat m\wat n} +\frac{\ell'(\theta)}{r^2 \ell(\theta)} \, D_{\theta} \Phi^{\wat m \wat n} + {[A_{\mu'},\partial^{\mu'}\Phi^{\wat m \wat n}]} + {[A_{\mu'},[A^{\mu'},\Phi^{\wat m \wat n}]]} \\
D_{\mu'} \rho^{\wat m}
&=\partial_{\mu'}\rho^{\wat m} +{[A_{\mu'},\rho^{\wat m}]}\\
(M_{\Phi})^{\wat m\wat n}_{\wat r\wat s}
&= {2 \ell''(\theta) \over 5r^2 \ell(\theta)} 
\,\delta^{\wat m}_{[\wat r}\delta^{\wat n}_{\wat s]}+\frac{1}{2 \ell(\theta)^2} \left(S^{ \wat m}_{[\wat r}S_{ \wat s]}^{\wat n}-S^{\wat n}_{\wat t} S^{\wat t}_{[\wat r} \delta^{\wat m}_{\wat s]}\right)-\frac{1}{15}D^{\wat m\wat n}_{\wat r\wat s} \\
(M_\rho)^{ \wat m\wat n}{}^{ m'}_{n' }
&= {1\over \ell(\theta)  }\left(\half S^{\wat m \wat n}\delta^{m'}_{n'} + \frac{i\ell'(\theta)}{2r} \,  \Omega^{\wat m \wat n} (\gamma_5)^{m'}_{n'} \right) \,,
\ea
\end{equation}
where the five dimensional Ricci scalar vanishes, because we have a flat metric on the interval.
In the non-abelian case we can add  the following interaction terms
\be\ba 
\label{5dSint}
S_{\rm int} 
=\int d^5x \sqrt{|g|} \text{Tr}\left( \frac{\ell(\theta)^3}{64 }[\Phi_{\wat m \wat n},\Phi^{\wat n \wat r}][\Phi_{\wat r \wat s},\Phi^{\wat s \wat m}]
 + \frac{\ell(\theta)}{24} S_{\wat m \wat n}\Phi^{\wat m \wat r}[\Phi^{\wat n \wat s},\Phi_{\wat r \wat s}] 
- \ell(\theta)^2 \rho_{m' \wat m}[\Phi^{\wat m \wat n},\rho^{m'}_{\wat n}] 
 \right) \,,
\ea\ee
where the non-vanishing background fields are
\be \label{SUSY5d}
\ba 
S^{\wat m}_{\wat n} &= - { \ell'(\theta)\over r} (\Gamma^{\wat 4 \wat 5})^{\wat m}_{\wat n} \\
D^{\wat m \wat n}_{\wat r \wat s} &= 
\frac{3\ell''(\theta)}{2r^2\ell(\theta)} \left[ 5 (\Gamma^{\wat 4 \wat 5})^{[\wat m}_{\wat r} (\Gamma^{\wat 4 \wat 5})^{\wat n]}_{\wat s} 
- \delta^{[\wat m}_{\wat r} \delta^{\wat n]}_{\wat s} - \Omega^{\wat m \wat n} \Omega_{\wat r \wat s} 
\right]  \,,
\ea
\ee
where $\ell '$ and $\ell ''$ denote first and second derivatives of $\ell$ with respect to $\theta$. The action is invariant under the following supersymmetry transformations\footnote{Note that the spinor variation would have a $1/16$ instead of $1/8$ in the naive dimensional reduction from the abelian 6d theory. However, the non-abelianized version is only invariant under the variation as given in the following equation. This coefficient is not fixed in the abelian theory, but is required to be $1/8$ in the non-abelian one. This is also consistent with \cite{Cordova:2013bea}.}
\be\label{SUSY5dGENERAL}
\ba
\delta A_{\mu'} &= \ell(\theta) \, \epsilon_{\wat m} \gamma_{\mu'} \rho^{ \wat m} \cr
\delta \Phi^{\wat m \wat n}&=-4i\epsilon^{[\wat m}\rho^{\wat n]} -i 
\Omega^{\wat{m}\wat{n}} \epsilon^{\wat{r}} \, \rho_{\wat{r}}\cr 
\delta \rho^{ \wat m} & = \frac{i}{8 \ell(\theta) } F_{\mu'\nu'}\gamma^{\mu' \nu'} \epsilon^{\wat m}+\frac{1}{4}\slashed{D}\Phi^{\wat m \wat n} \epsilon_{\wat n}+ \frac{i}{4 \ell(\theta)}S^{[\wat m}_{\wat r}\Phi^{\wat n] \wat r}\epsilon_{\wat n} 
-\frac{i}{8} \ell(\theta) {\Omega_{\wat n \wat r}[\Phi^{\wat m\wat n},\Phi^{\wat r \wat s}]\epsilon_{\wat s}}\,.
\ea
\ee
Note that the Killing spinor  $\epsilon^{m'}_{ \wat m}$ satisfies the relation (\ref{ProjectionCond}) which now reads
\be \label{KSrelation5d}
(\Gamma^{\wat 4 \wat 5})^{\wat m\wat n}\epsilon^{m'}_{ \wat n} = -i (\gamma_5)^{m'}_{n'} \epsilon^{n' \wat m} \,.
\ee
So far we have kept the $\mathfrak{sp}(4)_R$ R-symmetry indices explicit. However the background breaks the R-symmetry to $\su(2)_R \oplus \so(2)_R$. To make the symmetry of the theory manifest, we decompose 
the  scalar fields $\Phi^{\wat m \wat n}$ into a  triplet of scalars $\varphi^{\wat a}$, transforming in the ${\bf 3}_{{\bf 0}}$ of $\su(2)_R \oplus\so(2)_R $,  and the complex field $\varphi$, which is a singlet ${\bf 1}_{{\bf 1}}$. This can be achieved as follows
\be
\ba
\varphi^{\wat a} &= \frac{1}{4} (\Gamma^{\wat a})_{\wat m \wat n} \Phi^{\wat m \wat n} \,, \quad \wat a = 1,2,3 \cr 
 \varphi = \varphi^4 + i\varphi^5 &= \frac{1}{4} \left(\Gamma^4 + i \Gamma^5\right)_{\wat m \wat n} \Phi^{\wat m \wat n}  \,.
\ea
\ee
The spinors $\rho_{\wat m}$ decompose into the two doublets $\rho^{(1)}_{\wat p}$, $\rho^{(2)}_{\wat p}$, transforming in $({\bf 2})_{{\bf 1}} \oplus ({\bf 2})_{-{\bf 1}}$, as detailed in appendix \ref{app:SpinorDecomp}.
We also split the gauge field (singlet of the R-symmetry) into the components $A_\mu$ along $\R^4$ and the component $A_\theta$ along the interval. 

The spinor $\epsilon_{\wat n}$ parametrizing supersymmetry transformations decomposes under the R-symmetry subalgebra $\su(2)_R \oplus \so(2)_R$ into two $\su(2)_R$ doublets of opposite $\so(2)_R$ charge: $\epsilon_{\wat m} \to \epsilon^{(1)}_{\wat p} , \epsilon^{(2)}_{\wat p}$ (see appendix \ref{app:SpinorDecomp}). The projection condition \eqref{KSrelation5d} becomes
\begin{align}
\epsilon^{(1)}_{\wat p} - \gamma^5 \epsilon^{(1)}_{\wat p} = 0 \,, \quad  \epsilon^{(2)}_{\wat p} + \gamma^5 \epsilon^{(2)}_{\wat p} = 0 \,.
\label{KSrelation5d2}
\end{align}
For any 5d spinor $\chi$ we define 
\be
\chi_\pm = \frac 12 (\chi \pm \gamma^5 \chi)\,,
\ee
 as the four-dimensional chirality. 
The action for the gauge field is 
\begin{equation}
\ba
S_F&=-\frac{1}{8}\int d^5x \sqrt{|g|}\,\,\frac{1}{\ell(\theta)} \Tr \left(F_{\mu\nu} F^{\mu\nu} +2 F_{\mu\theta}F^{\mu\theta}\right)\,,
\ea
\end{equation}
and for the scalars we find
\begin{equation}
\ba
& S_{\rm scalar}\cr 
& =-\frac{1}{4}\int d^5x\sqrt{|g|}\ell(\theta) \Tr\left(
D^{\mu}\varphi^{\wat a}D_{\mu}\varphi_{\wat a}+D^{\mu}\varphi D_{\mu}\bar{\varphi} 
+ {1\over r^2} D_\theta \varphi^{\wat{a}} D_\theta \varphi_{\wat{a}}
+ {1\over r^2} D_\theta \varphi  D_\theta \bar\varphi 
+ m_{\varphi}^2 \varphi\bar\varphi\right)  \,,
\ea
\end{equation}
with the mass term 
\be
m_{\varphi}(\theta)^2= \frac{\ell'(\theta)^2 - \ell(\theta)\ell''(\theta)}{r^2\ell(\theta)^2} \,,
\ee
which for the round sphere is $m_{\varphi}^2 = \cot(\theta)^2/r^2$ and diverges at the endpoints of the interval. 
We will return to this matter when discussing the boundary conditions. 
The action for the fermions is
\begin{equation}
\ba
\S_\rho &= -
2i\int d^5x \sqrt{|g|}\,\,\ell(\theta) \Tr \left(\rho_{\wat p +}^{(1)}\gamma^{\mu}D_\mu\rho^{(2)\wat p }_-+\rho_{\wat p -}^{(1)}\gamma^{\mu}D_\mu\rho^{(2) \wat p }_+ + \frac{1}{r}\rho^{(1)}_{\wat p +}D_\theta\rho_+^{(2)\wat p } - \frac{1}{r}\rho^{(2)}_{\wat p -}D_\theta\rho_-^{(1)\wat p } \right)\,.
\ea
\end{equation}
Finally, the interaction terms in this decomposition read as follows
\begin{equation}
\ba
S_{\rm Yukawa}&=-\int d^5x \sqrt{|g|}\,\ell(\theta)^2 \Tr\left[2 (\sigma^{\wat a})^{\wat p \wat q} \rho^{(2)}_{\wat p -}\left[\varphi_{\wat a},\rho_{\wat q -}^ {(1)}\right] + 2 (\sigma^{\wat a})^{\wat p \wat q}\rho^{(2)}_{\wat p +}\left[\varphi_{\wat a},\rho_{\wat q +}^ {(1)}\right]  \right. \\
&\qquad \quad \left.+i\left(    \rho^{(1)}_{\wat p -}\left[\bar{\varphi},\rho_{ -}^ {\wat p(1)}\right] +\rho^{(1)}_{\wat p +}\left[\bar{\varphi},\rho_{ +}^ {\wat p(1)}\right]-\rho^{(2)}_{\wat p -}\left[\varphi,\rho_{ -}^ {\wat p(2)}\right] -\rho^{(2)}_{\wat p +}\left[\varphi,\rho_{ +}^ {\wat p(2)}\right]   \right)\right]
\cr 
S_{\rm quartic}&=-{1\over 4}\int d^5x \sqrt{|g|}\,\ell(\theta)^3\Tr\left( [\varphi_{\wat a},\varphi][\varphi^{\wat a},\bar{\varphi}]+ \frac{1}{2} [\varphi_{\wat a},\varphi_{\wat b}][\varphi^{\wat a},\varphi^{\wat b}] -\frac{1}{4}[\varphi,\bar{\varphi}][\varphi,\bar{\varphi}]\right) 
\cr
S_{\rm cubic}&=-\frac{1}{6}\int d^5x \sqrt{|g|}\,{\ell(\theta)\ell'(\theta)\over r}\,  \epsilon^{\wat a\wat b\wat c} \Tr \left(\varphi_{\wat a}[\varphi_{\wat b},\varphi_{\wat c}] \right) \,.
\ea
\end{equation}
The complete 5d action  is 
\be\label{5dFinal}
S_{5d} =  S_F + S_{\rm scalar} + S_\rho + S_{\rm Yukawa} + S_{\rm quartic} + S_{\rm cubic}\,,
\ee
and the supersymmetry variations for this action, decomposed with regards to the R-symmetry,  are summarized in appendix \ref{app:5dSUSY}.
The action above should be supplemented with appropriate boundary terms, which ensure that supersymmetry is preserved and that the action is finite. This will be addressed subsequently.



We need to determine the boundary conditions of the 5d fields at the endpoints of the $\theta$ interval. 
To proceed we first notice that the complex scalar $\varphi$ has a mass term $m(\theta)^2$ which diverges at the boundaries $\theta=0, \pi$\, \footnote{This follows from the regularity conditions (\ref{ellAsump}) on $\ell$.} 
\begin{equation}
m(\theta)^2 \simeq \left\lbrace
\begin{array}{cc}
\frac{1}{\theta^2}  & \,, \quad \theta \rightarrow 0 \,,\\
\frac{1}{(\pi-\theta)^2}  & \,, \quad \theta \rightarrow \pi  \,.
\end{array}
\right.
\end{equation}
 Finiteness of the action requires that $\varphi$ behaves as 
 \begin{equation}
\varphi = \left\lbrace
\begin{array}{cc}
O(\theta)  & \,, \quad \theta \rightarrow 0 \,, \\
O(\pi-\theta)  & \,, \quad \theta \rightarrow \pi \,.
\end{array}
\right.
\end{equation}
The boundary conditions on the other fields are most easily determined by the requirement of preserving supersymmetry under the transformations generated by $\epsilon^{(1)}_{\wat p}$ and $\epsilon^{(2)}_{\wat p}$ presented in appendix \ref{app:5dSUSY}. We obtain at $\theta=0$:
\begin{equation}
\ba
 \rho^{(1)}_{\wat p \, +} &=  O(\theta) \,, \quad \rho^{(2)}_{\wat p \, -} =  O(\theta) \,, \quad
 A_{\mu} =  O(\theta^{2}) \,, 
 \ea
\end{equation}
and the counterpart at $\theta =\pi$.

The fields $\varphi^{\wat a}, A_\theta$ are constrained by supersymmetry to obey modified Nahm's equations as they approach the boundaries, given by
\be\label{GenNahmEqn}
D_\theta \varphi^{\wat a} -  \frac 12 r \ell(\theta) \epsilon^{\wat a}{}_{\wat b \wat c} [\varphi^{\wat b} ,\varphi^{\wat c}] = 0 \,.
\ee
These equations are compatible with a singular boundary behaviour of the fields at the endpoints of the $\theta$-interval. For simplicity let us assume the gauge $A_\theta = 0$ in a neighborhood of $\theta=0$, then the above modified Nahm's equations are compatible with the polar behavior at $\theta =0$
\be \label{GeneralPole}
\varphi^{\wat a} = \frac{2 \varrho( \tau^{\wat a})}{r^2 \theta^2} + O(1)   \,,
\ee
where 
\be
\varrho: \quad \su(2) \, \rightarrow\,  \mathfrak{u}(k) 
\ee 
denotes a Lie algebra homomorphism from $\su(2)$ to $\mathfrak{u}(k)$, see e.g. in \cite{Witten:2011zz, Gaiotto:2011xs} and $\tau^{\wat a}$ are related to the Pauli matrices $\sigma^{\wat a}$ as follows
\be\label{taudef}
\tau^{\wat a}=  \frac{i}{2} \sigma^{\wat a}\,.
\ee
Moreover the $O(1)$ term  is constrained to be in the commutant of $\varrho$ in $\mathfrak{u}(k)$. 
The reduction that we study, from a smooth two-sphere to the interval, corresponds to $\varrho$ being an irreducible embedding \cite{Gaiotto:2011xs}. 

More generally the Nahm pole boundary condition \eqref{GenNahmEqn} is compatible with any homomorphism $\varrho$ and is associated with the presence of `punctures' -- or field singularities -- at the poles of the two-sphere in the 6d non-abelian theory \cite{Gaiotto:2009we}. An embedding $\varrho$ can be associated to a decomposition of the fundamental representation ${\bf k}$ under $\su(2)$ and can be recast into a partition $[n_1, n_2, \cdots]$ of $k$. The irreducible embedding is associated to the partition $\varrho = [k]$ and corresponds to the absence of punctures in 6d, and is therefore the sphere reduction that we consider here. 
The boundary conditions at $\theta = \pi$ are symmetric to the ones at  $\theta=0$ and are also characterized by Nahm pole behaviour with irreducible embedding $\varrho = [k]$.

The remaining fermions $\rho^{(1)}_-, \rho^{(2)}_+$ appear in the supersymmetry variations of $\varphi^{\wat a}$ and hence are of order $O(1)$ at $\theta=0$ 
\be
\rho^{(1)}_{\wat p \, -} =  O(1) \,, \qquad  \rho^{(2)}_{\wat p \, +}  =  O(1)   \,,
\ee
and similarly at $\theta =\pi$. 

The boundary condition \eqref{GeneralPole} for the scalars $\varphi^{\wat a}$ introduces two difficulties: the supersymmetry variation of the action results in a non-vanishing boundary term and the poles of the scalar fields make the action diverge. These two problems are cured by the addition of the following boundary term
\be\label{Sbdry}
\ba 
S_{\rm bdry} &= \left[\frac{\ell(\theta)^2}{12} \int d^4x \sqrt{|g_4|} \tr\left(\epsilon^{\wat a \wat b \wat c} \varphi_{\wat a}[\varphi_{\wat b}, \varphi_{\wat c}]\right)\right]^{\pi}_0 \\
&=\int d\theta  \,\partial_{\theta}\left[\frac{\ell(\theta)^2}{12} \int d^4x \sqrt{|g_4|} \tr\left(\epsilon^{\wat a \wat b \wat c} \varphi_{\wat a}[\varphi_{\wat b}, \varphi_{\wat c}]\right)\right]  \,,
\ea
\ee
The second line gives $S_{\rm bdry}$ as a total $\theta$-derivative and we shall take this as the definition of the boundary term. This additional term ensures supersymmetry and makes the 5d action finite. In particular, taking the derivative along $\theta$ we find,
\be 
S_{\rm bdry} = \int d^5x \sqrt{|g|} \left[\frac{\ell(\theta) \ell'(\theta)}{6r} \epsilon^{\wat a \wat b \wat c}\tr\left( \varphi_{\wat a}[\varphi_{\wat b}, \varphi_{\wat c}]\right)+ \frac{\ell(\theta)^2}{4r }\epsilon^{\wat a \wat b \wat c}\tr \left(\partial_{\theta}\varphi_{\wat a}[\varphi_{\wat b}, \varphi_{\wat c}]\right) \right]\,,
\ee 
where the first piece cancels the cubic scalar interaction in the 5d action and the second term combines to give
\be \label{NahmPlusBoundary} \ba 
-\frac{1}{4r^2}\int d^5x \sqrt{|g|} \ell(\theta)&  \Tr \left( D_{\theta}\varphi^{\wat a}D_{\theta}\varphi_{\wat a} + r^2 \ell(\theta)^2 \half[\varphi_{\wat a}, \varphi_{\wat b}][\varphi^{\wat a}, \varphi^{\wat b}] 
- r\ell(\theta) \epsilon^{\wat a \wat b \wat c} \partial_{\theta}\varphi_{\wat a}[\varphi_{\wat b}, \varphi_{\wat c}]\right) \\
= &-\frac{1}{4r^2}\int d^5x \sqrt{|g|} \Tr \left(D_{\theta} \varphi_{\wat a} -\half r \ell(\theta)\epsilon_{\wat a \wat b \wat c}[\varphi^{\wat b}, \varphi^{\wat c}]\right)^2 \,,
\ea \ee
which is the square of modified Nahm's equations. The 5d action is finite since the scalar fields $\varphi^{\wat a}$ obey modified Nahm's equations at the boundaries. 

We notice that the modified Nahm's equations \eqref{GenNahmEqn} can be recast into the form of standard Nahm's equations by a change of coordinate to
\be
\ti\theta = \frac 1 r \int_0^\theta dx \, \ell(x) \,.
\ee
One obtains 
\be\ba
& D_{\ti\theta} \varphi^{\wat a} -  \frac 12 r^2 \epsilon^{\wat a}{}_{\wat b \wat c} [\varphi^{\wat b} ,\varphi^{\wat c}] = 0 \,, \cr
& r^2 \varphi^{\wat a} = \frac{\varrho( \tau^{\wat a})}{\ti\theta} + O(\ti\theta^0) \,,
\ea\ee
and a similar Nahm pole behavior at the other end of the $\ti\theta$ interval. We conclude then that the moduli space of solutions of the modified Nahm's equations is the same as the moduli space of solution of the standard Nahm's equations.



\subsection{Cylinder Limit}

For general hyperbolic Riemann surfaces, with a half-topological twist, the dimensional reduction depends only on the complex structure moduli \cite{Gaiotto:2009we}. 
The two-sphere has no complex structure moduli, however, there will be a metric-dependence in terms of the area of the sphere, which enters as the coupling constant of the 4d sigma-model \cite{Gaiotto:2011xs}.  
We do not expect the reduction to depend on the function $\ell(\theta)$, except through the area of the sphere. 
This can be checked explicitly by performing the reduction keeping $\ell(\theta)$ arbitrary. However, for simplicity we consider here the special singular limiting case, when the two-sphere is deformed to a thin cylinder. This is achieved by taking the metric factor $\ell(\theta)$ as follows
\be
\ba
\ell(\theta) &=\ell = \hbox{constant} \, \ \,  \quad  \quad \textrm{ for } \ \epsilon < \theta < \pi -\epsilon \,, \nn\\
\ell(\theta) & \to \ \text{smooth caps}   \quad  \textrm{ for } \ \theta < \epsilon \, , \   \pi -\epsilon < \theta \,,
\ea\ee
and then taking the limit $\epsilon \to 0$. The limit is singular at the endpoints of the $\theta$-interval, since at finite $\epsilon$, the two-sphere has smooth caps, $\ell(\theta) \sim r\theta$, while at $\epsilon =0$, $\ell(\theta) = \ell$ is constant on the whole $\theta$ interval and describes the metric on a cylinder, or a sphere with two punctures.  
One may worry that such a singular limit is too strong and would change the theory itself. We will argue below in section \ref{subsec:NahmCylinder} that the reduction of the theory with $\ell$ constant leads to the same four dimensional sigma model as for arbitrary $\ell(\theta)$. The reason for choosing $\ell$ constant is only to simplify the derivation.

We rescale the fields as follows
\be
 \varphi^{\wat a}  \to  \frac{1}{r \ell} \varphi^{\wat a} \,, \quad  \varphi \to  \frac{1}{r \ell} \varphi \,, \quad 
\rho^{(1)}_{\pm} \to  \frac{1}{r \ell} \rho^{(1)}_{\pm}  \,, \quad  \rho^{(2)}_{\pm} \to  \frac{1}{r \ell} \rho^{(2)}_{\pm}  \,.
\label{rescaling}
\ee
The action in this limit simplifies to
\be \label{RescaledAction}
\ba
S_F
&=-\frac{r}{8\ell}\int d\theta d^4 x  
\gf \, \text{Tr} \left( F_{\mu\nu} F^{\mu\nu}+ \frac{2}{r^2}(\partial_\mu A_\theta-\partial_\theta A_\mu+[A_\mu,A_\theta])^2\right) \cr 
S_{\rm scalar}
&=-\frac{1}{4r\ell}\int d\theta d^4 x  \gf\, 
\text{Tr}\left(D^{\mu}\varphi^{\wat a}D_{\mu}\varphi_{\wat a}+D^{\mu}\varphi D_{\mu} 
\bar{\varphi} + {1\over r^2} D_\theta \varphi^{\wat{a}} D_\theta \varphi_{\wat{a}}
+ {1\over r^2} D_\theta \varphi  D_\theta \bar\varphi  \right)
\cr 
S_\rho 
&= -\frac{2 i}{r\ell} \int d\theta d^4 x  \gf \,
\text{Tr}\left(\rho_{\wat p +}^{(1)}\gamma^{\mu}D_\mu\rho^{(2)\wat p }_-+\rho_{\wat p -}^{(1)}\gamma^{\mu}D_\mu\rho^{(2) \wat p }_+ + \frac{1}{r}\rho^{(1)}_{\wat p +}D_\theta\rho_+^{(2)\wat p }-\frac{1}{r}\rho^{(1)}_{\wat p -}D_\theta\rho_-^{(2)\wat p }\right)
\cr 
S_{\rm Yukawa}
&=-{1\over r^2 \ell}\int d\theta d^4 x  \gf \, \text{Tr} \left(2\rho^{(2)}_{\wat p -}\left[\varphi^{\wat p \wat q},\rho_{\wat q -}^ {(1)}\right] + 2\rho^{(2)}_{\wat p +}\left[\varphi^{\wat p \wat q},\rho_{\wat q +}^ {(1)}\right]  \right. \cr 
& \qquad \qquad \qquad \left.+i\left(    \rho^{(1)}_{\wat p -}\left[\bar{\varphi},\rho_{ -}^ {\wat p(1)}\right] +\rho^{(1)}_{\wat p +}\left[\bar{\varphi},\rho_{ +}^ {\wat p(1)}\right]-\rho^{(2)}_{\wat p -}\left[\varphi,\rho_{ -}^ {\wat p(2)}\right] -\rho^{(2)}_{\wat p +}\left[\varphi,\rho_{ +}^ {\wat p(2)}\right]   \right)\right)
\cr
S_{\rm quartic}
&=-{1\over 4r^3 \ell}
\int d\theta d^4 x  \gf \,\text{Tr} \left( \frac{1}{2} [\varphi_{\wat a},\varphi_{\wat b}][\varphi^{\wat a},\varphi^{\wat b}] 
+[\varphi_{\wat a},\varphi][\varphi^{\wat a},\bar{\varphi}]-\frac{1}{4}[\varphi,\bar{\varphi}][\varphi,\bar{\varphi}]\right) \\
S_{\text{bdry}} &= \frac{1}{6r^3\ell}\int d\theta d^4x \gf \partial_{\theta} \tr\left(\epsilon^{\wat a \wat b \wat c} \varphi^{\wat a} \varphi^{\wat b} \varphi^{\wat c}\right)\,.
\ea
\ee
The supersymmetry variations of the 5d action summarized in appendix \ref{app:5dSUSY} simplify in the cylinder limit and for the bosonic fields are 
\be
\ba 
\delta A_{\mu} 
&= - {1\over r} \lp \epsilon^{(1)}{}^{\wat p} \gamma_\mu \rho^{(2)}_{\wat p \, -} + \epsilon^{(2)}{}^{\wat p} \gamma_\mu \rho^{(1)}_{\wat p \, +} \rp  \cr 
\delta A_{\theta}  
&= -  \lp \epsilon^{(1)}{}^{\wat p} \rho^{(2)}_{\wat p \, +} - \epsilon^{(2)}{}^{\wat p}  \rho^{(1)}_{\wat p \, -} \rp \cr 
\delta \varphi^{\wat a} 
&=   i \lp \epsilon^{(1)}{}_{\wat p}  (\sigma^{\wat a})^{\wat p\wat q} \rho^{(2)}_{\wat q \, +} - \epsilon^{(2)}{}_{\wat p}  (\sigma^{\wat a})^{\wat p\wat q} \rho^{(1)}_{\wat q \, -}  \rp  \cr 
\delta \varphi &= - 2 \epsilon^{(1)}{}^{\wat p}  \rho^{(1)}_{\wat p \, +}  \cr 
  \delta \bar\varphi &= + 2 \epsilon^{(2)}{}^{\wat p}  \rho^{(2)}_{\wat p \, -}  \cr 
\ea
\ee
and for the fermions
\be
\ba
\delta \rho^{(1)}_{\wat p \, +} 
&= {i r  \over 8} F_{\mu\nu} {\gamma}^{\mu\nu} \epsilon^{(1)}_{\wat p}  
 - \frac{i}{4} D_\mu \varphi {\gamma}^\mu \epsilon^{(2)}_{\wat p}  
 + \frac{1}{4r} D_\theta \varphi^{\wat q}_{\wat p}\epsilon^{(1)}_{\wat q}  - \frac{1}{8r } \lp \epsilon^{\wat a \wat b \wat c} [\varphi_{\wat a},\varphi_{\wat b}](\sigma_{\wat c})^{\wat q}_{\wat p} \epsilon^{(1)}_{\wat q} -i [\varphi, \bar\varphi ] \epsilon^{(1)}_{\wat p} \rp  \\
\delta \rho^{(1)}_{\wat p \, -}  
&=  \frac{i}{4} F_{\mu\theta} {\gamma}^{\mu} \epsilon^{(1)}_{\wat p}  
 +\frac 14  D_\mu \varphi^{\wat q}_{\wat p}\,{\gamma}^\mu\epsilon^{(1)}_{\wat q}  
 + \frac{i}{4r} D_\theta \varphi \epsilon^{(2)}_{\wat p}   
- \frac{1}{4r}  [\varphi, \varphi^{\wat q}_{\wat p}] \epsilon^{(2)}_{\wat q}  \\
\delta \rho^{(2)}_{\wat p \, +} 
&= -\frac{i}{4} F_{\mu\theta} {\gamma}^{\mu} \epsilon^{(2)}_{\wat p}  
 -\frac 14 D_\mu \varphi^{\wat q}_{\wat p} \,{\gamma}^\mu \epsilon^{(2)}_{\wat q}  
 + \frac{i}{4r} D_\theta \bar\varphi \epsilon^{(1)}_{\wat p}
- \frac{1}{4r}  [\bar\varphi, \varphi^{\wat q}_{\wat p}] \epsilon^{(1)}_{\wat q}\\
\delta \rho^{(2)}_{\wat p \, -} 
&= \frac{ir}{8} F_{\mu\nu} {\gamma}^{\mu\nu} \epsilon^{(2)}_{\wat p}  
 + \frac i4 D_\mu \bar\varphi {\gamma}^\mu \epsilon^{(1)}_{\wat p}  
 + \frac{1}{4r} D_\theta \varphi^{\wat q}_{\wat p} \epsilon^{(2)}_{\wat q} - \frac{1}{8r} \lp \epsilon^{\wat a \wat b \wat c} [\varphi_{\wat a},\varphi_{\wat b}] (\sigma_{\wat c})^{\wat q}_{\wat p} \epsilon^{(2)}_{\wat q} +i [\varphi, \bar\varphi ] \epsilon^{(2)}_{\wat p} \rp    \,.
\ea
\ee
The theory we obtain is nothing else than the maximally supersymmetric $N=2$ SYM in 5d. A similar reduction of the 6d (0,2) theory on a cigar geometry was considered in \cite{Witten:2011zz}.
This five-dimensional SYM theory is defined on a manifold with boundaries, which are at the end-points of the $\theta$-interval and  half of the supersymmetries are broken by the boundary conditions. It is key to study the boundary terms and boundary conditions in detail, which will be done in the next subsection.


\subsection{Nahm's Equations and Boundary Considerations}
\label{subsec:NahmCylinder}

The boundary conditions at the two ends of the $\theta$ interval  are affected by the singular cylinder limit. They can be worked out in the same way as in section \ref{subsec:5dGenf} by enforcing supersymmetry at the boundaries.
In the cylinder limit of the two-sphere $\ell(\theta) \rightarrow \ell$ the mass term $m(\theta)^2$ goes to zero everywhere along the $\theta$-interval except at the endpoints $\theta =0,\pi$ where it diverges, forcing the scalar $\varphi$ to vanish at the boundary, as before. The other boundary conditions are found by requiring supersymmetry under the eight supercharges. This requires that the scalars $\varphi^{\wat a}$ obey the standard Nahm's equations close to the boundaries
\be
\label{NahmEq}
 D_\theta\varphi^{\wat{a}} -\frac{1}{2} \epsilon^{\wat a}{}_{\wat b \wat c} [\varphi^{\wat b},\varphi^{\wat c}] =0\,.
\ee
Furthermore, the boundary behavior of the fields in the gauge $A_\theta=0$ around $\theta=0$ are (although this is not the gauge we will choose later)
\begin{equation} \label{BdryCondCylinder}
\ba
& \varphi = O(\theta) \,, \quad A_{\mu} = O(\theta)  \,,\quad  \varphi^{\wat a} = \frac{\varrho(\tau^{\wat a})}{\theta} + \varphi^{\wat a}_{(0)} + O(\theta)  \,, \\
& \rho^{(1)}_{\wat p \, -}  =  O(1) \,, \quad  \rho^{(2)}_{\wat p \, +}  = O(1) \,, \quad 
\rho^{(1)}_{\wat p \, +}  =  O(\theta) \,, \quad  \rho^{(2)}_{\wat p \, -}  = O(\theta)  \,,
\ea
\end{equation}
where $\varrho: \su(2) \rightarrow \mathfrak{u}(k)$ is an irreducible embedding of $\su(2)$ into $\mathfrak{u}(k)$, with $\tau$ as in (\ref{taudef}) . There are similar boundary conditions at $\theta =\pi$.  The constant term $\varphi^{\wat a}_{(0)}$  in the $\varphi^{\wat a}$-expansion is constrained to be in the commutant of embedding $\varrho$. With $\varrho=[k]$ the irreducible embedding, this commutant is simply the diagonal $\mathfrak{u}(1) \subset \mathfrak{u}(k)$, so $\varphi^{\wat a}_{(0)}$ is a constant diagonal matrix. This condition propagates by supersymmetry to the other fields.

The maximally supersymmetric configurations are vacua of the theory preserving eight supercharges and are given by solutions to the  BPS equations
\be
\ba
 D_\theta\varphi^{\wat{a}} -\frac{1}{2} \epsilon^{\wat a}{}_{\wat b \wat c} [\varphi^{\wat b},\varphi^{\wat c}] & =0  \cr 
 \varphi = \bar\varphi = F_{\mu \nu} = F_{\mu\theta} &=0  \cr 
 D_{\mu}\varphi_{\wat{a}} &=0 \,,
\ea
\label{SusyVac5d}
\ee
with all  fermions vanishing. The 5d action is minimized and vanishes for supersymmetric field configurations \eqref{SusyVac5d}. Moreover there is the additional constraint that the scalars $\varphi^{\wat a}$ have poles at $\theta= 0, \pi$ both characterized by the partition/embedding $\varrho =[k]$.
The first equation in \eqref{SusyVac5d} is Nahm's equation for the fields $(\varphi^{\wat a}, A_{\theta})$ 
and the boundary behaviour of $\varphi^{\wat a}$ are standard Nahm poles.

We can now address the validity of the singular cylinder limit $\ell(\theta)=\ell$ constant. In the following we will reduce the theory on the interval and find that the dominant field configurations are given by solutions of Nahm's equations. The resulting four-dimensional theory will be a sigma model into the moduli space of solutions of Nahm's equations. It is easy to see that for arbitrary $\ell(\theta)$ describing a smooth two-sphere metric, the same dimensional reduction will be dominated by field configurations satisfying the modified Nahm's equations \eqref{GenNahmEqn}. We can then reasonably expect that the reduction will lead to a four-dimensional sigma model into the moduli space of the modified Nahm's equations. However we argued at the end of section \ref{subsec:5dGenf} that this moduli space is the same as the moduli space of standard Nahm's equations, so the reduction for arbitrary $\ell(\theta)$ would lead to the same sigma model.

Finally, let us comment on generalizations of the Nahm pole boundary conditions with two arbitrary partitions $\varrho_0$ and $\varrho_\pi$ for the scalar fields at the two boundaries $\theta =0, \pi$, respectively, as described in \cite{Gaiotto:2011xs}.
The polar boundary behavior at $\theta=0$ is given by $\eqref{BdryCondCylinder}$ with $\varrho \to \varrho_0$ and the subleading constant piece $\varphi^{\wat a}_{(0)}$ takes values in the commutant of $\varrho_0$ (i.e. matrices commuting with the image of $\varrho_0$). 
 These boundary conditions preserve the same amount of supersymmetry and admit global symmetry groups $H_0 \times H_\pi \subset SU(k) \times SU(k)$ acting by gauge transformations at the end-points of the $\theta$-interval. $H_0$ and $H_\pi$ are the groups, whose algebras $\mathfrak{h}_0$, $\mathfrak{h_\pi}$ are respectively the commutants of $\varrho_0$ and $\varrho_\pi$ in $\su(k)$.
 These global transformations leave the $\varrho_0$ and $\varrho_\pi$ boundary conditions invariant. In the reduction to 4d, only a subgroup of $H_0 \times H_\pi$ can be preserved (see the discussion  in section 2 of \cite{Gaiotto:2011xs}).
 
The general $(\varrho_0,\varrho_\pi)$ boundary conditions correspond to inserting  singularities or `punctures' of the type $\varrho_0$ at one pole of the two-sphere and of the type $\varrho_\pi$ at the other pole in the 6d $(0,2)$ theory. All our results can be directly generalized to having general $(\varrho_0, \varrho_\pi)$ Nahm poles at the boundaries of the $\theta$-interval. In this case we would obtain sigma-models into a different moduli space: the moduli space of Nahm's equations with $(\varrho_0,\varrho_\pi)$ boundary conditions. 

For the sphere with two punctures labeled by two arbitrary partitions $\varrho_0$, $\varrho_\pi$, it is very natural to consider the metric describing a cylinder, since this is the topology of a sphere with two punctures, and the reduction, whether with the sphere or the cylinder metric, is expected to lead to the same four-dimensional theory. From this point of view, the sphere without punctures, or ``trivial punctures", is simply a subcase corresponding to the specific partitions $\varrho_0=\varrho_\pi=[k]$, and we may take the cylinder metric, as for any other choice of punctures.


\section{Nahm's Equations and 4d Sigma-Model}
\label{sec:SigmaModNahm}

In the last section we have seen that the 5d SYM in the background corresponding to the $S^2$ reduction of the 6d $(0,2)$ theory requires the scalars $\varphi^{\wat{a}}$ to satisfy Nahm's equations, and the supersymmetric boundary conditions require them to have Nahm poles \eqref{BdryCondCylinder} at the boundary of the interval. The four-dimensional theory is therefore dependent on solutions to Nahm's equations. To dimensionally reduce the theory, we pass to a description in terms of coordinates on the moduli space $\mathcal{M}_k$ of solutions to Nahm's equations and find the theory to be a four-dimensional sigma-model into $\mathcal{M}_k$ with the action 
\be 
\ba 
S_{4d} =  \frac{1}{4r\ell}\int d^4x \gf \left[
G_{IJ}\left(\partial_{\mu}X^I \partial^{\mu} X^J -2 i\xi^{(1)I\wat p} \sigma^{\mu}  \mathcal{D}_{\mu}\xi^{(2)J}_{\wat p}\right)  - \frac 12  R_{IJKL}\xi^{(1)I\wat p} \xi^{(1)J}_{\wat p}\xi^{(2)K \wat q}\xi^{(2)L}_{\wat q}  \right] 
\ea \ee                                                                                                                                                                                                                                                                                                                                                                                                                                                                                                                                                                                                                                                                                                                                                                                                                                                                                                                                                                                                                                                                                                                                                                                                                                                                                                                                                                                                                                                                                                                                                                                                                                                                                                                                                                                                                                                                                                                                                                                                                                                                                                                                                                                                                                                                                                                                                                                                                                                                                                                                                                                                                                                                                                                                                                                                                                                                                                                                                                                                                                                                                                                                                                                                                                                                                                                                                                                                                                                                                                                                                                                                                                                                                                                                                                                                                                                                                                                                                                                                                                                                                                                                                                                                   
with $X^I$ the coordinates on the moduli space 
\be\label{XM4M}
X: \qquad M_4 \ \rightarrow \ \mathcal{M}_k \,,
\ee
and  $\xi^{(i)}$, where $i = 1,2$, Grassmann-valued sections of the pull-back of the  tangent bundle to $\mathcal{M}_k$
\be
\xi^{(1,2)} \in \Gamma(X^* T\mathcal{M}_k \otimes \mathcal{S}^{\pm})\,,
\ee
where $\mathcal{S}^{\pm}$ is the spin bundle of $\pm$ chirality on $M_4$.
The sigma-model for $M_4 = \mathbb{R}^4$ is supersymmetric, with $N=2$ supersymmetry in 4d. 
The coupling constant for the sigma-model is proportional to the area of the two-sphere, which is $\sim r \ell$, as anticipated.


\subsection{Poles and Monopoles}
\label{ssec:PolesMonopoles}

Before studying the dimensional reduction to 4d, we summarize a few well-known useful properties of the moduli space $\cM_k$. The moduli space $\cM_k$ of solutions to Nahm's equations, on an interval with  Nahm pole boundary conditions given by the irreducible embedding $\varrho=[k]$, is well-known to be isomorphic to the moduli space of (framed) $SU(2)$ magnetic monopoles of charge $k$ \cite{hitchin1982, hitchin1983, Donaldson:1985id, Atiyah:1988jp}, which is $4k$-dimensional and has a Hyper-K\"ahler structure. The metric of the spaces $\cM_k$ is not known in explicit form, other than for the cases $\cM_1 \simeq \bbR^3\times S^1$ (which is the position of the monopole in $\bbR^3$ and the large gauge transformations parametrized by $S^1$) and for the case 
\be
\cM_2 \simeq \bbR^3 \times \frac{ S^1 \times \cM_{\rm AH} }{\bbZ_2}\,,
\ee where $\cM_{\rm AH}$ is the Atiyah-Hitchin manifold \cite{Atiyah:1988jp}. A detailed description of the metric in  the latter case will be given in section \ref{sec:AHCase}. 
Hitchin showed the equivalence of $SU(2)$ monopoles of charge $k$ with solutions of Nahm's equations \cite{hitchin1983} 
\be
{dT_i \over d\theta} - {1\over 2 }\epsilon_{ijk} [T_j ,T_k ]=0 \,,\qquad i=1, 2, 3\,,
\ee
where $T_i$ are matrix-valued, depending on $\theta\in [0, \pi]$ and have poles at the endpoints of the interval, the residues of which define representations of $\su(2)$. 
Furthermore, Donaldson \cite{Donaldson:1985id} identified Nahm's equations in terms of the anti-self-duality equation $F_A = -\star F_A$ of a connection 
\be
A= T_\theta d\theta+ \sum_i T_i dx_i\,,
\ee 
on $\mathbb{R}^4$, where $T_\theta$, the gauge field along the interval, can be gauged away and the $T_i$ are taken independent of the $x^i$ coordinates. The metric of the solution-space (modulo gauge transformations) has a Hyper-K\"ahler structure \cite{Hitchin:1986ea, dancer1993}.

This Nahm moduli space (or monopole moduli space) takes the form  \cite{Atiyah:1988jp}
\be
\cM_k \simeq \bbR^3 \times \frac{ S^1 \times \cM_k^0 }{\bbZ_k}\,,
\ee
where $\bbR^3$ parameterizes the center of mass of the $k$-centered monopole.
A particularly useful characterization of the reduced Nahm moduli space $\cM_k^0$ is in terms of Slodowy-slices. Kronheimer has shown that the solutions of Nahm's equations with no poles at the boundaries have a moduli space given by the cotangent bundle of the complexified gauge group, $T^* G_{\mathbb{C}} \equiv \mathfrak{g}_\mathbb{C} \times G_\mathbb{C}$, which has a natural Hyper-K\"ahler structure. Furthermore, 
Bielawski showed in \cite{MR2309936, Bielawski01041997}, that the moduli space of solutions with Nahm pole boundary conditions for $k$-centered $SU(2)$ monopoles is given in terms of
\be
\mathcal{M}_k^0 \cong \{ (g, X) \, \in SU(N)_{\bbC} \times \su(N)_{\bbC}; \ X \in S_{[k]} \cap  g^{-1} S_{[k]} g \} \subset T^* SU(k)_{\mathbb{C}} \,,
\ee
where the Slodowy slice for an embedding $\rho:\su(2) \to \mathfrak{u}(k)$  is  
\be
S_{\rho} = \{\rho (\tau^+) + x  \, \in \su(k)_{\bbC}; \, [\rho (\tau^-), x]=0 \}\,.
\ee
Here $\tau^{\pm} \equiv \tau^1 \pm i \tau^2$ are the raising/lowering operators of $\mathfrak{su}(2)$.
The Hyper-K\"ahler metric on $\mathcal{M}_k$ will play a particularly important role in section \ref{sec:TwoForms}, where this will be discussed in more detail.


\subsection{Reduction to the 4d Sigma-Model}
 \label{sec:4dUntwistedRedux}
 
To proceed with the reduction on the $\theta$-interval to four dimensions, we take the limit where the size of the interval, $r$, is small.\footnote{By $r$ small, we mean that we consider the effective theory at energies small compared to $\frac 1 r$.}
The terms in the action \eqref{RescaledAction} are organized in powers of $r$, and in the limit, the divergent terms which are of order $r^{-n}$, $n=2,3$, must vanish separately. The terms of order $r^{-1}$ contain the four-dimensional kinetic terms and lead to the 4d action. The terms of order $r^{n}$, $n \ge 0$ are subleading and can be set to zero. To perform this reduction we must expand the fields in powers of $r$, $\Phi = \Phi_0 + \Phi_1 r + \Phi_2 r^2 + \cdots$, and compute the contribution at each order. We find that only the leading term $\Phi_0$ contributes to the final 4d action for each field, except for the `massive' scalars $\varphi, \bar\varphi$ and spinors $\rho^{(1)}_{+\wat{p}},\rho^{(2)}_{-\wat{p}}$,  whose leading contribution arise at order $r$. The final 4d action will arise with the overall coupling $\frac{1}{r \ell}$.

Let us now proceed with detailing the dimensional reduction.
At order $r^{-3}$ we find the term
\be
\ba
S & = - \frac{1}{4r^3\ell}   \int d\theta d^4x \gf  \cr 
& \qquad \qquad  \Tr \left[ \left( D_\theta\varphi^{\wat{a}} -\frac{1}{2} \epsilon^{\wat a}{}_{\wat b \wat c} [\varphi^{\wat b},\varphi^{\wat c}] \right)^2 +  [\varphi_{\wat a}, \varphi][\varphi^{\wat a}, \bar \varphi]
 + D_{\theta}\varphi D_{\theta} \bar \varphi   -\frac{1}{4}[\varphi, \bar\varphi][\varphi, \bar\varphi] \right] \,.
\ea
\ee
This term is minimized (and actually vanishes)\footnote{To avoid possible confusions about the positivity of the action, we remind that our conventions are such that the fields are anti-hermitian.}, up to order $O(r^{-1})$ corrections, upon imposing the following constraints: $\varphi, \bar\varphi$ vanish at order $r^0$,
\be
\varphi  = \bar\varphi =  O(r) \,,
\ee
 and the fields $\varphi^{\wat a}$ and $A_\theta$ obey Nahm's equations, up to order $O(r)$ corrections,
\be
\ba
D_\theta\varphi^{\wat{a}} -{\epsilon^{\wat a}{}_{\wat b \wat c}\over 2} [\varphi^{\wat b},\varphi^{\wat c}] &= 0 \,, 
 \ea
\ee
with Nahm pole behaviour $\varrho = [k]$  at the two ends of the interval. The four-dimensional theory then localizes onto maps $X: \mathbb{R}^4 \rightarrow \mathcal{M}_{k}$, where $\mathcal{M}_{k}$ is the moduli space of $\mathfrak{u}(k)$ valued solutions of Nahm's equations on the interval with $\varrho$-poles at the boundaries, or equivalently the moduli space of $k$-centered $SU(2)$ monopoles, as reviewed in section \ref{ssec:PolesMonopoles}. 
The fields satisfying  Nahm's equations can be written in terms of an explicit dependence on the point $X^I$ in the moduli space $\mathcal{M}_{k}$
\begin{equation}
\varphi^{\wat a}(\theta, x^\mu) = \varphi^{\wat a}(\theta, X(x^\mu)) \,, \quad  A_\theta(\theta, x^\mu) = A_\theta (\theta, X(x^\mu)) \,.
\label{NahmEqnBis}
\end{equation}
Furthermore, we choose the gauge fixing
\be
\partial_\theta A_\theta =0 \,.
\ee
The terms at $O(r^{-2})$ vanish by imposing the spinors $\rho^{(1)}_{\wat p \, +} , \rho^{(2)}_{\wat p \, -}$ to have no $O(r^0)$ term
\be
\rho^{(1)}_{\wat p \, +} = O(r) \,, \quad   \rho^{(2)}_{\wat p \, -} = O(r) \,.
\ee
The kinetic term of these spinors becomes of order $r$ and can be dropped in the small $r$ limit. The fermions $\rho^{(1)}_{\wat p \, +} , \rho^{(2)}_{\wat p \, -}$  become Lagrange multipliers and can then be integrated out, leading to the constraints on the fermions $\rho^{(1)}_{\wat p \, -} , \rho^{(2)}_{\wat p \, +} $
\be
\ba
D_\theta \rho^{(2)}_{+\wat{p}} + i [\varphi_{\wat{q}}^{\wat{p}} , \rho^{(2)}_{+\wat{q}}] &= 0 \cr 
D_{\theta} \rho^{(1)}_{-\wat{p}} + i  [\varphi_{\wat{q}}^{\wat{p}} , \rho^{(1)}_{-\wat{q}}] &= 0 \,,
\ea
\label{SpinorNahmEqn}
\ee
which are supersymmetric counterparts to Nahm's equations (\ref{NahmEq}). We will use these localizing equations below to expand the fermionic fields in terms of vectors in the tangent space to the moduli space of Nahm's equations, $\mathcal{M}_{k}$.

Finally we drop the order $r$ kinetic terms of the 4d gauge field  and scalars $\varphi, \bar\varphi$ (which contribute only at order $r$), and we are left with the terms of order $\frac{1}{r}$ which describe the 4d action. The remaining task is to express this action in terms of the fields $X = \{X^{I}\}$ and the massless fermionic  degrees of freedom, and to integrate out the 4d components of the gauge field $A_\mu$ and the scalars $\varphi, \bar\varphi$, which appear as auxiliary fields in the 4d action. The subleading terms (at order $r$) in the $\varphi^{\wat a}$ expansion can similarly be integrated out without producing any term in the final 4d action, so we ignore these contributions in the rest of the derivation.

In addition one should integrate over the one-loop fluctuations of the fields around their saddle point configurations. We will assume here that the bosonic and fermionic one-loop determinants cancel, as is frequently the case in similar computations \cite{Bershadsky:1995vm}, and now turn to deriving the 4d action. Some of the technical details have been relegated to appendix \ref{app:4dFlat}.


\subsubsection{Scalars}

We will now describe the 4d  theory in terms of `collective coordinates' $X^I$, similar to the approach taken in e.g. \cite{Bershadsky:1995vm} for the dimensional reduction of 4d SYM theories on a Riemann surface resulting in a 2d sigma-model into the Hitchin moduli space. Related work can also be found in \cite{Harvey:1991hq, Gauntlett:1993sh}. The resulting theory is a (supersymmetric) sigma-model (\ref{XM4M}), where for this part of the paper we will consider $M_4= \mathbb{R}^4$.
The three scalar fields $\varphi_{\wat a}$  and $A_\theta$ are expanded in the collective coordinates as follows
 \begin{equation}\label{eqn:coordexpansion}
\ba
\delta\varphi^{\wat a} = & \Upsilon_{I}^{\wat{a}} \delta X^{I} \cr 
\delta A_{\theta}=& \Upsilon_{I}^{\theta}  \delta X^{I} \,,
\ea
\end{equation} 
where $I = 1,\ldots,  4k$. Here, the basis of the cotangent bundle of $\mathcal{M}_{k}$ is given by
\be\label{eqn:cotangentbasis}
\ba
\Upsilon_{I}^{\wat a} &= \frac{\partial \varphi^{\wat a}}{\partial X^{I}} + [E_{I},\varphi^{\wat a}]\\
 \Upsilon_{I}^{\theta} &=\frac{\partial A_{\theta}}{\partial X^{I}}  - D_{\theta}E_{I}\,,
\ea
\ee
where $E_{I}$ defines a $\mathfrak{u}(k)$ connection $\nabla_I \equiv \p_I + [E_I,.]$ on $\mathcal{M}_k$.
The  $\Upsilon_{I}^{\wat a},\Upsilon_{I}^{\theta}$ satisfy linearized Nahm's equations
\be \label{CotangentCondition}
D_\theta \Upsilon_{I}^{\wat a} + \left[\Upsilon_{I}^{\theta}, \varphi^{\wat a} \right]=\epsilon^{\wat a \wat b\wat c} \left[ \Upsilon_{I\wat b}, \varphi_{\wat c} \right]\,.
\ee
The metric on  $\mathcal{M}_k$ can be expressed in terms of these one-forms as 
\be\label{eqn:metricmodspace}
G_{I J}= -\int d\theta\, \text{Tr}( \Upsilon_{I}^{\wat a} \Upsilon_{J \wat a} +\Upsilon_{I}^{\theta} \Upsilon_{J}^{\theta})\,.
\ee
The Hyper-K\"ahler structure on $\mathcal{M}_k$ can be made manifest in this formulation, by defining the three symplectic forms (see for instance \cite{Gaiotto:2008sa})
\begin{equation}\label{SympForm}
\omega^{\wat a}_{IJ} = \int d\theta \, \text{Tr}( \epsilon^{\wat a \wat b\wat c} \Upsilon_{I \wat b} \Upsilon_{J \wat c} + \Upsilon^{\wat a}_I \Upsilon^{\theta}_J - \Upsilon^{\theta}_I\Upsilon^{\wat a}_J)\,.
\end{equation}
Some useful properties of these are summarized in appendix \ref{app:SigmaModel}.
Using the expansion (\ref{eqn:cotangentbasis}) we obtain 
\begin{equation} \label{SigmaScalars}
S_{\rm scalars} = - \frac{1}{4r\ell}\int d\theta d^4x  \gf\,\text{Tr} \left( \partial_IA_{\theta} \partial_J A_{\theta} + \partial_I \varphi^{\wat a} \partial_J \varphi_{\wat a} \right) \partial_{\mu} X^{I}\,\partial^{\mu}X^{J} \,.
\end{equation}
This will combine with terms arising from integrating out the gauge field to give the usual sigma-model kinetic term.


\subsubsection{Fermions}

The fermions satisfy the equation (\ref{SpinorNahmEqn}), which is the supersymmetry variation of Nahm's equations. The spinors therefore take values in the cotangent bundle to the moduli space $\mathcal{M}_k$ and we can expand them in the basis that we defined in (\ref{eqn:cotangentbasis})
\be
\label{SpinorDecomp}
\ba
\rho^{(1)}_{-\wat{p}}&=\Upsilon^{\wat a}_{I} (\sigma_{\wat a})_{\wat p}^{\wat q}\lambda^{(1)I}_{\wat q} + i \Upsilon_{I}^{(\theta)} \lambda^{(1)I}_{\wat p} \cr 
\rho^{(2)}_{+\wat{p}}&=\Upsilon^{\wat a}_{I} (\sigma_{\wat a})_{\wat p}^{\wat q}\lambda^{(2)I}_{\wat q} + i \Upsilon_{I}^{(\theta)} \lambda^{(2)I}_{\wat p} \, ,
\ea
\ee
where $\lambda^{(1)I}_{\wat p}, \lambda^{(2)I}_{\wat p} $ are spacetime spinors, valued in $T\cM_k$. The identities \eqref{CSUpsilonRel} imply that the fermionic fields obey the constraints 
\be \label{FermionUntwistedCon}
\omega^{\wat aI}{}_J\lambda^{(i) J}_{\wat p} = i(\sigma^{\wat a})_{\wat p}^{\wat q}\lambda^{(i)I}_{\wat q}\,.
\ee
The expansion in \eqref{SpinorDecomp} can be seen to satisfy the equation of motion for the spinors \eqref{SpinorNahmEqn} by making use of \eqref{CotangentCondition} and the gauge fixing condition \eqref{GaugeFixing}. 
Then substituting in the kinetic term for the spinors and making use of the expression for the metric on $\mathcal{M}_k$ (\ref{eqn:metricmodspace}), the symplectic forms $\omega_{IJ}^{\wat a}$ and the constraint \eqref{FermionUntwistedCon}, we find 
\be \label{SigmaSpinors}
\ba
S_{\rho_{\rm kin}} &=  \frac{8i}{rl}\int  d^4 x  \gf \, \Big[\, G_{IJ} \lambda^{(1)I\wat{p}} \gamma^\mu\partial_\mu  \lambda^{(2)J}_{\wat{p}} \cr
&\qquad \qquad \qquad  \qquad   - \int d\theta    \,\text{Tr}  \Big(\Upsilon_I^{\wat a} \partial_J \Upsilon_{K \wat a}+\Upsilon_I^{(\theta)}\partial_J\Upsilon^{(\theta)}_K\Big)  \lambda^{(1)I\wat{p}} \gamma^\mu \lambda^{(2)K}_{\wat{p}}\partial_{\mu}X^J \Big] \,.
\ea
\ee


\subsection{4d Sigma-Model into the Nahm Moduli Space}

Finally, we need to integrate out the gauge field and the scalars $\varphi, \bar\varphi$, which is done in appendix \ref{app:GaugeField2}. The conclusion is that, in addition to giving the standard kinetic term for the scalars, this covariantizes the fermion action and results in a quartic fermion interaction that depends on the Riemann tensor of the moduli space. In summary we find the action
\be \label{4dTwistedLambdas}
\ba 
S = \frac{1}{r\ell}\int d^4x \gf \left[\frac{1}{4}
G_{IJ}\partial_{\mu}X^I \partial^{\mu} X^J + \right. & 8iG_{IJ}\lambda^{(1)I\wat p} \gamma^{\mu}  \mathcal{D}_{\mu}\lambda^{(2)J}_{\wat p}  \\
&\left.
- 32 R_{IJKL} \left(\lambda^{(1)I\wat p}\lambda^{(1)J}_{\wat p}\right) \left(\lambda^{(2)K\wat q}\lambda^{(2)L}_{\wat q}\right)  \right] \,,
\ea \ee                                                                                                                                                                                                                                                                                                                                                                                                                                                                                                                                                                                                                                                                                                                                                                                                                                                                                                                                                                                                                                                                                                                                                                                                                                                                                                                                                                                                                                                                                                                                                                                                                                                                                                                                                                                                                                                                                                                                                                                                                                                                                                                                                                                                                                                                                                                                                                                                                                                                                                                                                                                                                                                                                                                                                                                                                                                                                                                                                                                                                                                                                                                                                                                                                                                                                                                                                                                                                                                                                                                                                                                                                                                                                                                                                                                                                                                                                                                                                                                                                                                                                                                                                                                                   
where $\mathcal{D}_{\mu}\lambda^{(2)I}_{ \wat p} = \partial_{\mu}\lambda^{(2)I}_{ \wat p} + \lambda^{(2) J}_{\wat p} \Gamma^{I}_{JK}\partial_{\mu}X^K$. 
The final step is to decompose the spinors  $\lambda^{(i)}$, as explained in appendix \ref{app:ConvSpinors}, into 4d Weyl spinors 
\be \label{LambdaPsi}
\lambda^{(1) I}_{\wat p} =  \frac{1}{4}\binom{\xi_{\wat p}^{(1) I }}{0} \,, \quad \lambda^{(2) I}_{\wat p} =  \frac{1}{4} \binom{0}{\xi_{\wat p}^{(2) I }}\,,
\ee
obeying the reality conditions
\be 
(\xi^{(1) p})^* = \xi^{(2)}_{\dot{ p}} \, , \quad (\xi^{(2) \dot p})^* = \xi^{(1)}_p \,,
\ee
and the constraint 
\be \label{FermionUntwistedConXi}
\omega^{\wat aI}{}_J\xi^{(i) J}_{\wat p} = i(\sigma^{\wat a})_{\wat p}^{\wat q}\xi^{(i)I}_{\wat q}\,.
\ee
The 4d sigma-model action from flat $M_4$ into the monopole moduli space $\mathcal{M}_k$ is then given by
\be \label{4dUntwisted}
\ba 
S_{4d} =  \frac{1}{4r\ell}\int d^4x \gf &
 \left[ 
G_{IJ}\left(\partial_{\mu}X^I \partial^{\mu} X^J -2 i\xi^{(1)I\wat p} \sigma^{\mu}  \mathcal{D}_{\mu}\xi^{(2)J}_{\wat p}\right)  \right. \cr 
& \left.\qquad \qquad \qquad 
-\frac 12 R_{IJKL}\xi^{(1)I\wat p} \xi^{(1)J}_{\wat p}\xi^{(2)K}{}^{\wat q}\xi^{(2)L}_{\wat q}  \right] \,.
\ea \ee                                                                                                                                                                                                                                                                                                                                                                                                                                                                                                                                                                                                                                                                                                                                                                                                                                                                                                                                                                                                                                                                                                                                                                                                                                                                                                                                                                                                                                                                                                                                                                                                                                                                                                                                                                                                                                                                                                                                                                                                                                                                                                                                                                                                                                                                                                                                                                                                                                                                                                                                                                                                                                                                                                                                                                                                                                                                                                                                                                                                                                                                                                                                                                                                                                                                                                                                                                                                                                                                                                                                                                                                                                                                                                                                                                                                                                                                                                                                                                                                                                                                                                                                                                                                   
The supersymmetry transformations are
\be \ba 
\delta X^I &=- i\left( \epsilon^{(2) \wat p} \xi^{(1)I}_{\wat p} + \epsilon^{(1) \wat p}\xi^{(2)I}_{\wat p}\right) \\
\delta\xi^{(1) I}_{\wat p} &=  \frac{1}{4}\left(\partial_{\mu}X^I\sigma^{\mu}\epsilon^{(1)}_{\wat p}-i\omega^{\wat a I}{}_{J}(\sigma_{\wat a})_{\wat p}^{\wat q}\partial_{\mu}X^J \sigma^{\mu}\epsilon_{\wat q}^{(1)}\right) -\Gamma^I_{JK}\delta X^J\xi^{(1)K}_{\wat p} \\
\delta\xi^{(2) I}_{\wat p} &= -\frac{1}{4}\left(\partial_{\mu}X^I\bar{\sigma}^{\mu}\epsilon^{(2)}_{\wat p}-i\omega^{\wat a I}{}_{J}(\sigma_{\wat a})_{\wat p}^{\wat q}\partial_{\mu}X^J \bar{\sigma}^{\mu}\epsilon_{\wat q}^{(2)}\right) -\Gamma^I_{JK}\delta X^J\xi^{(2)K}_{\wat p} \,.
\ea \ee
We have thus shown, that  the M5-brane theory reduced on an $S^2$ gives rise to a four-dimensional sigma-model with $N=2$ supersymmetry, based on maps from $\bbR^4$ into the moduli space  $\cM_k$ of Nahm's equations (with $\varrho= [k] $ boundary conditions).


\subsection{Relation to the Bagger-Witten Model}

There is an equivalent description of the sigma-model in \eqref{4dUntwisted}, which relates it to the models 
in \cite{AlvarezGaume:1981hm,Bagger:1983tt}. In this alternative description we make use of the reduced holonomy of the Hyper-K\"ahler target $\cM_k$. We will consider an $(Sp(k) \times Sp(1))/\mathbb{Z}_2$ subgroup of $SO(4k)$, under which the complexified tangent bundle of a Hyper-K\"ahler space decomposes into a rank $2k$ vector bundle $V$ and a rank 2 trivial bundle $S$. The index $I$ decomposes under this into $i\wat{p}$, where $i=1,\cdots,2k$ labels the ${2k}$-dimensional representation of $\mathfrak{sp}(k)$ and $\wat{p}=1,2$ is the doublet index of $\mathfrak{sp}(1) = \mathfrak{su}(2)_R$. The map $I \rightarrow i \wat p$ is realized by the invariant tensors $f^{i \wat p}{}_I$ \cite{MR2920151}, which satisfy
\be 
f^{i \wat p}{}_I f^J{}_{i \wat p} = \delta^I_J\,, \quad f^{i \wat p}{}_I \, f^I{}_{j \wat q} = \delta^i_j \delta^{\wat p}_{\wat q}\,, \quad 
2f^{i \wat p}{}_I \, f^J{}_{i \wat q} = \delta^{I}_{J} \delta^{\wat p}_{\wat q} + i \omega^{\wat a}_I{}^J (\sigma_{\wat a})_{\wat q}^{\wat p}\,.
\ee
The alternative description of the sigma-model is obtained by defining the fields
\be \label{UntwistedMap}
\xi^{(1) i}  \equiv \half f^{i \wat p}{}_{I}\,  \xi^{(1) I}_{\wat p}\,, \qquad \xi^{(2) i}  \equiv \half f^{i \wat p}{}_{I} \, \xi^{(2) I}_{\wat p}\,.
\ee 
which can be inverted, by using the constraint on the fermions \eqref{FermionUntwistedConXi} 
\be \label{XiConstraints}
\xi^{(1) I}_{\wat p} = f^{I}{}_{i \wat p} \, \xi^{(1) i}\,, \qquad \xi^{(2) I}_{\wat p} = f^{I}{}_{i \wat p} \, \xi^{(2) i}\,.
\ee
Using this decomposition the 4d untwisted sigma-model action into the monopole moduli space $\mathcal{M}_k$ can be re-expressed in terms of the fermionic fields \eqref{UntwistedMap}
 \be\label{Final4dUntwisted}
S = \frac{1}{r\ell}\int d^4x  \gf \, \left[\frac{1}{4}G_{IJ } \partial_{\mu} X^{I}\,\partial^{\mu}X^{J} - ig_{ij} \xi^{(1)i} \sigma^\mu D_\mu  \xi^{(2)j} - \frac{1}{4} W_{ijkl} (\xi^{(1)i}\xi^{(1)j})(\xi^{(2)k}\xi^{(2)l})  \right]  \,,
 \ee
where the covariant derivative is
\be 
D_{\mu}\xi^{(2)i} = \partial_{\mu}\xi^{(2)i}+ \xi^{(2) j} w_{Ij}{}^{i}\partial_{\mu}X^I \,.
\ee
The tensors $w_{I j}{}^i$ and $W_{ijkl}$ are the $Sp(k)$ connection on $V$ and the totally symmetric curvature tensor, respectively. These are expressed in terms of the Christoffel connection and Riemann tensor as
\be \ba
w_{I i}{}^j &= \half f^{j \wat p}{}_{J}\left(\partial_I f^J{}_{ i\wat p} + \Gamma_{IK}^J  f^K{}_{i\wat p}\right)\\
W_{ijkl} &= \half f^{I \wat p}{}_i f^J{}_{\wat p j} f^{K \wat q}{}_k f^L{}_{ \wat q l} R_{IJKL}\,.
\ea \ee
The supersymmetry transformations are 
\be \ba 
\delta X^I &= -i\epsilon^{(2) \wat p} f^I{}_{i \wat p} \xi^{(1) i} - i \epsilon^{(1) \wat p} f^I{}_{i \wat p} \xi^{(2)i}\\
\delta \xi^{(1) i} &=\half f^{i \wat p}{}_I \partial_{\mu} X^I \sigma^\mu \epsilon_{\wat p}^{(1)} - w_{I j}{}^i \delta X^I \xi^{(1)j} \\
\delta \xi^{(2) i} &=-\half f^{i \wat p}{}_I \partial_{\mu} X^I \bar{\sigma}^\mu \epsilon_{\wat p}^{(2)} - w_{I j}{}^i \delta X^I \xi^{(2)j}  \,.
\ea \ee
It is natural to ask how this sigma-model can be extended to general, oriented four-manifolds $M_4$. 
Using the topological twist 1 in section \ref{ssec:TwistsM4}, we will now consider this generalization.

\section{4d Topological Sigma-Models: Hyper-K\"ahler $M_4$}
\label{sec:4dHK}

So far we have discussed the five-dimensional theory on flat $I \times \bbR^4 $, where $I$ is the $\theta$ interval, reducing it to a sigma-model in four-dimensional flat space. 
The goal in the following is to define a 4d topological sigma-model on a general four-manifold. 
We first describe the twist in terms of the 4d theory in section \ref{subsec:TopoTwist}. 

As we shall see, for the target space a Hyper-K\"ahler manifold, as is the case for the Nahm moduli space, and general gauge group, we determine a general form of the sigma-model for the case of Hyper-K\"ahler $M_4$. For compact $M_4$, this comprises $T^4$ and $K3$ varieties. We will discuss the special reductions for the abelian case and the two-monopole case for general $M_4$ later on. 


\subsection{Topological Twist}
\label{subsec:TopoTwist}

Twist 1 in section \ref{ssec:TwistsM4} was formulated for the 6d theory. We now briefly summarize how this twist acts in 4d.  From now on we switch to Euclidean signature \footnote{For this twist we change from Lorentzian to Euclidean signature. In what follows $\gamma_0$ as defined in appendix \ref{app:ConvSpinors} is replaced with $\gamma_{0'} = i\gamma_0$, where the prime will be omitted.}. 

Recall, that in 6d, we twist the $\su(2)_\ell \subset \su(2)_\ell \oplus \su(2)_r$ of the 4d Lorentz algebra with the $\mathfrak{su}(2)_R \subset \su(2)_R \oplus\so(2)_R \subset \mathfrak{sp}(4)_R$. 
From the point of view of the 4d theory, we start with the R-symmetry $\mathfrak{su}(2)_R$ and twist this with the Lorentz symmetry of $M_4$, which generically is $\mathfrak{so}(4)_L \cong \mathfrak{su}(2)_\ell \oplus \mathfrak{su}(2)_r$, resulting in
\be
\mathfrak{g}_{4d}= \mathfrak{su}(2)_R \oplus \mathfrak{so}(4)_L \quad \rightarrow \quad \mathfrak{g}_{\rm twist} \ = \ \mathfrak{su}(2)_{\rm twist} \oplus \mathfrak{su}(2)_r  \,.
\ee
In terms of 4d representations, $\epsilon^{(1)}_{\wat p}$ and $\epsilon^{(2)}_{\wat p}$ are Weyl spinors of positive and negative chirality respectively. We adopt the convention that negative/positive chirality spinors correspond to doublets of $\su(2)_\ell / \su(2)_r$ respectively. After the twisting, $\epsilon^{(2)}_{\wat p}$ has one scalar component under $\su(2)_{\rm twist} \oplus \su(2)_r$, which is selected by the projections
\be 
\ba
& (\gamma_{0a} \delta^{\wat q}_{\wat p} +  i (\sigma_{\wat a})^{\wat q}_{\wat p})\epsilon^{(2)}_{\wat q} = 0 \,, \quad a\simeq \wat a = 1,2,3 \,,
\ea
\ee
where the indices $a$ and $\wat a$ are identified in the twisted theory. The spinor $\epsilon^{(2) \wat p}$ parametrizes the preserved supercharge and can be decomposed as 
\be 
\epsilon^{(2) \wat p}  = u \, \tilde{\epsilon}^{\wat p} \,,
\ee
where $u$ is a complex Grassmann-odd parameter and $\tilde{\epsilon}^{\wat p}$ is a Grassmann-even spinor normalized so that
\be 
\tilde{\epsilon}^{\wat p}\tilde{\epsilon}_{\wat p} = 1\,.
\ee
We can associate the $\mathfrak{u}(1)_R$ charge $-1$ to the parameter $u$ and consider $\tilde{\epsilon}^{\wat p}$ as uncharged.

The $\mathfrak{su}(2)_R$ R-symmetry with which we twist rotates the complex structures of the target and therefore is identified with the $\mathfrak{sp}(1) \subset \mathfrak{so}(4k)$ of the Hyper-K\"ahler target. This means that $SU(2)_R/\bbZ_2$ is mapped to an $SO(3)$ isometry of the metric on $\cM_k$. In order to do the twist one needs to know how the coordinates $X^I$ transform under this $\mathfrak{sp}(1) \equiv \mathfrak{su}(2)_R$. For the monopole moduli space with charge $1$ and $2$, $\cM_1$ and $\cM_2$, where the explicit metric on the moduli space is known, the coordinates split into two sets transforming respectively in the trivial and adjoint representation of $\mathfrak{su}(2)_R$. This suggests that this property could hold for moduli spaces $\cM_k$, with $k > 2$. Under the twist, the coordinates transforming in the adjoint of $\su(2)_R$ become self-dual two forms on $M_4$ and the resulting theory is a sigma-model, whose bosonic fields are maps into a reduced target space and self-dual two-forms. We shall study the $\cM_1$ and $\cM_2$ cases in section \ref{sec:TwoForms}.

A simplification occurs when the bundle of self-dual two-forms on $M_4$ is trivial i.e. when $M_4$ is Hyper-K\"ahler. In this case, all the coordinates transform as scalars on $M_4$ after the twist and therefore the twist can be performed without knowledge of the metric on $\mathcal{M}_k$. In this situation, the twisting procedure is simply a re-writing of the theory, making manifest the transformation of the fields under the new Lorentz group. This is done in the next section and gives a topological sigma-model on Hyper-K\"ahler $M_4$.


\subsection{Topological Sigma-Model for Hyper-K\"ahler $M_4$}
\label{sec:TopSigmaHK}

The 4d sigma-model into the Nahm moduli space (\ref{4dUntwisted}) can be topologically twisted for Hyper-K\"ahler $M_4$. We now show that this reduces to the 
4d topological theory by Anselmi and Fr\`e \cite{Anselmi:1993wm}, for the special target space given by the moduli space of Nahm's equations. This topological theory describes {\it tri-holomorphic maps} from $M_4$ into $\cM_k$
\be
X = \{X^I \} : \qquad M_4 \ \to\  \cM_k\,,
\ee 
which satisfy the triholomorphicity constraint 
\begin{align}
\p_\mu X^I - (j^a)_\mu{}^{\nu} \p_\nu X^J \omega^a{}_J{}^I = 0 \,,
\end{align}
where the index $a=1,2,3$ is summed over and $j^a$ and $\omega^a$ are triplets of complex structures on $M_4$ and $\cM_k$ respectively, which define the Hyper-K\"ahler structures.
We will also comment in section \ref{subsec:Relto5d} on how this can be obtained by first topologically twisting the 5d SYM theory, and then dimensionally reducing this to 4d. This alternative derivation from the twisted 5d SYM theory can be found in appendix \ref{sec:TopSigM4}.


We now turn to the topological twisting of the 4d sigma-model into the Nahm moduli space (\ref{4dUntwisted}), by the twist  of section \ref{subsec:TopoTwist}. The fields of the 4d sigma-model become forms on $M_4$, with the degree depending on their transformations under $\mathfrak{g}_{\rm twist} $
\be
\begin{array}{c|c|c|c}
\hbox{Field} & \hbox{$\mathfrak{g}_{\rm 4d}$} &  \hbox{$\mathfrak{g}_{\rm twist}$} & \hbox{Twisted Field}\cr \hline
X^I &  ({\bf 1}, {\bf 1}, {\bf 1})& ({\bf 1}, {\bf 1}) & X^I \cr 
\xi_{\wat p}^{(1) I p} & ({\bf 2}, {\bf 2}, {\bf 1})&({\bf 1} \oplus {\bf 3}, {\bf 1}) & \lambda^I, \chi^I_{\mu\nu}\cr 
\xi_{\wat p}^{(2) I \dot{p}} &({\bf 2}, {\bf 1}, {\bf 2}) &({\bf 2}, {\bf 1}) & \kappa_\mu^I
\end{array}
\ee
Despite the fact that the index $I$ transforms non-trivially under the R-symmetry $SO(3)_R$, this will not play a role in the twist for the Hyper-K\"ahler four-manifold $M_4$: the holonomy is reduced to $\su(2)_r$ and the $\su(2)_\ell$ connection that we twist with vanishes. 
To be even more concrete, the covariant derivatives acting on fields with an index $I$ will not pick up any $\su(2)_{\rm twist}$ connection because the connection vanishes, so we may treat $I$ as an external index. This is of course not true for non-Hyper-K\"ahler $M_4$.

The most general decomposition of the spinors into twisted fields is given by
\be \ba 
\xi^{(1) I}_{\wat p} &= \left( \lambda^I + \frac{1}{4} \sigma^{\mu \nu} \chi_{\mu \nu}^I\right) \tilde{\epsilon}_{\wat p} \\
\xi^{(2) I}_{\wat p} &= \bar{\sigma}^{\mu} \kappa_{\mu}^I \tilde{\epsilon}_{\wat p} \,,
\ea \ee
where the Grassmann-odd fields $\lambda^I, \chi_{\mu \nu}^I, \kappa_{\mu}^I$ are respectively a scalar, a self-dual two-form and a one-form, valued in the pull-back of the tangent bundle of the target space $X^*T\cM_k$.
However the components of $\xi^{(i)I}_{\wat p}$ are not all independent as they satisfy the constraint \eqref{FermionUntwistedConXi}. This constraint on the components of  $\xi^{(i) I}_{\wat p}$ translates into 
\be \ba \label{TwistingCons}
\omega_{\mu \nu}{}^I{}_J \lambda^J &= \chi_{\mu \nu}^I\,,\\ 
\omega_{\mu \nu}{}^I{}_J \kappa^{\nu J} &= -3\kappa_{\mu}^I\,,
\ea \ee
where  $\omega_{\mu \nu}{}^I{}_J \equiv - (j^{\wat a})_{\mu \nu} \omega^{\wat a I}{}_J$. As the self-dual two-form $\chi_{\mu \nu}^I$ is not an independent degree of freedom we shall consider the decomposition of $\xi^{(1) I}_{\wat p}$ just in terms of the fermionic scalar $\lambda^I$, with a convenient normalization,
\be \ba 
\xi^{(1) I}_{\wat p} &= i\left( \lambda^I + \frac{1}{4} \sigma^{\mu \nu} \omega_{\mu \nu}{}^I{}_J \lambda^J \right) \tilde{\epsilon}_{\wat p} \\
\xi^{(2) I}_{\wat p} &= -\frac{1}{4}\bar{\sigma}^{\mu} \kappa_{\mu}^I \tilde{\epsilon}_{\wat p} \,.
\ea \ee
Note that this decomposition of $\xi^{(1) I}_{\wat p}$ solves the constraint \eqref{FermionUntwistedConXi} automatically, and thus all components of $\lambda^I$ are independent. However, this is not the case for $\xi^{(2) I}_{\wat p}$ and we need to impose upon the fermionic one-form $\kappa_{\mu}^I$ the constraint \eqref{TwistingCons}, which can be re-expressed as
\be
\kappa^I_\mu + \frac 13 (j^a)_\mu{}^\nu \kappa^J_\nu (\omega^a){}_{J}{}^I = 0 \,.
\label{kappaConstr}
\ee
The action in terms of the twisted fields takes the form
\be \label{4dHKSigmaModel}
\ba 
S_{HK} = \frac{1}{4r\ell}\int d^4x \gf \left[
G_{IJ}\partial_{\mu}X^I \partial^{\mu} X^J -2 G_{IJ}g^{\mu \nu}\lambda^{I} \mathcal{D}_{\mu}\kappa_{\nu}^J  
+\frac{1}{8} R_{IJKL}\kappa^I_{\mu} \kappa^J_{\nu} \lambda^K \lambda^L\right] \,,
\ea \ee                                                                                                                                                                                                                                                                                                                                                                                                                                                                                                                                                                                                                                                                                                                                                                                                                                                                                                                                                                                                                                                                                                                                                                                                                                                                                                                                                                                                                                                                                                                                                                                                                                                                                                                                                                                                                                                                                                                                                                                                                                                                                                                                                                                                                                                                                                                                                                                                                                                                                                                                                                                                                                                                                                                                                                                                                                                                                                                                                                                                                                                                                                                                                                                                                                                                                                                                                                                                                                                                                                                                                                                                                                                                                                                                                                                                                                                                                                                                                                                                                                                                                                                                                                                                   
and is invariant under the supersymmetry transformations
\be \ba 
\delta X^I &= u\lambda^I \\
\delta \lambda^I &= 0 \\
\delta \kappa_{\mu}^I &= u\left( \partial_{\mu}X^I - \omega_{\mu \nu}{}^I{}_J \partial^{\nu}X^J\right) - \Gamma^I_{JK}\delta X^J \kappa_{\mu}^K \,.
\ea \ee
This is precisely the form of the topological sigma-model of \cite{Anselmi:1993wm} for  Hyper-K\"ahler $M_4$. The action takes a simpler form than in the model presented in  \cite{Anselmi:1993wm} since the target space $\mathcal{M}_{k}$ is also Hyper-K\"ahler (i.e. has a  covariantly constant quaternionic structure).

The topological BRST transformation $Q$ (with $\delta_u = uQ$) squares to zero $Q^2 = 0$ on-shell. To make the algebra close off-shell, we can introduce an auxiliary one-form $b^I_\mu$ valued in the pull-back of the tangent space to $\mathcal{M}_{k}$, $b \in \Gamma(X^* T\mathcal{M} \otimes \Omega^1)$ and satisfying the constraint
\be
b^I_\mu + \frac 13 (j^a)_\mu{}^\nu b^J_\nu (\omega^a){}_{J}{}^I = 0 \,.
\ee
We then define the BRST transformation to be
\be \ba 
Q X^I &=  \lambda^I \\
Q \lambda^I &= 0 \\
Q \kappa_{\mu}^I &=  b^I_\mu -     \Gamma^I_{JK}\lambda^J\kappa_{\mu}^K \\
Q b_\mu^I &=  \frac 12 R_{JK}{}^I{}_L \lambda^J \lambda^K \kappa^L_\mu - \Gamma^I_{JK} \lambda^J b^K_\mu \,.
\ea \ee
The action \eqref{4dHKSigmaModel} can then be recast in the form
 \be \label{FinalHK}
S_{HK}^{\rm off-shell} = S' - S_T\,.
 \ee
 where $S'$ and $S_T$ are $Q$-exact and topological, respectively, given by
\be \label{TopoAction2} \ba 
S' &=  Q \lp \frac{1}{2 r\ell}\int d^4x \gf \, G_{IJ} g^{\mu\nu} \kappa^{I}_\mu \lp \p_\nu X^J - \frac 18 b^J_\nu  \rp \rp \cr 
S_T &= \frac{1}{4r\ell} \int d^4x \gf \, (j^{a})^{\mu \nu} \omega^a_{IJ} \partial_{\mu}X^I \partial_{\nu}X^J\,.
\ea \ee
Integrating out $b^I_\mu$
\begin{align}
b^I_\mu &= \partial_{\mu}X^I - (j^a)_{\mu}{}^\nu \partial_{\nu} X^J \omega^{a}{}_J{}^I \,,
\end{align}
we recover the on-shell action \eqref{4dHKSigmaModel}.
The term $S_T$ is `topological', in the sense that it is invariant under Hyper-K\"ahler deformations, and can be written as 
\be
S_T= {1\over 2 r \ell} \int_{M_4} j^a \wedge X^*\omega^a \,,
\ee
where $X^*\omega^a$ is the pull-back of the K\"ahler forms on $\mathcal{M}_k$, and for Hyper-K\"ahler $M_4$, $j^a$ are the K\"ahler forms. From this form it is clear that the term is invariant under Hyper-K\"ahler deformations, but not deformations, that break the Hyper-K\"ahlerity.

Finally, to show that the theory is topological, meaning independent of continuous deformations of the metric (which preserve the Hyper-K\"ahler structure),  we must check that the energy-momentum tensor $T_{\mu\nu}$ associated with $S'$ part of the action is $Q$-exact. We find
\be
\ba
T_{\mu\nu} &\equiv \frac{2}{\sqrt g} \frac{\delta S'}{\delta g^{\mu\nu}} \ = \ 
G_{IJ} b_\mu^I (\p_\nu X^J - \frac 18 b^J_\nu ) + G_{IJ} b_\nu^I (\p_\mu X^J - \frac 18 b^J_\mu )  - g_{\mu\nu} \mathcal{L}' \,,
\ea
\ee
where $\mathcal{L}$ is the Lagrangian density in \eqref{TopoAction2}. This can be expressed as
\be
\ba
T_{\mu\nu} &= Q \left\lbrace 
G_{IJ} \kappa^{I}_\mu \lp \p_\nu X^J - \frac 18 b^J_\nu  \rp + G_{IJ} \kappa^{I}_\nu \lp \p_\mu X^J - \frac 18 b^J_\mu  \rp  - g_{\mu\nu} G_{IJ} \kappa^{I}{}^\rho \lp \p_\rho X^J - \frac 18 b^J_\rho  \rp \right\rbrace  \,.
\ea
\ee
Clearly it is of interest to study further properties of these theories, in particular observables, which will be postponed to future work. 
Some preliminary results for sigma-models that localize on tri-holomorphic maps have appeared in \cite{Anselmi:1993wm}, however only in terms of simplified setups, where the target is the same as $M_4$. 


\subsection{Relation to topologically twisted 5d SYM}
\label{subsec:Relto5d}

The topological sigma-model (\ref{4dHKSigmaModel}) for the Hyper-K\"ahler case, can also be obtained by first topologically twisting the 5d SYM theory on an interval obtained in section \ref{sec:5dTheory}, with the twist described in section \ref{subsec:TopoTwist}. The derivation is quite similar to the analysis in section \ref{sec:SigmaModNahm}, and we summarize the  salient points here. The details are provided for the interested reader in appendix \ref{sec:TopSigM4}. There, we also discuss the topological twist  1 in the context of the 5d SYM theory. The action for the bosonic fields, and some analysis of the boundary conditions in terms of Nahm data, has appeared in \cite{Witten:2011zz}. The supersymmetric version has appeared in \cite{Anderson:2012ck}, albeit without the supersymmetric boundary conditions.

The topologically twisted 5d SYM theory can be written in terms of the fields $B_{\mu\nu}$, which is a self-dual two-form defined in (\ref{Bdef}), a complex scalar field $\varphi$, the gauge field $A_\mu$ and fermions, which in terms of the twisted fields have the following decomposition 
\be 
\ba 
\rho^{(1)}_{+\wat p} &=\gamma^{\mu}\psi^{(1)}_{\mu}\tilde{\epsilon}_{\wat p} \cr 
\rho^{(1)}_{-\wat p} &=\left(\eta^{(1)} + \frac{1}{4} \gamma^{\mu \nu} \chi^{(1)}_{\mu \nu}\right)\tilde{\epsilon}_{\wat p} 
\ea\qquad 
\ba
\rho^{(2)}_{+\wat p} &=\gamma^{\mu}\psi^{(2)}_{\mu}\tilde{\epsilon}_{\wat p} \cr 
\rho^{(2)}_{-\wat p} &=\left(\eta^{(2)} + \frac{1}{4} \gamma^{\mu \nu} \chi^{(2)}_{\mu \nu}\right)\tilde{\epsilon}_{\wat p} \,.
\ea 
\ee
Nahm's equations in terms of the self-dual two-forms are
\be\label{BNahm}
D_\theta B_{\mu\nu} - \frac 12 [B_{\mu\rho}, B_{\nu}{}^{\rho}] = 0 \,.
\ee
The supersymmetric vacuum configurations which satisfy this, are again characterized in terms of maps into the moduli space of solutions to
the equations (\ref{BNahm}), which is the $k$-centered monopole moduli space, when $M_4$ is Hyper-K\"ahler. 
The 4d topological theory is obtained by expanding the fields $B_{\mu\nu}$, $A_\theta$ and the fermions in terms of coordinates on the moduli space, much like in section \ref{sec:SigmaModNahm}, and the resulting 4d topological 
sigma-model is precisely the one we obtained by twisting the flat space sigma-model in \eqref{4dHKSigmaModel}.



\section{Sigma-models with Self-dual Two-forms}
\label{sec:TwoForms}

Having understood the Hyper-K\"ahler $M_4$ case, we can finally turn to the case of general $M_4$. The reduction proceeds in the same way as for the Hyper-K\"ahler case, but the situation is somewhat complicated by the fact that part of the coordinates $X^I$ become sections of $\Omega^+_2(M_4)$, namely self-dual two-forms. 
We consider in detail the abelian case with target space $\cM_1 \simeq \bbR^3 \times S^1$ and the first non-trivial case, corresponding to the reduction of the 5d $U(2)$ theory, with target space $\cM_2 \simeq \bbR^3 \times \frac{S^1 \times \cM^0_2}{\bbZ_2}$, where $\cM^0_2$ is the Atiyah-Hitchin manifold. 

In the case of an arbitrary (oriented) four-manifold $M_4$, there is no Hyper-K\"ahler structure, only an almost quaternionic structure \cite{SalamonBook}. One could anticipate dimensionally reducing the twisted 5d SYM theory, as discussed in section \ref{subsec:Relto5d} and appendix \ref{subsec:TopoTwist5d}.  However, this requires that Nahm's equations for the self-dual two-forms $B_{\mu\nu}$
\be
D_\theta B_{\mu\nu} - \frac 12 [B_{\mu\rho}, B_{\nu}{}^{\rho}] = 0 \,,
\ee
to be solved locally on patches in $M_4$ and the patching must be defined globally, according to the transformation of $B$ on overlaps. Generically this means that part of the mapping coordinates $X^I$ will transform from one patch to the other and therefore belong to non-trivial $SU(2)_{\ell}$ bundles over $M_4$. A similar situation appears in \cite{Bershadsky:1995vm} appendix B, when twisting the sigma-model into the Hitchin moduli space.  To understand precisely, which coordinates $X^I$ become sections of $SU(2)_{\ell}$ bundles on $M_4$, we require a detailed understanding of the metric on $\cM_k$ and the action of the $SU(2)_{\ell}$ isometries. In the following, we will address this in the case of $k=1,2$, where the metrics are known. 

We provide here the analysis in the case of the reduction of the abelian theory, as a warm-up, and then the reduction of the $U(2)$ theory, which is the first non-trivial case. In these cases we find that the four-dimensional theory is a topological sigma-model with part of the coordinates $X^I$ on the target space transforming as self-dual two-forms on $M_4$.


\subsection{Abelian Theory} 

Recall that the dimensional reduction on $S^2$ of the untwisted single M5-brane theory gives a free hyper-multiplet in $\mathbb{R}^{1,3}$. We shall now discuss this in the context of the topologically twisted theory on $S^2\times M_4$ and determine the sigma-model into the one-monopole moduli space $\mathcal{M}_{k=1} \cong \bbR^3 \times S^1$, with $\bbR^3$ the position of the center and $S^1$ parametrizing a phase angle. As the metric is known, we can identity the coordinates parametrising the position of the center as those which transform under the $\mathfrak{su}(2)_R$ and the twist gives a topological model for general $M_4$. In fact, we find the abelian version of a model in \cite{Kapustin:2010ag} in the context of 4d topological A-models. 
The 4d field content is the self-dual two-form $B_{\mu\nu}$, the scalar $\phi$ and (twisted) for the fermions, a scalar $\eta$, a vector $\psi_\mu $, and a self-dual two-form $\chi_{\mu\nu}$.

We begin by decomposing the target space index $I \rightarrow (a, \phi)$, with $a= 1, 2, 3$. 
Under this decomposition the constraints on the spinors $\xi^{(i) I}_{\wat p}$ can be solved as
\be \label{AbelianFermionSol}
\xi^{(i) \wat a}_{\wat p} = i (\sigma^a)_{\wat p}^{\wat q}\, \xi^{(i) \phi}_{\wat q} \,,
\ee
leaving only $\xi^{(i) \phi}_{\wat q} $ as the unconstrained fermions in the theory. 
Under the twist the fields become
\be
\begin{array}{c|c|c|c}
\hbox{Field} & \hbox{$\mathfrak{g}_{\rm 4d}$} &  \hbox{$\mathfrak{g}_{\rm twist}$} & \hbox{Twisted Field}\cr \hline
X^\phi &  ({\bf 1}, {\bf 1}, {\bf 1})& ({\bf 1}, {\bf 1}) & \phi \cr 
X^{ a}  &  ({\bf 3}, {\bf 1}, {\bf 1})& ({\bf 3}, {\bf 1}) & B_{\mu \nu} \cr
\xi^{(1) \phi}_{\wat p} & ({\bf 2}, {\bf 2}, {\bf 1})& ({\bf 1}\oplus {\bf 3}, {\bf 1}) & \eta, \chi_{\mu \nu} \cr
\xi^{(2) \phi}_{\wat p} & ({\bf 2}, {\bf 1}, {\bf 2})& ( {\bf 2}, {\bf 2}) & \psi_{\mu}
\end{array}
\ee
where the twisted fermions are obtained from the decompositions
\be \ba 
\xi^{(1) \phi}_{\wat p} &= i\left( \eta + \frac{1}{4} \sigma^{\mu \nu} \chi_{\mu \nu}\right) \tilde{\epsilon}_{\wat p} \\
\xi^{(2) \phi}_{\wat p} &= -\frac{1}{4}\bar{\sigma}^{\mu} \psi_{\mu} \tilde{\epsilon}_{\wat p} \,.
\ea \ee
The scalars $X^{a}$ are decomposed in terms of the self-dual two-form $B_{\mu \nu}$ by making use of the invariant tensors $j^a_{\mu \nu}$
\be
B_{\mu \nu} = -j^a_{\mu \nu} \varphi^a \equiv \frac{i}{2}(\sigma_{\mu \nu})_{\wat p \wat q} \varphi^{\wat p \wat q} \,.
\ee
The action for the $k=1$ topological sigma-model from flat space into the monopole moduli space  $\mathcal{M}_1$ is then
 \be 
 S_{\mathcal{M}_1} =  \frac{1}{4r\ell}\int d^4x \gf ( \partial_{\mu}\phi \partial^{\mu}\phi + \frac{1}{4}\partial_{\mu} B_{\rho \sigma} \partial^{\mu} B^{\rho \sigma} - 2 \psi^{\mu}\partial_{\mu}\eta +2 \psi_{\mu} \partial_{\nu} \chi^{\mu \nu} )\,,
\ee 
and it is invariant the supersymmetry transformations 
\be \ba 
 \delta \phi &= u \eta \cr 
 \delta B_{\mu\nu} &= u \chi_{\mu\nu} \cr 
 \delta \eta = \delta \chi& =0 \cr 
 \delta \psi_\mu &= u (\partial_\mu \phi + \partial^\nu B_{\nu\mu})  \,.
\ea \ee
To show that this action is topological we introduce the auxiliary field 
\be
P_\mu = \partial_\mu \phi + \partial^\nu B_{\nu\mu} \,,
\ee
so that $\delta P_\mu =0$ and $\delta \psi_\mu = u P_\mu$.
The action can be written as the sum of a $Q$-exact term and a topological term by noting that $\delta_u = u Q$ 
\be\label{AbelianTerm}
S_{\mathcal{M}_1} = Q \mathcal{V} +{1\over 2 r \ell }\int_{M_4}  d\phi \wedge d B \,,
\ee
where
\be
\mathcal{V}=  {1\over 4 r \ell}  \int_{M_4} d^4x \gf \left(- \psi^\mu P_\mu + 2 \psi^\mu (\partial_\mu \phi + \partial^\nu B_{\nu \mu}) \right)  \,.
\ee
For $M_4$ without boundary, the second term in (\ref{AbelianTerm}) vanishes upon integrating by parts. This action can then be generalised to arbitrary $M_4$ by covariantising the derivatives, and add curvature terms 
\be 
 \mathcal{R}_{\mu \nu \rho \sigma} B^{\mu \nu} B^{\rho \sigma}\,, \quad \mathcal{R}_{\mu \nu} B^{\mu \nu}. 
\ee

The resulting theory is a (free) topological sigma-model based on the map $\phi: M_4 \to U(1)$, together with a self-dual two-form $B$ and fermionic fields and is given by
\be\label{4dAbelian}
S_{\mathcal{M}_1} = {1\over 4 r \ell} \int \left(\star d\phi \wedge d\phi + \star dB \wedge dB + 2 \psi \wedge (\star d\eta - d\chi) \right) \,.
\ee
The supersymmetric vacua, which are the saddle points of the action, satisfy 
\be
d\phi + \star d B=0 \,,
\ee
which implies that $\phi$ and $B$ are harmonic, and in particular then $d\phi=0$ and $dB=0$. Thus, $\phi$ is a constant scalar, and $B$ is a self-dual 2-form in a cohomology class of $H^{2,+} (M_4)$. 

Note, likewise one can obtain the same abelian theory starting with the 5d twisted theory for curved $M_4$ as discussed in section \ref{subsec:Relto5d} and appendix \ref{subsec:TopoTwist5d}. The reduction can be done straight forwardly, integrating out the fields $\psi^{(1)}$, $\chi^{(2)}$ and $\eta^{(2)}$, and taking the leading $1/r$ terms in the action. The match to the action in \eqref{4dAbelian} can be found by defining the fields in the 4d reduction as 
\be 
A_{\theta} \equiv \phi \,, \quad \eta\equiv \eta^{(1)}\,, \quad \psi_\mu \equiv 4 i \psi^{(2)}_\mu \,, \quad \chi_{\mu\nu}\equiv \chi_{\mu\nu}^{(1)} \,.
\ee
The scalar $\phi$ is actually defined in a gauge invariant way as $\phi = \int_0^\pi d\theta A_\theta$. Moreover it takes values in $i \bbR/\bbZ = U(1)$ \footnote{The factor $i$ is due to our conventions in which $A_\theta$ is purely imaginary.}, where the $\bbZ$-quotient is due to the large gauge transformations $\delta (\int A_\theta) = 2 \pi i n$, $n\in \bbZ$ \footnote{These transformations correspond to gauge group elements $g = e^{i \alpha(\theta)}$ with $\alpha(0) = 0$ and $\alpha(\pi) = 2\pi n$. The quantization of $n$ is required for $g$ to be trivial at the endpoints of the $\theta$ interval.}.


\subsection{$U(2)$ Theory and Atiyah-Hitchin Manifold}
\label{sec:AHCase}

In this section we study the simplest non-abelian case, corresponding to two M5-branes wrapped on $S^2$, or equivalently we study the reduction of the 5d $U(2)$ theory to 4d on an interval with Nahm pole boundary conditions. The flat 4d theory is given by a map into the 2-monopole moduli space $\cM_{2}$, with the action given in (\ref{4dUntwisted}). For the curved space theory we find a description in terms of a sigma-model into $S^1 \times \bbR_{\ge 0}$ supplemented by self-dual two-forms obeying some constraints.
We provide a detailed analysis of the geometrical data entering the sigma-model and we give the bosonic part of the topological sigma-model on an arbitrary four-manifold $M_4$.

The 2-monopole moduli space has been studied extensively in the literature (see for instance \cite{Atiyah:1988jp, Gibbons:1986df, Ivanov:1995cy, Dorey:1997ij, Hanany:2000fw}), starting with the work of Atiyah and Hitchin \cite{Atiyah:1988jp}. It has the product structure 
\be
\cM_2 = \bbR^3 \times \frac{S^1 \times \cM_{\rm AH}}{\bbZ_2}\,,
\label{M2space}
\ee
where $\bbR^3$ parametrizes the position of the center of mass of the 2-monopole system, and $\cM_{\rm AH}$ is the Atiyah-Hitchin manifold, which is a four-dimensional Hyper-K\"ahler manifold.
The metric on $\bbR^3 \times S^1$ is flat, it is associated to the abelian part of the theory $U(1) \subset U(2)$. The non-trivial geometry is carried by the Atiyah-Hitchin (AH) manifold \cite{Atiyah:1988jp}, whose Hyper-K\"ahler metric (AH metric) is given by
\be 
ds_{\rm AH}^2 = f(r)^2 dr^2 + a(r)^2 \sigma_1^2 + b(r)^2 \sigma_2^2 + c(r)^2 \sigma_3^2 \,,
\label{AHmetric1}
\ee
where $f,a,b,c$ are functions of $r \in \bbR_{\ge 0}$ and $\sigma_i$ are $SO(3)$ left invariant one-forms
\be
\ba
\sigma^1 &= - \sin \psi d\theta + \cos(\psi) \sin(\theta) d\phi \\
\sigma^2 &= \cos\psi d\theta + \sin(\psi) \sin(\theta) d\phi \\
\sigma^2 &= \cos(\theta) d\phi + d\psi \,,
\ea
\ee
with $0 \le \theta \le \pi$, $0 \le \phi \le 2\pi$ and $0 \le \psi < 2\pi$, with $\psi \sim \psi + 2\pi$. In addition the coordinates are subject to the following identifications \cite{Gibbons:1986df},
\be
(\theta,\phi,\psi) \sim (\pi - \theta, \phi + \pi, -\psi) \,, \quad  (\beta, \psi) \sim (\beta+ \pi , \psi + \pi) \,,
\label{Z2Quotients}
\ee
where the second identification accounts for the $\bbZ_2$ quotient in \eqref{M2space}, $\beta \in [0,2\pi]$ being the angle coordinate on the $S^1$.
The one-forms obey
\begin{align}
d\sigma^1 = \sigma^2 \wedge \sigma^3 \,,
\end{align}
and cyclic permutations of $1,2,3$.
The metric has an $SO(3) \equiv SO(3)_{\rm AH}$ isometry (leaving the one-form $\sigma^{1,2,3}$ invariant).
The function $f$ can be fixed to any desirable value by a reparametrization of $r$ (usual choices are $f = abc$ or $f= -b/r$). The functions $a,b,c$ obey the differential equation
\be 
\frac{da}{dr} = \frac{f}{2bc} \lp b^2 + c^2 -a^2 - 2bc \rp \,,
\ee
and cyclic permutations of $a,b,c$.
More details on the geometry of $\cM_{\rm AH}$, including the explicit Riemann tensor, can be found in \cite{Dorey:1997ij}.

The geometry is Hyper-K\"ahler and therefore possesses three complex structures $J^a$, $a=1,2,3$. These three complex structures transform as a triplet of the $SO(3)_{\rm AH}$ isometry. They extend naturally to complex structures on the full $\cM_2$ geometry and then transform as a triplet of $SO(3)_{\cM_2} = $diag$(SO(3)_{\rm AH} \times SO(3)_{\rm abel})$, where $SO(3)_{\rm abel}$ is the rotation group of $\bbR^3$.
 In the untwisted sigma-model \eqref{4dUntwisted}, this $SO(3)_{\cM_2}$ isometry is identified with the $SO(3)_R$ R-symmetry of the 4d theory,
\be
\underline{\textrm{Untwisted theory:}} \quad  SO(3)_{\cM_2} \simeq  SO(3)_R \,.
\ee
In the twisted sigma-model $ SO(3)_{\cM_2}$ gets identified with the $SO(3)_\ell$ left Lorentz rotations on the base manifold $M_4$,
\be
\underline{\textrm{Twisted theory:}} \quad  SO(3)_{\cM_2} \simeq  SO(3)_\ell \,.
\ee
Because of this identification, some coordinates on $\cM_2$ acquire $SO(3)_\ell$ Lorentz indices and become forms on $M_4$. To make the action of $SO(3)_\ell$ on the $\cM_2$ coordinates explicit and manageable, we need to choose appropriate coordinates. 

The treatment of the $\bbR^3 \times S^1$ coordinates is identical to the abelian case. We have coordinates $\phi^a$, $a=1,2,3$, parametrizing $\bbR^3$, transforming as a triplet of  $SO(3)_{\cM_2}$, and $\beta$ parametrizing $S^1$, scalar under $SO(3)_{\cM_2}$. Here and in the rest of the section we identify the indices $\wat a$ and $a$, namely we implement the 4d twisting which identifies $SO(3)_R$ and $SO(3)_\ell$.

The treatment of the coordinates on $\cM_{\rm AH}$ is more involved. Here we propose to introduce the coordinates $y^{i,a} \equiv y^a{}_{i}$, with $a,i=1,2,3$, forming an $SO(3)$ matrix $ (y^a{}_{i})  \in SO(3)$
\be
(y^a{}_{i}) = \left(
\begin{array}{ccc}
- \sin\psi\sin\phi + \cos\theta\cos\phi\cos\psi & - \cos\psi\sin\phi - \cos\theta\cos\phi\sin\psi &\cos\phi\sin\theta  \\
- \sin\psi\cos\phi - \cos\theta\sin\phi\cos\psi& - \cos\psi\cos\phi + \cos\theta\sin\phi\sin\psi &  -\sin\phi\sin\theta \\
 \cos\psi\sin\theta & -\sin\theta\sin\psi &-\cos\theta
\end{array}
\right) \,.
\label{yMatrix}
\ee
The $SO(3)_{\cM_2}$ isometries act on the matrix $(y^a{}_{i})$ by left matrix multiplication, so that the three vectors $y^{1,a}, y^{2,a},y^{3,a}$ transform as three triplets of $SO(3)_{\cM_2}$. The identifications \eqref{Z2Quotients} become 
\be
(\beta, y^{1,a},y^{2,a},y^{3,a}) \sim (\beta, y^{1,a}, - y^{2,a}, -y^{3,a}) \,, \quad (\beta, y^{1,a},y^{2,a},y^{3,a}) \sim (\beta + \pi , -y^{1,a}, - y^{2,a}, y^{3,a}) \,.
\label{Z2Quotients2}
\ee
We can express the AH metric in terms of the $y^{i,a}$ coordinates by using the relations
\be \ba 
(\sigma_1)^2 &= \frac{1}{2}(-dy^{1,a} dy^{1,a} + dy^{2,a} dy^{2,a} + dy^{3,a} dy^{3,a}) \\
(\sigma_2)^2 &= \frac{1}{2}(dy^{1,a} dy^{1,a} - dy^{2,a} dy^{2,a} + dy^{3,a} dy^{3,a}) \\
(\sigma_3)^2 &= \frac{1}{2}(dy^{1,a} dy^{1,a} + dy^{2,a} dy^{2,a} - dy^{3,a} dy^{3,a})  \,,
\ea \ee
where the index $a$ is summed over.
The AH metric \eqref{AHmetric1} is then understood as the pull-back of the metric
\be \label{AHmetric2}
\ti{ds}_{\rm AH}^2 = f^2 dr^2 + v_1 dy^{1,a} dy^{1,a} + v_2 dy^{2,a} dy^{2,a} + v_3 dy^{3,a} dy^{3,a} \,,
\ee
where 
\be
 v_1 = \frac{1}{2}(-a^2 + b^2 +c^2) \,, \ v_2 = \frac{1}{2}(a^2 - b^2 + c^2) \,, \ v_3 = \frac{1}{2}(a^2 + b^2 -c^2) \,.
\ee
As already mentioned the AH manifold $\cM_{\rm AH}$ admits three complex structures $J^a$, $a=1,2,3$, preserved by the above metric, and satisfying the quaternionic relations
\begin{align}
(J^a)^{I}{}_{J} (J^b)^{J}{}_{K} &= - \delta^{ab} \delta_K^I + \epsilon^{abc} (J^c)^{I}{}_{K} \,,
\end{align}
where the indices $I,J,K$ run over the four coordinates of the AH metric \footnote{This is a small abuse of notation compared to the convention of previous sections where $I,J,K$ run over all the coordinates on $\cM_k$.}.
Lowering an index with the AH metric $G_{IJ}$ \eqref{AHmetric1}, we define the three K\"ahler forms $(\Omega^a)_{IJ} = G_{IK} (J^a)^{K}{}_{J}$. These forms can be nicely expressed as the pull-back of the forms $\ti\Omega^a$ on the space parametrized by the $r,y^{i,a}$ coordinates:\footnote{We found the expression of one complex structure in \cite{Ivanov:1995cy} in terms of the Euler angles $\theta,\phi,\psi$ and worked out the re-writing in terms of $y^{i,a}$. The other two complex structures were easily obtained by cyclic permutation of the $y^{i,a}$ coordinates.}
\be
\ba
\ti\Omega^a = \frac 12 \epsilon^{a}{}_{bc} \Big[ & (-a + b+c) f y^{1,b} dr \wedge dy^{1,c} +  (a - b + c) f y^{2,b} dr \wedge dy^{2,c} +  (a + b-c) f y^{3,b} dr \wedge dy^{3,c}  \\
& - bc \, dy^{1,b}\wedge dy^{1,c} - ac \, dy^{2,b}\wedge dy^{2,c} - ab \, dy^{3,b}\wedge dy^{3,c} \Big] \,.
\label{Omega1}
\ea
\ee
These forms can be further simplified by using the functions $w_1 = bc$, $ w_2 = ca$, $w_3=ab$, which obey
\be 
\frac{dw_1}{dr} = - f \lp -a + b +c \rp \,, \quad \frac{dw_2}{dr} = - f \lp c + a - b \rp \,, \quad \frac{dw_3}{dr} = - f \lp b- c +a \rp \,.
\ee
 We obtain the nice expression
\be
\ti\Omega^a = - \frac 12 \epsilon^{a}{}_{bc}  \sum_{i=1,2,3} d(w_i y^{i,b}) \wedge dy^{i,c} \,.
\label{Omega2}
\ee
The pull-backs $\Omega^a$ are complex structures on $\cM_{\rm AH}$, hence they obey $d\Omega^a=0$.
This description of the complex structures is convenient, because it is much simpler than the expression in terms of the Euler angles $\theta,\phi,\psi$, but more importantly because it makes manifest the fact that the three K\"ahler forms $\Omega^a$, or the three complex structures $J^a$, transform as a triplet under the $SO(3)_{\cM_2}$ isometry.

After this preliminary work we can express the bosonic part of the flat space sigma-model action (\ref{4dUntwisted}) in terms of the new coordinates $\beta, \phi^a, r, y^{i,a}$, describing the maps $M_4 \to \cM_2$. Fixing $f(r)=1$ for simplicity, we obtain
\be
\ba
S_{\cM_2 ,{\rm bos}} &=  \frac{1}{4r\ell}\int d^4 x \gf \, \lp \p^\mu\beta \p_\mu\beta + \delta_{ab}\p^\mu\phi^a \p_\mu\phi^b + \p^\mu r \p_\mu r + \sum_{i=1}^3  v_i(r) \delta_{ab} \p^\mu y^{i,a} \p_\mu y^{i,b} \rp \,,
\ea
\label{S2monopR4}
\ee
where the sigma-model coordinates $y^{i,a}$ are constrained to form an $SO(3)$ matrix \eqref{yMatrix} and to obey \eqref{Z2Quotients2}. These constraints can be stated explicitly
\be
 \delta _{ab}y^{i,a} y^{j,b} = \delta^{ij}\,, \qquad  \epsilon_{abc} y^{1,a} y^{2,b} y^{3,c} = 1 \,. 
 \label{yConstr}
\ee
The coordinate $r$ is also constrained to be positive $r\geq 0$. 

Having described the (bosonic) action of the twisted theory on flat space we can easily derive the (bosonic) action on an arbitrary $M_4$. The fields $\beta, r$ are scalars on $M_4$, so their kinetic term is unchanged. The fields $\phi^a, y^{i,a}$ are triplets of $SO(3)_\ell$. They  are mapped to self-dual two-forms
\be
b_{\mu\nu} = - j^a_{\mu\nu} \phi^a \,, \quad y^i_{\mu\nu} = - j^a_{\mu\nu} y^{i,a} \,.
\ee
Their kinetic term gets covariantized by adding suitable curvature terms and we obtain
\be
\ba
S_{\cM_2 ,{\rm bos}} &=-  \frac{1}{4r\ell}\int  d\beta \wedge \star d\beta + db\wedge \star db  + dr\wedge \star dr + \sum_{i=1}^3 v_i(r) dy^i\wedge \star dy^i  \,.
\ea
\label{S2monopM4}
\ee
The constraints \eqref{yConstr} become $y^i_{\mu\nu} y^j{}^{\mu\nu} = 4 \delta^{ij}$ and $y^1{}_\mu{}^\nu y^2{}_\nu{}^{\rho} y^3{}_\rho{}^{\mu} = 4$.

The fermionic part of the action $S_{\cM_2 ,{\rm ferm}}$ that is obtained from the untwisted action (\ref{4dUntwisted}), is somewhat more involved, due to the presence of the four-Fermi interaction and the constraint \eqref{FermionUntwistedConXi} on the fields $\xi^{(i)I}$. From the abelian part of the $U(2)$ theory we obtain the fermionic field content of the abelian model \eqref{4dAbelian}. In the following we describe only the fermions related to $\mathcal{M}_{AH}$. 
Explicitly we can define the push-forward of the fermionic fields 
\be
\xi^{(1) \ti I p}_{\wat q} = \p_{I} y^{\ti I} \xi^{(1)  I p}_{\wat q}\,,\qquad 
\xi^{(2) \ti I \dot p}_{\wat q} = \p_{I} y^{\ti I} \xi^{(2)  I \dot p}_{\wat q}\,,
\ee
 where the index $\ti I$ runs over $r, (i,a)$.
In the twisted theory we identify the $\su(2)_\ell$ and $\su(2)_R$ doublet indices $q$ and $\wat q$ and the fermionic fields of the resulting sigma model are a vector $\kappa_\mu$, a scalar $\eta$ and  self-dual two-forms $\eta^{i,a} \sim \eta^i_{\mu\nu}$ satisfying the constraints
\be
 \delta_{ab} y^{i,a} \eta^{j,b} = -  \delta_{ab} y^{j,a} \eta^{i,b}\,, \qquad \sum_j y^{j,a} \eta^{j,b} = - \sum_j  y^{j,b} \eta^{j,a}\, .
 \ee
  The other fields appearing after the twisting are expect to be expressed in terms the above fields by solving the constraints \eqref{FermionUntwistedConXi}.  However the computation is rather involved and we do not provide an explicit expression here. 

The sigma-model we obtain seems to be different from the sigma-models studied in the literature so far. It is a sigma-model with target $S^1 \times \bbR_{\ge 0}$ with constrained self-dual two-forms.
To study this sigma-model, and in particular to show that it defines a topological theory, one would need to work out the details of the fermionic part of the Lagrangian and the action of the preserved supersymmetry (or BRST) transformation on the fields. We leave this for future work.

To conclude we can see how the bosonic action \eqref{S2monopM4} compares with the bosonic action of the topological model that we obtained for Hyper-K\"ahler $M_4$ \eqref{4dHKSigmaModel}. More precisely we would like to know how the action \eqref{S2monopM4} decomposes into $Q$-exact plus topological terms as in \eqref{TopoAction2}. For this we simply evaluate $S_T$ for the sigma-model into $\cM_2$, using the explicit form of the $\Omega^a$ \eqref{Omega2}. The terms involving the fields $\phi$ and $b$ vanish upon integration by parts as in the abelian case, assuming $M_4$ has no boundary. 
When the theory is defined on an generic four-manifold $\cM_4$, the remaining contribution is  
\begin{align}
S_T &=\frac{1}{2 r\ell}\int j^a \wedge X^\ast(\Omega^a) = \frac{1}{4 r\ell} \int j^a \wedge dx^{\mu} \wedge dx^{\nu} (\Omega^a)_{IJ}  D_{\mu} X^I D_\nu X^J + {\rm curv.} \,,
\end{align}
where $D_\mu$ is covariant with respect to the Christoffel connection and $SU(2)_\ell$ Lorentz rotations (in the tangent space), and ``+curv." denotes extra curvature terms, which appear when we consider a general curved $M_4$ and covariantize $S_T$.  Replacing $X^I \rightarrow r, y^{i,a}$ we obtain
\be
\ba
S_T &= -  \sum_{i=1}^3 \frac{1}{16 r\ell} \int  dx^{\mu} \wedge dx^{\nu} \wedge  dx^{\rho} \wedge dx^{\sigma} \epsilon^{abc} (j^a)_{\rho\sigma} D_\mu (w_i y^{b,i}) D_\nu y^{c,i}  + {\rm curv.}\\
&= - \sum_{i=1}^3 \frac{1}{16 r\ell} \int  d^4x \sqrt{g} \epsilon^{\mu\nu\rho\sigma} \, (j^b)_{\rho}{}^{\tau} (j^c)_{\tau\sigma}  D_\mu (w_i y^{b,i}) D_\nu y^{c,i}  + {\rm curv.} \\
&= - \sum_{i=1}^3 \frac{1}{16 r\ell} \int  d^4x \sqrt{g} \epsilon^{\mu\nu\rho\sigma}   D_\mu (w_i y^{i}_{\rho}{}^{\tau}) D_\nu y^{i}_{\tau\sigma}  + {\rm curv.} \\
&= \sum_{i=1}^3 \frac{1}{16 r\ell} \int  d^4x \sqrt{g} \epsilon^{\mu\nu\rho\sigma}   (w_i y^{i}_{\rho}{}^{\tau}) D_{[\mu} D_{\nu]} y^{i}_{\tau\sigma}  + {\rm curv.} \\
&= 0 \,.
\ea
\ee
From the third to the fourth line we have integrated by parts assuming $M_4$ has no boundary. The result on the fourth line can be recognized as containing only curvature terms (no derivatives on the fields $r, y^{i}_{\mu\nu}$) which must cancel each-other. This is necessary for supersymmetry to be preserved (since this term must be supersymmetric by itself).
We conclude that the sigma-model action \eqref{S2monopM4} must be $Q$-exact, without an extra topological term. Clearly, studying topological observables and further properties of this model are interesting directions for future investigations.

\section{Conclusions and Outlook}
\label{sec:Conc}

 In this paper we determined the dimensional reduction of the 6d $N=(0,2)$ theory on $S^2$, and found this to be a 4d sigma-model into the moduli space $\mathcal{M}_k$ of $k$-centered $SU(2)$ monopoles.  There are several exciting follow-up questions to consider: 
\begin{enumerate}

\item 4d-2d Correspondence: \\
Let us comment now on the proposed correspondence between 2d $N=(0,2)$ theories with a half-topological twist, and four-dimensional topological sigma-models into $\mathcal{M}_k$.
The setup we considered, much like the AGT and 3d-3d correspondences, implies a dependence of the 2d theory on the geometric properties of the four-manifold. In \cite{Gadde:2013sca} such a dictionary was setup in the context of the torus-reduction, which leads to the Vafa-Witten topological field theory in 4d. It would be very important to develop such a dictionary  in the present case.  From the point of view of the 2d theory, the twist along $M_4$ is the same, and thus the dictionary developed between the topological data of $M_4$ and matter content of the 2d theory will apply here as well. The key difference is that we consider this theory on a two-sphere, and the corresponding `dual' is not the Vafa-Witten theory, but the topological sigma-model into the Nahm moduli space. 

\item Observables in 2d $(0,2)$ theories:\\
Recently much progress has been made in 2d $(0,2)$ theories, both in constructing new classes of such theories 
\cite{Gadde:2013sca, Franco:2015tna, Schafer-Nameki:2016cfr, Apruzzi:2016iac} as well as studying anomalies \cite{Benini:2013cda} and computing correlation functions using localization \cite{Closset:2015ohf}. In particular, the localization results are based on deformations of $N=(2,2)$ theories and the associated localization computations in \cite{Closset:2015rna, Benini:2015noa}. The theories obtained in this paper from the compactification of the M5-brane theory do not necessarily have such a $(2,2)$ locus and thus extending the results on localization beyond the models studied in  \cite{Closset:2015ohf} would be most interesting. 

\item Observables in the 4d topological sigma-model: \\
An equally pressing question is to develop the theory on $M_4$, determine the cohomology of the twisted supercharges, and compute topological observables. For the case of Hyper-K\"ahler $M_4$, with the target also given by $M_4$, some observables of the topological sigma-model were discussed in \cite{Anselmi:1993wm}. However, we find ourselves in a more general situation, where the target is a  specific $4k$ dimensional Hyper-K\"ahler manifold. 
For the general $M_4$ case we clearly get a new class of theories, which have scalars and self-dual two-forms. The only place where a similar theory has thus far appeared that we are aware of, is in \cite{Kapustin:2010ag} in the context of 4d topological A-models. We have studied the topological sigma-models for $k=1,2$, and the explicit topological sigma-models for $k\ge 3$ remain unknown. It would  certainly be one of the most interesting directions to study these.  

\item Generalization to spheres with punctures:\\
The analysis in this paper for the sphere reduction can be easily generalized to spheres with two (general) punctures, i.e. with different boundary conditions for the scalars in the 5d SYM theory. We expect the 4d theory to be again a topological sigma-model, however, now into the moduli space of Nahm's equations with modified boundary conditions. 
Studying this case may provide further interesting examples of 4d topological field theories, which seem to be an interesting class of models to study in the future. 

\item Reduction to three-dimensions and 3d duality:\\
The four-dimensional sigma-model that we found by compactification of the 6d (0,2) theory on a two-sphere, can be further reduced on a circle $S^1$ to give rise to a three-dimensional sigma-model into the same $\cM_k$ target space. Similarly the twisted sigma-model on a manifold $S^1 \times M_3$ reduces along $S^1$ to a twisted sigma-model on $M_3$. On the other hand the compactification of the twisted 6d (0,2) $A_k$ theory on $S^2 \times S^1\times M_3$ can be performed by reducing first on $S^1$, obtaining 5d $\cN=2$ SYM theory on $S^2 \times M_3$, and then reducing on $S^2$. We expect this reduction to yield a different three-dimensional theory, which would be dual to the 3d sigma model into $\cM_k$, for $M_3=\bbR^3$, or twisted sigma model, for general $M_3$, that we studied in this paper. This new duality would be understood as an extension of 3d mirror symmetry \cite{Intriligator:1996ex} to topological theories. 
To our knowledge the reduction of 5d SYM on the topologically twisted $S^2$ has not been studied \footnote{Note that the reduction of 5d SYM on a two-sphere, but in a different supersymmetric background, has been considered in \cite{Lee:2013ida, Yagi:2013fda}, in relation with the 3d-3d correspondence \cite{Dimofte:2011ju,Dimofte:2011py}, and leads to an $SL(k,\bbC)$ Chern-Simons theory on $M_3$ with a complex Chern-Simons coupling.}. It would be very interesting to study it and to further investigate these ideas in the future.


\end{enumerate}


\subsection*{Acknowledgements}

We thank  Cyril Closset, Clay Cordova, Stefano Cremonesi, Tudor Dimofte, Neil Lambert, Simon Salamon, 
Owen Vaughan, Cristian Vergu and Timo Weigand for many insightful discussions. We also thank Damiano Sacco for collaboration at an earlier stage of the project. B.A. acknowledges support by the ERC Starting Grant N. 304806, ``The Gauge/Gravity Duality and Geometry in String Theory''. 
The work of SSN and JW is supported in part by STFC grant ST/J002798/1. 
SSN thanks the Aspen Center for Physics for hospitality during the course of this work. The Aspen Center for Physics is supported by National Science Foundation grant PHY-1066293.


\appendix

\section{Conventions and Spinor Decompositions}
\label{app:conventions}

\subsection{Indices}
\label{app:ConvIndices}

Our index conventions, for Lorentz and R-symmetry representations, which are used throughout the paper are summarized in the following  tables. Note that R-symmetry indices are always hatted. 
Furthermore, note that $\underline{m} = 1, \cdots, 8$, however only four components are independent for Weyl spinors in 6d.  

\begin{table}[ht]
\vspace{4mm}
\centering 
\begin{tabular}{c | c c c c c}
 Lorentz indices &  6d & 5d & 4d & 3d & 2d \\
 [0.5ex]
\hline
Curved vector & $\un{\mu}, \un{\nu}$ & $\mu', \nu'$ & $\mu , \nu$ & . & . \\
 [0.5ex]
 \hline
Flat vector  & $\un A, \un B$ & $A', B'$ & $A, B$ & $a, b$ \quad &$x, y$ \\
 [0.5ex]     
 \hline
 Spinors  & $\un{m}, \un{n}$ & $m', n'$ & $p, q ; \,  \dot p, \dot q$ & . & . \\
  &  ($\bf{4}$ of $\su(4)_L$) &  ($\bf{4}$ of $\mathfrak{sp}(4)_L$) & ($\bf{2}$  of $\su(2)_\ell$; {\bf 2} of $\su(2)_{r}$) & & 
\end{tabular}
\caption{Spacetime indices in various dimensions.
\label{table:IndicesLorentz}}
\end{table}

\begin{table}[ht]
\vspace{.5cm}
\centering 
\begin{tabular}{c | c c cc c}
&  $\so(5)_R$ &  $\mathfrak{sp}(4)_R$  & $\so(3)_R$ & $\su(2)_R$ & $\so(2)_R$  \\
 [0.5ex]
\hline
Index for the fundamental rep& $\wat A, \wat B$ & $\wat m, \wat n$ & $\wat a, \wat b$ & $\wat p, \wat q$ & $\wat x, \wat y $
\end{tabular}
\caption{R-symmetry indices.\label{table:IndicesRsym}}
\end{table}

\subsection{Gamma-matrices and Spinors: 6d, 5d and 4d}
\label{app:ConvSpinors}

We work with the mostly $+$ signature $(-,+,\cdots,+)$. 
The gamma matrices $\Gamma^{\un A}$ in 6d, $\gamma^{A'}$ in 5d and $\gamma^A$ in 4d, respectively, are defined as follows:
\begin{align}
\Gamma_1 &=  \ i \sigma_2 \otimes \bold{1}_2 \otimes \sigma_1  \ \  \equiv \ \gamma_1 \otimes \sigma_1  \nn\\
\Gamma_2 &= \ \sigma_1 \otimes \sigma_1 \otimes \sigma_1  \quad  \equiv \ \gamma_2 \otimes \sigma_1  \nn\\
\Gamma_3 &= \  \sigma_1 \otimes \sigma_2 \otimes \sigma_1  \quad  \equiv \ \gamma_3 \otimes \sigma_1  \nn\\
\Gamma_4 &= \ \sigma_1 \otimes \sigma_3 \otimes \sigma_1  \quad  \equiv \ \gamma_4 \otimes \sigma_1  \nn\\
\Gamma_5   &=  \  - \sigma_3 \otimes \bold{1}_2 \otimes \sigma_1  \,  \equiv \ \gamma_5\otimes\sigma_1  \nn\\
\Gamma_6 &=  \ \bold{1}_2 \otimes \bold{1}_2 \otimes\sigma_2\,,
\end{align}
with the Pauli matrices 
\begin{equation}
\sigma_{1}=\left(\begin{array}{cc} 0 & 1 \\ 1 & 0\end{array}\right)\,, \qquad  
\sigma_{2}=\left(\begin{array}{cc} 0 & -i \\ i & 0\end{array}\right)\,,\qquad 
\sigma_{3}=\left(\begin{array}{cc} 1 & 0 \\ 0 & -1\end{array}\right) \,.
\end{equation}
The 6d gamma matrices satisfy the Clifford algebra
\be 
\{ \Gamma_{\un A}, \Gamma_{\un B} \} = 2 \eta_{\un A \un B} \,,
\ee
and similarly for the 5d and 4d gamma matrices.

Futhermore we define
\be\Gamma^{\un A_1 \un A_2 \dots \un A_n} \equiv \Gamma^{[\un A_1 \un A_2 \dots \un A_n]} =  \frac{1}{n!} \sum_{w \in \CS_n} (-1)^{w} \Gamma^{\un A_{w(1)}}\  \Gamma^{\un A_{w(2)}} \dots \Gamma^{\un A_{w(n)}}\,,
\ee
and similarly for all types of gamma matrices.

The chirality matrix in 4d is $\gamma_5= - \sigma_3 \otimes \bold{1}_2$ and in 6d is defined by
\begin{align}
\Gamma_7 &= \Gamma^1 \Gamma^2 \cdots \Gamma^6 \ = \  \bold{1}_2 \otimes \bold{1}_2 \otimes\sigma_3 \,.
\end{align}
The charge conjugation matrices in 6d, 5d and 4d are defined by
\begin{align}
C_{\rm (6d)} &= \sigma_3 \otimes \sigma_2 \otimes \sigma_2 \quad \equiv \un C  \nn\\
C_{\rm (5d)} &= \ C_{\rm (4d)} \ = \ -i \, \sigma_3 \otimes \sigma_2  \quad \equiv C  \,.
\end{align}
They obey the identities
\begin{align}
\lp \Gamma^{\un A} \rp^T &=  \ -  \un C \Gamma^{\un A} \un C^{-1} \,, \quad  \un A = 1, \cdots , 6. \nn\\
\lp \gamma^{A'} \rp^T &= \ C \gamma^{A'} C^{-1} \,, \quad  A' = 1, \cdots , 5. \nn\\
\lp \gamma^{A} \rp^T &= \ C \gamma^{A} C^{-1} \,, \quad  A= 1, \cdots , 4.
\end{align}

To define irreducible spinors we also introduce the B-matrices
\begin{align}
B_{\rm (6d)} &= i \sigma_1 \otimes \sigma_2 \otimes \sigma_3  \nn\\
B_{\rm (5d)} &= \ B_{\rm (4d)} \ = \ i \, \sigma_1 \otimes \sigma_2 \,,
\end{align}
which satisfy
\begin{align}
\lp \Gamma^{\un A} \rp^\ast &=  \  B_{\rm (6d)} \Gamma^{\un A} B_{\rm (6d)}^{-1} \,, \quad  \un A = 1, \cdots , 6. \nn\\
\lp \gamma^{A'} \rp^\ast &= \ - B_{\rm (5d)} \gamma^{A'} B_{\rm (5d)}^{-1} \,, \quad  A' = 1, \cdots , 5. \nn\\
\lp \gamma^{A} \rp^\ast &= \ - B_{\rm (4d)} \gamma^{A} B_{\rm (4d)}^{-1} \,, \quad  A= 1, \cdots , 4.
\end{align}

The 6d Dirac spinors have eight complex components. Irreducible spinors have a definite chirality and have only four complex components. For instance a spinor $\rho$ of positive chirality satisfies $\Gamma_7 \rho = \rho$. 
Similarly Dirac spinors in 4d have four complex components and Weyl spinors obey a chirality projection, for instance $\gamma_5 \psi = \psi$ for positive chirality, and have two complex components. The components of positive and negative, chirality spinors in 4d are denoted with the index $\dot p=1,2$ and  $p = 1,2$, respectively.

The indices of Weyl spinors in 6d can be raised and lowered using the SW/NE (South-West/North-East) convention:
\begin{align}\label{Craise}
\rho^{\un m} = \rho_{\un n} \un C^{\un n \un m} \,, \quad \rho_{\un m} = \un C_{\un m \un n} \rho^{\un n} \,,
\end{align}
with $(\un C^{\un m \un n}) = (\un C_{\un m \un n} ) = \un C$. There is a slight abuse of notation here: the indices $\un m, \un n$ go from 1 to 8 here (instead of 1 to 4), but half of the spinor components are zero due to the chirality condition.
When indices are omitted the contraction is implicitly SW/NE. For instance
\be\label{InvariantContraction}
\rho \ti\rho = \rho_{\un m} \ti\rho^{\un m} \,, \quad \rho \Gamma^{\un A}\ti\rho = \rho_{\un n} (\Gamma^{\un A})^{\un n} {}_{\un m} \ti\rho^{\un m} \,,
\ee
with $(\Gamma^{\un A})^{\un n} {}_{\un m}$ the components of $\Gamma^{\un A}$ as given above.

The conventions on 5d and 4d spinors are analogous: indices are raised and lowered using the SW/NE convention with $(C^{m' n'})=  (C_{m' n'}) = C$ in 5d and with the epsilon matrices $\epsilon^{pq} = \epsilon_{pq} = \epsilon^{\dot p \dot q} = \epsilon_{\dot p \dot q}$, with $\epsilon^{12} = 1$. They are contracted using the SW/NE convention.

We also introduce gamma matrices $\Gamma^{\wat A}$ for the $\mathfrak{sp}(4)_R = \so(5)_R$ R-symmetry
\be
\ba
\Gamma^{\hat{1}} = \sigma_1 \otimes \sigma_3  \quad , \ 
\Gamma^{\hat{2}} = \sigma_2 \otimes \sigma_3  \quad , \ 
\Gamma^{\hat{3}} = \sigma_3 \otimes \sigma_3  \quad , \ 
\Gamma^{\hat{4}} = {\bf 1}_2 \otimes \sigma_2  \quad , \ 
\Gamma^{\hat{5}} = {\bf 1}_2 \otimes \sigma_1  \quad . \ 
\ea
\label{RsymGammas}
\ee
For the R-symmetry indices we use the opposite convention compared to the Lorentz indices, namely indices are raised and lowered with the NW/SE convention:
\be
\rho_{\wat m} = \rho^{\wat n} \Omega_{\wat n \wat m} \,, \quad \rho^{\wat m} = \Omega^{\wat m \wat n} \rho_{\wat n} \,,
\ee
with $(\Omega_{\wat m \wat n}) = (\Omega^{\wat m \wat n})  \ = \  i \sigma_2 \otimes \sigma_1$.
When unspecified, R-symmetry indices are contracted with the NW/SE convention, so that we have for instance $\rho \ti\rho = \rho^{\wat m}_{m} \ti\rho^{m}_{\wat m}$.

A collection of Weyl spinors $\rho_{\wat m}$ in 6d transforming in the $\bf{4}$ of $\mathfrak{sp}(4)_R$ can further satisfy a  Symplectic-Majorana condition (which exists in Lorentzian signature, but not in Euclidean signature)
\begin{align}
\lp \rho_{\wat m} \rp^{\ast} &=   B_{\rm (6d)} \rho^{\wat m}  \,.
\label{6dSMCond}
\end{align}
In 5d the Symplectic-Majorana condition on spinors is similarly
\begin{align}
\lp \rho_{\wat m} \rp^{\ast} & =   B_{\rm (5d)} \rho^{\wat m}  \,.
\label{5dSMCond}
\end{align}
In 4d the Weyl spinors are irreducible, however 4d Dirac spinor can obey a Symplectic-Majorana condition identical to \eqref{5dSMCond}.

Let us finally comment on the conventions for the supersymmetries and their chiralities in 6d. The fermions and supercharges have the same chirality, which we will chose to be $\overline{\bf 4}$ of $\mathfrak{so}(6)_L$, and we consider an $N=(0,2)$ theory in  6d. Subsequently, from the invariant contraction of spinors (\ref{InvariantContraction}) and (\ref{Craise}), it follows since $\{\Gamma_7, \underline{C} \} =0$ and $\underline{C}^T=\underline{C}$, that the supersymmetry transformation parameters are of opposite chirality, i.e. left chiral spinors transforming in ${\bf 4}$.

\subsection{Spinor Decompositions}
\label{app:SpinorDecomp}

\subsubsection*{\underline{6d to 5d :}}

A Dirac spinor in 6d  decomposes into two 5d spinors. A 6d spinor $\un\rho =(\un\rho^{\un m})$ (eight components) of positive chirality  reduces to a single 5d spinor $\rho = (\rho^{m'})$, with the embedding 
\begin{align}
\un\rho &= \  \rho \otimes \binom{1}{0}   \,.
\end{align}
For a 6d spinor of negative chirality, the 5d spinor is embedded in the complementary four spinor components.
The 6d Symplectic-Majorana condition \eqref{6dSMCond} on $\un\rho_{\wat m}$ reduces to the 5d Symplectic-Majorana condition \eqref{5dSMCond} on $\rho_{\wat m}$ if $\un\rho_{\wat m}$ has positive chirality, or reduces to the opposite reality condition (extra minus sign on the right hand side of \eqref{5dSMCond}), if  $\un\rho_{\wat m}$ has negative chirality.

\subsubsection*{\noindent\underline{5d to 4d :}}

A 5d spinor $\rho = (\rho^{m'})$ decomposes into two 4d Weyl spinors $\psi_+, \psi_-$ of opposite chiralities, with the embedding
\begin{align}
\rho &= \  \binom{0}{1} \otimes \psi_+  +  \binom{1}{0} \otimes \psi_-  \  = \  \binom{\psi_-}{\psi_+} \,. 
\end{align}
If $\rho^{\wat m}$ obeys the 5d Symplectic-Majorana condition \eqref{5dSMCond}, the spinors $\psi_+^{\wat m}, \psi_-^{\wat m}$ are not independent. They form four-component spinors which obey a 4d Symplectic-Majorana condition:
\begin{align}
\binom{\psi_-{}_{\wat m}}{\psi_+{}_{\wat m}}^\ast = B_{\rm (4d)} \binom{\psi_-{}^{\wat m}}{\psi_+{}^{\wat m}} \,.
\end{align}
With these conventions, we obtain for two 5d spinors $\rho, \ti\rho$ the decomposition of bilinears
\begin{align}
& \rho \ti\rho = \rho_{m'} \ti\rho^{m'} = \psi_+{}_{p} \ti \psi_+^{p} - \psi_-{}_{\dot p} \ti \psi_-^{\dot p} = \psi_+ \ti\psi_+ - \psi_- \ti\psi_-  \,, \nn\\
& \rho \gamma^5 \ti\rho = \rho_{m'} (\gamma^5)^{m'}{}_{n'} \ti\rho^{n'} = \psi_+{}_{p} \ti \psi_+^{p} + \psi_-{}_{\dot p} \ti \psi_-^{\dot p} = \psi_+ \ti\psi_+ + \psi_- \ti\psi_-  \nn\\
&\rho \gamma^\mu \ti\rho =  \psi_+{}_{p} (\tau^\mu)^p {}_{\dot p} \ti \psi_-^{\dot p} + \psi_-{}_{\dot p} (\bar \tau^\mu)^{\dot p} {}_{p} \ti \psi_+^{p} = \psi_+ \tau^\mu \ti\psi_-  + \psi_- \bar\tau^\mu \ti\psi_+ \,, 
\end{align}
with $(\tau_1, \tau_2, \tau_3, \tau_4) = (-{\bf 1}_2, \sigma_1, \sigma_2, \sigma_3)$ and $(\bar\tau_1, \bar\tau_2, \bar\tau_3, \bar\tau_4) = (-{\bf 1}_2, -\sigma_1, -\sigma_2, -\sigma_3)$.

\subsubsection*{\noindent\underline{R-symmetry reduction :}}

In this paper we consider the reduction of the R-symmetry group 
\be
\mathfrak{sp}(4)_R \ \rightarrow \ \mathfrak{su}(2)_R \oplus \mathfrak{so}(2)_R \,.
\ee
The fundamental index $\wat m$ of $\mathfrak{sp}(4)_R$ decomposes into the index $(\wat p, \wat x)$ of $\su(2)_R \oplus \mathfrak{so}(2)_R$.
A (collection of) spinors $\rho_{\wat m}$ in any spacetime dimension can be gathered in a column four-vector $\rho$ with each component being a full spinor. The decomposition is then
\begin{align}
\rho &= \  \rho^{(1)} \otimes \binom{1}{0} + \rho^{(2)} \otimes \binom{0}{1} \,, 
\end{align}
with $\rho^{(1)} = (\rho^{(1)}{}_{\wat p})$ transforming in the $({\bf 2})_{+1}$ of $\mathfrak{su}(2)_R \oplus \mathfrak{so}(2)_R$ and $\rho^{(2)} = (\rho^{(2)}{}_{\wat p})$  transforming in the $({\bf 2})_{-1}$.
So the four spinors $\rho_{\wat m}$ get replaced by the four spinors $\rho^{(1)}{}_{\wat p}, \rho^{(2)}{}_{\wat p}$. 
From the $\mathfrak{sp}(4)_R$ invariant tensor $\Omega_{\wat m \wat n}$, with $\Omega = \epsilon \otimes \sigma_1$, and the explicit gamma matrices \eqref{RsymGammas} we find the bilinear decompositions. For instance
\begin{align}
 \rho^{\wat m} \ti\rho_{\wat m}&  =   \rho^{(1)}{}^{\wat p} \ti\rho^{(2)} _{\wat p} + \rho^{(2)}{}^{\wat p} \ti\rho^{(1)} _{\wat p}  \cr
  \rho \Gamma^{\wat a} \ti\rho  &\equiv  \rho^{\wat m} (\Gamma^{\wat a})_{\wat m}{}^{\wat n} \ti\rho_{\wat n} \ = \ 
  \rho^{(2)}{}^{\wat p} (\sigma^{\wat a}) _{\wat p}{}^{\wat q} \ti\rho^{(1)} _{\wat q} - \rho^{(1)}{}^{\wat p} (\sigma^{\wat a}) _{\wat p}{}^{\wat q} \ti\rho^{(2)} _{\wat q}  \cr 
  & \equiv \  \rho^{(2)} \sigma^{\wat a} \ti\rho^{(1)}  - \rho^{(1)} \sigma^{\wat a} \ti\rho^{(2)}  \,, \quad \wat a =1,2,3\,. \nn
\end{align}
Another useful identity is
\be
(\Gamma^{\wat A})^{\wat m \wat n} (\Gamma_{\wat A})_{\wat r \wat s} = 4 \delta^{[\wat m}{}_{\wat r} \delta^{\wat n]}{}_{\wat s} - \Omega^{\wat m \wat n} \Omega_{\wat r \wat s}  \, .
\ee


\section{Killing Spinors for the $S^2$ Background}
\label{app:Kill}

In this appendix we determine the solutions to the Killing spinor equations for the $S^2$ background of section \ref{sec:Ansaetze}. 

\subsection{$\dd\psi^{\wat m}_A=0$}
\label{sec:Killing1}

The supersymmetry transformations of conformal supergravity are parametrized by two complex eight-component spinors $\epsilon^{\wat m}, \eta^{\wat m}$, of positive chirality and negative chirality, respectively,\footnote{In Lorentzian signature these spinors obey a Symplectic-Majorana condition, leaving 16+16 real supercharges.} with an index $\wat m$ transforming in the $\bf 4$ of $\mathfrak{sp}(4)_R$. The first Killing spinor equation is
\be
0 = \dd\psi^{\wat m}_{\un{A}} = \CD_{\un{A}} \epsilon^{\wat  m} + \frac{1}{24} \, (T^{ \wat  m \wat n})^{\un{BCD}} \Gamma_{\un{BCD}} \Gamma_{\underline{A}} \epsilon_{\wat n} + \Gamma_{\underline{A}} \eta^{\wat m}
\label{KSE1}
\ee
with
\be
\ba
 \quad \CD_{\underline{\mu}} \epsilon^{\wat m }&= \p_{\underline{\mu}} \epsilon^{\wat m} + \half b_{\underline{\mu}} \epsilon^{\wat m} + \frac 1 4 \ti \omega_{\underline{\mu}}^{\un{BC}} \Gamma_{\un{BC}} \epsilon^{\wat m} - \half V^{\wat m}_{\underline{\mu}}{}_{\wat n}\epsilon^{\wat n} \\
\ti \omega_{\underline{\mu}}^{\un{AB}} &= 2 e^{\underline{\nu} [\un{A}}\p_{[\underline{\mu}} e_{\underline{\nu}]}{}^{\un{B}]} - e^{\underline{\rho} [\un{A}} e^{\un{B}] \underline{\sigma}} e_{\underline{\mu}}^{\un C} \p_{\underline{\rho}} e_{\underline{\sigma} \un{C}} + 2 e^{[\un{A}}_{\underline{\mu}} b^{\un{B}]} \ = \ \omega_{\underline{\mu}}^{\un{AB}} + 2 e^{[\un{A}}_{\underline{\mu}} b^{\un{B}]} \ ,
\ea
\ee
where the background fields have been converted to $\mathfrak{sp}(4)_R$ representations with
\be
\ba 
& V^{\wat m}_{\un{A} \ \wat n} = V_{\un{A} \wat B \wat C} (\Gamma^{\wat B \wat C})^{\wat m}_{\ \ \wat n}
\quad , \quad  T^{\wat m \wat n}_{\un{BCD}} = T_{\wat A \un{BCD}} (\Gamma^{\wat A})^{\wat m \wat n} \quad , \quad  D^{\wat m \wat n}{}_{\wat r \wat s} = D_{\wat A \wat B} (\Gamma^{\wat A})^{\wat m \wat n} (\Gamma^{\wat B})_{\wat r \wat s} \, .
\ea
\label{sp4indices}
\ee
We choose to set $\eta=0$.
After inserting our ansatz, in particular $T^{\wat m \wat n}_{\un{BCD}} = b_{\un{A}} =0$, we obtain
\be
\ba
0 &=  \p_\phi \epsilon^{\wat m}  - \frac{1}{2r }  \ell'(\theta) \, \Gamma^{56} \epsilon^{\wat m} - \half \, v(\theta) \, (\Gamma^{\wat 4 \wat 5})^{\wat m}{}_{\wat n} \epsilon^{\wat n}  \\
0 &=    \p_{\mu'}  \epsilon^{\wat m}  \quad , \quad {\mu'} = x^1,x^2,x^3,x^4,\theta \ ,
\ea
\ee
We find solutions for constant spinors  $\epsilon^{\wat m}$ subject to the constraint 
\be
0 \ = \ -\Gamma^{56} \epsilon^{\wat m} + (\Gamma^{\wat 4 \wat 5})^{\wat m}{}_{\wat n} \epsilon^{\wat n} \,,
\label{ProjectionCondApp}
\ee
with
\be
v(\theta) = - {\ell'(\theta)\over r} \,.
\ee
The condition \eqref{ProjectionCondApp} projects out half of the components of a constant spinor, leaving eight real supercharges in Lorentzian signature, or eight complex supercharges in Euclidean signature.


\subsection{$\dd\chi^{\wat m \wat n}_{\wat r}=0$}
\label{sec:Killing2}

The second Killing spinor equation is given by
\be
\ba
0 &= \dd\chi^{\wat m \wat n}_{\wat r} \cr 
&= \ \frac{5}{32} \lp \CD_{\un{A}} T^{\wat m \wat n}_{\un{BCD}} \rp \Gamma^{\un{BCD}} \Gamma^{\un{A}} \epsilon_{\wat r} - \frac{15}{16}  \Gamma^{\un{BC}} R_{\un{BC} \, \wat r}^{[\wat m} \epsilon^{\wat n]}  - \frac{1}{4} D^{\wat m \wat n}{}_{\wat r \wat s} \epsilon^{\wat s}  + \frac{5}{8} T^{\wat m \wat n}_{\un{BCD}}\Gamma^{\un{BCD}} \eta_{\wat r}
- {\rm traces} \,,
\label{ChiEqn}
\ea
\ee
with
\be\ba
\CD_{\underline{\mu}} T_{\un{BCD}}^{\wat m \wat n} &= \p_{\underline{\mu}} T_{\un{BCD}}^{\wat m \wat n} + 3 \ti \omega_{\underline{\mu} [\un{B}}^{\un{E}} T^{\wat m \wat n}_{\un{CD}]\un{E}} - b_{\underline{\mu}} \, T_{\un{BCD}}^{\wat m \wat n} + V_{\un{\mu} \wat r}^{[\wat m} T_{\un{BCD}}^{\wat n] \wat r} \cr 
R_{\underline{\mu}\underline{\nu}}^{\wat m \wat n} &= 2 \p_{[\underline{\mu}} V_{\underline{\nu}]}^{\wat m \wat n} + V_{[\underline{\mu}}^{\wat r (\wat m} V_{\underline{\nu}]\wat r}^{\wat n)} \,.
\ea\ee
Here, `traces' indicates terms proportional to invariant tensors $\Omega^{\wat m \wat n}, \dd^{\wat m}_{\wat r}, \dd^{\wat n}_{\wat r}$. Again the background fields are converted to $\mathfrak{sp}(4)_R$ representations using~\eqref{sp4indices}.

With $T_{\un{BCD}}^{\wat m \wat n}=0$, we obtain the simpler conditions
\begin{align}
0 = - \frac{15}{4}  \Gamma^{\un{BC}} R_{\un{BC} \, \wat r}^{[\wat m} \epsilon^{\wat n]}  - D^{\wat m \wat n}{}_{\wat r \wat s} \epsilon^{\wat s} - {\rm traces} \ .
\label{KSE2simple}
\end{align}
The R-symmetry field strength has a single non-vanishing component, corresponding to a flux on $S^2$
\be
R_{\theta\phi}^{\wat m \wat n} = -R_{\phi \theta}^{\wat m \wat n} = -{\ell''(\theta) \over r}\, (\Gamma^{\wat 4 \wat 5})^{\wat m \wat n}\,.
\ee 
In flat space indices this becomes 
\be
R_{56}^{\wat m \wat n} = -R_{65}^{\wat m \wat n} = - \frac{\ell''(\theta)}{r^2 \ell(\theta)} \, (\Gamma^{\wat 4 \wat 5})^{\wat m \wat n}\,.
\ee 
Moreover our ans\"atze for $D_{\wat A \wat B}$ \eqref{Dansatz} can be re-expressed in $\mathfrak{sp}(4)_R$ indices as:
\begin{align}
D^{\wat m \wat n}{}_{\wat r \wat s} &= 
d \Big[ 5 (\Gamma^{\wat 4 \wat 5})^{[\wat m}{}_{\wat r} (\Gamma^{\wat 4 \wat 5})^{\wat n]}{}_{\wat s} 
- \delta^{[\wat m}{}_{\wat r} \delta^{\wat n]}{}_{\wat s} - \Omega^{\wat m \wat n} \Omega_{\wat r \wat s} 
\Big]  \ ,
\end{align}
where the two last terms lead only to ``trace" contributions in \eqref{KSE2simple} and hence drop from the equations.  We obtain
\begin{align}
0 = \frac{15}{2}  \frac{\ell''(\theta)}{r^2 \ell(\theta)}  \, \Gamma^{56} (\Gamma^{\wat 4 \wat 5})^{[\wat m}{}_{\wat r} \epsilon^{\wat n]}  - 5 d (\Gamma^{\wat 4 \wat 5})^{[\wat m}{}_{\wat r} (\Gamma^{\wat 4 \wat 5})^{\wat n]}{}_{\wat s} \epsilon^{\wat s}  \ .
\end{align}
Using \eqref{ProjectionCondApp}, we solve the equations without further constraints on $\epsilon^{\wat m}$ if 
\be
d = \frac{3}{2} \, \frac{\ell''(\theta)}{r^2 \ell(\theta)}  \, .
\ee
The background we found corresponds to the twisting $ \mathfrak{u}(1)_L \oplus \mathfrak{u}(1)_R \rightarrow \mathfrak{u}(1)$ on $S^2$. It preserves half of the supersymmetries (and no conformal supersymmetries) of the flat space theory, and corresponds to the topological half-twist of the 2d theory.


\section{6d to 5d Reduction for $b_{\mu}=0$}
\label{app:6dto5dGen}

In this appendix we detail the reduction of the six dimensional equations of motion on an $S^1$. This is done following \cite{Kugo:2000hn,Cordova:2013bea} however we choose to gauge fix $b_{\underline{\mu}} = 0$, which is possible without loss of generality. 

We start by decomposing the six dimensional frame as
\be \label{FrameDecomp6dto5d}
e^{\underline{\mu}}_{\underline{A}}=
\begin{pmatrix}
e^{\mu'}_{A'}  & &e^\phi_{A'}= -C_{A'}\\
e^{\mu'}_6 =0  & &e_6^\phi = \alpha
\end{pmatrix} \qquad 
e_{\underline{\mu}}^{\underline{A}}=
\begin{pmatrix}
e_{\mu'}^{A'}  & &e_{\mu'}^6 =\alpha^{-1} C_{\mu'} \\
  e_\phi^{A'} = 0& &e^6_\phi =\alpha^{-1}
\end{pmatrix}\,,
\ee
where the 5d indices are primed. We work in the gauge $b_{\underline{\mu}} = 0$, which is achieved by fixing the special conformal generators, $K_{\underline{A}}$. Note that this choice is different from the gauge fixing of $b_{\underline{\mu}}$ in \cite{Kugo:2000hn,Cordova:2013bea}, in particular $\alpha$ is not covariantly constant in our case. Furthermore, we fix the conformal supersymmetry generators to ensure $\psi_5 = 0$, which means that $e_6^{\underline{\mu}} = 0$ is invariant under supersymmetry transformations.
For a general background the bosonic supergravity fields descend to 5d fields as
\begin{equation} \ba
D^{\wat m \wat n}_{\wat r \wat s}&\quad \rightarrow \quad D^{\wat m \wat n}_{\wat r \wat s} \\
V_{\underline{A}}^{\wat m \wat n}&\quad \rightarrow\quad  \left\{ \begin{array}{c} V_{A'}^{\wat m \wat n} \quad \underline{A} \neq 6 \\ S^{\wat m \wat n} \quad \underline{A} = 6 \end{array} \right. \\
T^{\wat m \wat n}_{\underline{ABC}} &\quad \rightarrow \quad T_{A'B'6}^{\wat m \wat n} \equiv T_{A'B'}^{\wat m \wat n} \,.
\ea  \end{equation}  
The components of the spin connection along the $\phi$ direction are given by
\be 
\omega_{\phi}^{A'6} = \frac{1}{\alpha^2} e^{\mu'A'} \partial_{\mu'} \alpha \,,\qquad 
\omega_{\phi}^{A'B'} = - \frac{1}{2\alpha^2}G^{A'B'} \,,\qquad 
 \omega_{\mu'}^{A'6} =  \frac{1}{2\alpha}e^{\nu' A'}G_{\mu'\nu'} +\frac{1}{\alpha^2} C_{\mu '}  e^{A '}_{\nu '}\partial^{\nu '} \alpha \,,
\ee
where $G= dC$, 
and can be derived  from the six dimensional vielbein using
\begin{equation}
\omega_{\un{\mu}}^{\un{AB}}=2e^{\un{\nu}[\un{A}}\partial_{[\un{\mu}}e_{\un{\nu}]}^{\un{B}]}-e^{\un{\rho}[\un{A}}e^{\un{B}]\un{\sigma}} e_{\un{\mu}}^{\un{C}}\partial_{\un{\rho}} e_{\un{\sigma C}} \,.
\end{equation}


\subsection{Equations of Motion for $\mathcal{B}$}

The 6d equations of motion for the three-form $H$ are given by
\be \label{H5dEoM}
\ba 
 dH &=0 \\
 H^-_{\underline{ABC}} - \half \Phi_{\wat m \wat n} T^{\wat m \wat n}_{\underline{ABC}} &= 0 \,.
\ea \ee
We decompose $H$ into 5d components
\be
H={1\over 3!}
H_{A'B'C'}e^{A'}\wedge e^{B'}\wedge ^{C'}+{1\over 2} H_{D'E'6}e^{D'}\wedge e^{E'}\wedge e^6 \,.
\ee
We can solve the second equation of motion by setting
\be  \ba
H_{A'B'6} &= \alpha F_{A'B'} \\
H_{A'B'C'} &=  \half {\epsilon_{A'B'C'}}^{D'E'}\left( \alpha F_{D'E'}-\Phi_{\wat m \wat n} T^{\wat m \wat n}_{D'E'} \right)\,,
\ea \ee 
where $F_{\mu'\nu'}$ is a two-form in five dimensions. Substituting this into the expansion of $H$ and reducing to 5d we obtain
\be 
H = \alpha\star_{5d} \left( F-\frac{1}{\alpha}\Phi_{\wat m \wat n} T^{\wat m \wat n}\right) +   F \wedge C  + F\wedge d \varphi \,.
\ee
The equations of motion $dH = 0$ imply 
\be
dF =0 \,,\qquad 
 F\wedge dC  + d \left(\alpha \star_5 F - \Phi_{\wat m \wat n} \star_5 T^{\wat m \wat n} \right) \,,
\ee
which can be integrated to the 5d action
\be
S_F= -\int \left( \alpha\tilde{F}\wedge \star_{5d}\tilde{F} + C \wedge F \wedge F  \right)\,,
\ee
where 
\be 
\tilde{F} = F-\frac{1}{\alpha}\Phi_{\wat m \wat n} T^{\wat m \wat n} \,.
\ee
Together with the constraint $dF=0$, which identifies $F$ with the field strength of a five-dimensional connection $A$, given by $F_{\mu'\nu'} = \partial_{\mu'}A_{\nu'}-\partial_{\nu'}A_{\mu'}$.


\subsection{Equations of Motion for the Scalars}

The dimensionally reduced 6d scalar equations of motion are
\be
D^2 \Phi^{\wat m \wat n}+ 2F_{A'B'}T^{A'B'}_{\wat m \wat n}+(M_{\Phi})^{\wat m\wat n}_{\wat r\wat s}\Phi^{\wat r\wat s}=0 \,,
\ee
where
\be
\ba
D_{\mu'} \Phi^{\wat m\wat n}&=\partial_{\mu'}+V_{\mu' \wat r}^{[\wat m}\Phi^{\wat n]\wat r}\\
D^2\Phi^{\wat m\wat n}&=(\partial^{A'}+\omega_{B'}^{B'A'})D_{A'}
\Phi^{\wat m\wat n}+V_{\mu' \wat r}^{\wat [m}D^{\mu'}\Phi^{\wat n]\wat r}\\
(M_{\Phi})^{\wat m\wat n}_{\wat r\wat s}&= 
-\frac{R_{6d}}{5}\delta^{[\wat m}_{\wat r}\delta^{\wat n]}_{\wat s} {+\frac{1}{\alpha}C^{\mu'} \partial_{\mu'} \alpha S^{[\wat m}_{\wat r} \Phi^{\wat n] \wat r}} +\half\alpha^2(S^{[\wat m}_{\wat r} S^{\wat n]}_{\wat s}-S^{\wat t}_{\wat s}S^{[\wat m}_{\wat t}\delta^{\wat n]}_{\wat r})-\frac{1}{15}D^{\wat m\wat n}_{\wat r\wat s} -T^{A'B'}_{\wat r \wat s}T^{\wat m \wat n}_{A'B'} \,.
\ea
\ee
The 6d Ricci scalar $R_{6d}$ can be rewritten of course in terms of the 5d fields. This equation of motion can be integrated to the following action
\be
\S_\Phi= -\int d^5x \,\,\sqrt{|g|}\,\alpha^{-1}\left( D_{A'}\Phi^{\wat m\wat n}D^{A'}\Phi_{\wat m\wat n}+ 4 \Phi^{\wat m \wat n}F_{A'B'}T^{A'B'}_{\wat m \wat n}-\Phi_{\wat m\wat n}(M_{\Phi})^{\wat m\wat n}_{\wat r\wat s}\Phi^{\wat r\wat s}\right) \,.
\ee


\subsection{Equations of Motion for the Fermions}

The 6d fermions are decomposed as follows
\be
\rho^{ \un{m} \wat m}\rightarrow\begin{pmatrix}
0\\ i\rho^{m' \wat m}
\end{pmatrix} \,.
\ee
Then for a general background the six dimensional equation of motion reduces to 
\begin{equation}
i\slashed{D}\rho^{m'\wat m}+(M_\rho)^{m' \wat m}_{n' \wat n}\rho^{n' \wat n}=0\,,
\end{equation} 
where 
\begin{equation}
\ba
D_{\mu'} \rho^{m'\wat m}&=\left(\partial_{\mu'}+\frac{1}{4}\omega_{\mu'}^{A'B'}\gamma_{A'B'}\right)\rho^{m'\wat m}-\half V_{\mu' \wat n}^{\wat m}\rho^{\wat n} \cr 
(M_\rho)^{m' \wat m}_{n' \wat n}&=\alpha\left(-\half S^{\wat m}_{\wat  n} \delta^{m'}_{n'} +\frac{1}{8\alpha^2}G_{A'B'}(\gamma^{A'B'})^{m'}_{n'}\delta^{\wat m}_{\wat n} -\frac{i}{2\alpha^2} e^{\mu'A'} \partial_{\mu'} \alpha(\gamma_{A'})^{m'}_{n'}\delta^{\wat m}_{\wat n}\right) \cr 
&\quad + {\frac{1}{2\alpha^2}(\gamma^{\mu'} \gamma^{\nu '})^{m'}_{n'} \delta^{\wat m}_{\wat n} C_{\mu '}\partial_{\nu '} \alpha } +\half T_{A'B'}{}^{\wat m}{}_{ \wat n}(\gamma^{A'B'})^{m'}_{n'}  \,.
\ea
\end{equation}
From this we obtain the action
\begin{equation}
S_{\rho}= -\int d^5x \sqrt{|g|} \,\alpha^{-1}\rho_{m \wat m}\left(i \slashed{D}^m_n\rho^{n \wat m}+(M_\rho)^{m \wat m }_{n \wat n} \rho^{n \wat n}\right) \,.
\end{equation}


\section{Supersymmetry Variations of the 5d Action}
\label{app:5dSUSY}

The supersymmetry variations (\ref{SUSY5dGENERAL}), which leave the 5d action \eqref{5dFinal}} invariant, can be decomposed with respect to the R-symmetry, following  appendix \ref{app:SpinorDecomp}. This decomposition will be useful in further proceeding to four dimensions. 
The scalar and gauge field variations are then 
\be
\ba 
\delta A_{\mu} &= - \ell(\theta) \lp \epsilon^{(1)}{}^{\wat p} \gamma_\mu \rho^{(2)}_{\wat p \, -} + \epsilon^{(2)}{}^{\wat p} \gamma_\mu \rho^{(1)}_{\wat p \, +} \rp  \\
\delta A_{\theta}  &= - r\ell(\theta) \lp \epsilon^{(1)}{}^{\wat p} \rho^{(2)}_{\wat p \, +} - \epsilon^{(2)}{}^{\wat p}  \rho^{(1)}_{\wat p \, -} \rp \\
\delta \varphi^{\wat a} &=   i \lp \epsilon^{(1)}{}_{\wat p}  (\sigma^{\wat a})^{\wat p\wat q} \rho^{(2)}_{\wat q \, +} - \epsilon^{(2)}{}_{\wat p}  (\sigma^{\wat a})^{\wat p\wat q} \rho^{(1)}_{\wat q \, -}  \rp  \\
\delta \varphi &= - 2 \epsilon^{(1)}{}^{\wat p}  \rho^{(1)}_{\wat p \, +}  \,, \quad   \delta \bar\varphi \ = \ 2 \epsilon^{(2)}{}^{\wat p}  \rho^{(2)}_{\wat p \, -}  
\ea
\ee
and for the fermions we find
\be
\ba
\delta \rho^{(1)}_{\wat p \, +} &= \frac{i }{8\ell(\theta)} F_{\mu\nu} {\gamma}^{\mu\nu} \epsilon^{(1)}_{\wat p}  
 - \frac i4 D_\mu \varphi  {\gamma}^\mu \epsilon^{(2)}_{\wat p}  
 + \frac{1}{4r}D_\theta \varphi^{\wat q}_{\wat p}\epsilon^{(1)}_{\wat q} 
 - \frac{\ell(\theta)}{8} \lp \epsilon^{\wat a \wat b \wat c} [\varphi_{\wat a},\varphi_{\wat b}](\sigma_{\wat c})^{\wat q}_{\wat p} \epsilon^{(1)}_{\wat q} -i [\varphi, \bar\varphi ] \epsilon^{(1)}_{\wat p} \rp  \\
\delta \rho^{(1)}_{\wat p \, -}  &=  \frac{i}{4 r\ell(\theta)} F_{\mu\theta} {\gamma}^{\mu} \epsilon^{(1)}_{\wat p}  
 +\frac 14D_\mu \varphi^{\wat q}_{\wat p}\,{\gamma}^\mu\epsilon^{(1)}_{\wat q}  
 + \frac{i}{4r}\lp D_\theta \varphi  + \frac{\ell'(\theta)}{\ell(\theta)} \varphi \rp  \epsilon^{(2)}_{\wat p}  - \frac{\ell(\theta)}{4}  [\varphi, \varphi^{\wat q}_{\wat p}] \epsilon^{(2)}_{\wat q}  \\
\delta \rho^{(2)}_{\wat p \, +} &= -\frac{i}{4 r \ell(\theta)} F_{\mu\theta} {\gamma}^{\mu} \epsilon^{(2)}_{\wat p}  
 -\frac 14 D_\mu \varphi^{\wat q}_{\wat p} \,{\gamma}^\mu \epsilon^{(2)}_{\wat q}  
 + \frac{i}{4r} \lp D_\theta \bar\varphi  + \frac{\ell'(\theta)}{ \ell(\theta)} \bar\varphi\rp  \epsilon^{(1)}_{\wat p} - \frac{\ell(\theta)}{4}  [\bar\varphi, \varphi^{\wat q}_{\wat p}] \epsilon^{(1)}_{\wat q}\\
\delta \rho^{(2)}_{\wat p \, -} &= \frac{i}{8\ell(\theta)} F_{\mu\nu} {\gamma}^{\mu\nu} \epsilon^{(2)}_{\wat p}  
 + \frac i4 D_\mu \bar\varphi  {\gamma}^\mu \epsilon^{(1)}_{\wat p}  
 + \frac{1}{4r} D_\theta \varphi^{\wat q}_{\wat p} \epsilon^{(2)}_{\wat q}  - \frac{\ell(\theta)}{8} \lp \epsilon^{\wat a \wat b \wat c} [\varphi_{\wat a},\varphi_{\wat b}] (\sigma_{\wat c})^{\wat q}_{\wat p} \epsilon^{(2)}_{\wat q} +i [\varphi, \bar\varphi ] \epsilon^{(2)}_{\wat p} \rp    \,,
\ea
\ee
where $\varphi_{\wat p}{}^{\wat q} = \sum_{\wat a} \varphi^{\wat a} (\sigma^{\wat a})_{\wat p}{}^{\wat q}$.

\section{Aspects of the 4d Sigma-model}
\label{app:4dFlat}

In this appendix we collect several useful relations for the sigma-model reduction, as well as give details on integrating out the gauge field and the scalars $\varphi$ and $\bar\varphi$, which appear only algebraically in the $r\rightarrow 0$ limit of the 5d action.  

\subsection{Useful Relations}
\label{app:SigmaModel}

We now summarize properties of the sigma-model defined in section \ref{sec:SigmaModNahm}. 
The three symplectic structures  (\ref{SympForm}) of the Hyper-K\"ahler target 
can be used to define the three complex structures 
 $\omega^{\wat a}_K{}^I = \omega^{\wat a}_{KJ}G^{JI}$, which satisfy
 \be 
 \omega_{\wat a I}{}^J \omega_{\wat b J}{}^K = -\delta_{\wat a\wat b}\delta_I^K + \epsilon_{\wat a \wat b \wat c}\omega^{\wat c}_I{}^K\,.
 \ee
The complex structures exchange the cotangent vectors $\Upsilon_{I}^{\wat a}$ and $\Upsilon_I^{\theta}$  in the following fashion
\be \label{CSUpsilonRel}
\ba 
\omega^{\wat a}_I{}^J \Upsilon_J^{\theta} &= - \Upsilon_I^{\wat a} \\
\omega^{\wat a}_I{}^J \Upsilon_J^{\wat b} &=  \delta^{\wat a \wat b}\Upsilon^{\theta}_I +  \epsilon^{\wat a \wat b\wat c} \Upsilon_{I\wat c} \,.
\ea \ee
We introduce a complete set of functions, satisfying the completeness relations \cite{Gauntlett:1993sh}
\be \label{CompletenessRel}
\ba 
G^{IJ} \Upsilon_{I }^{{\wat a}\alpha}(\theta) \Upsilon_{J}^{{\wat b}\beta}(\tau) + \sum_i \Psi^{{\wat a} \alpha}_i(\theta) \Psi^{{\wat b}\beta}_i(\tau) &= \delta^{\wat a \wat b} \, \delta^{\alpha \beta}\,  \delta(\theta-\tau) \\
G^{IJ}  \Upsilon^{\theta\alpha}_I(\theta) \Upsilon_{J}^{\theta\beta}(\tau)+ \sum_i \Psi^{\theta\alpha}_i(\theta) \Psi^{\theta\beta}_i(\tau)  &= \delta^{\alpha\beta}\,  \delta(\theta-\tau)   \cr 
G^{IJ} \Upsilon_{I }^{{\wat a}\alpha}(\theta) \Upsilon_{J}^{\theta\beta}(\tau) + \sum_i \Psi^{{\wat a} \alpha}_i(\theta) \Psi^{\theta\beta}_i(\tau) & =0 \,.
\ea
\ee
Here, $\alpha, \beta$ are indices labeling the generators of the gauge algebra. 
These functions satisfy the orthogonality relations
\be
\int d\theta \Upsilon^{{\wat a} \alpha}_I(\theta) \Psi_{i}^{{\wat b} \beta}(\theta)  = 0 \,, \quad
\int d\theta \Upsilon^{\theta \alpha}_I(\theta) \Psi_i^{\theta \beta}(\theta) = 0\,.
\ee


\subsection{Integrating out Fields}
\label{app:GaugeField2}

In this appendix we discuss how the scalars $\varphi, \bar\varphi$ and the 4d gauge field $A_\mu$ are integrated out in the sigma-model reduction. The equation of motions for $\varphi, \bar\varphi$ and $A_\mu$ are
\be \label{GaugeFieldEoM}
\ba
 \mathcal{D}_{\theta}^2 \varphi + \left[\varphi_{\wat a}, \left[\varphi^{\wat a}, \varphi \right]\right] &= - 4i r [\rho_{-\wat p }^{(1)},  \rho^{(1)\wat p}_+]\cr
 \mathcal{D}_{\theta}^2 \bar\varphi + \left[\varphi_{\wat a}, \left[\varphi^{\wat a}, \bar\varphi \right]\right] &=  4i r [\rho_{+\wat p }^{(2)}, \rho^{(2)\wat p}_-]\cr 
 \mathcal{D}_{\theta}^2 A_{\mu} + \left[\varphi_{\wat a}, \left[\varphi^{\wat a}, A_{\mu}\right]\right] &= \left[A_{\theta}, \partial_{I} A_{\theta}\right]\partial_\mu X^I + \left[\varphi_{\wat a}, \partial_{I}\varphi^{\wat a}\right]\partial_\mu X^I - 4i [\rho_{-\wat p }^{(1)}, \gamma_\mu \rho^{(2)\wat p}_+]\,.
\ea
\ee
We adopt a convenient gauge for the connection $E_I$
\be \label{GaugeFixing} 
\mathcal{D}_{\theta}\Upsilon^\theta_I +[\varphi_{\wat a}, \Upsilon^{\wat a}_I] = 0\,,
\ee
which can be re-expressed as 
\be 
\mathcal{D}_{\theta}^2 E_I + [\varphi_{\wat a},[\varphi^{\wat a}, E_I]] = [A_{\theta}, \partial_I A_{\theta}] + [\varphi_{\wat a}, \partial_I \varphi^{\wat a}] \,,
\ee
where we have used the gauge fixing condition $\partial_{\theta} A_{\theta} = 0$. Using the expansion for the spinors \eqref{SpinorDecomp} and the constraints \eqref{FermionUntwistedCon}, we evaluate the spinor bilinears in \eqref{GaugeFieldEoM} to give
\be \ba
\left[\rho_{-\wat p }^{(1)}, \rho^{(1)\wat p}_- \right] &=  - 4 \lp \left[ \Upsilon_I^{\wat a}, \Upsilon_{J \wat a} \right] + \left[\Upsilon_I^{\theta}, \Upsilon_J^{\theta}\right] \rp \lambda_{\wat p}^{(1)I} \lambda^{(1)J\wat p}  \\
\left[\rho_{+\wat p }^{(2)}, \rho^{(2)\wat p}_+ \right] &=  - 4 \lp \left[ \Upsilon_I^{\wat a}, \Upsilon_{J \wat a} \right] + \left[\Upsilon_I^{\theta}, \Upsilon_J^{\theta}\right] \rp \lambda_{\wat p}^{(2)I} \lambda^{(2)J\wat p}  \\
\left[\rho_{-\wat p }^{(1)}, \gamma_\mu \rho^{(2)\wat p}_+\right] &= - 4\left(\left[\Upsilon_I^{\wat a}, \Upsilon_{J \wat a}\right] + \left[\Upsilon_I^{\theta}, \Upsilon_J^{\theta}\right] \right) \lambda_{\wat p}^{(1)I} \gamma_{\mu} \lambda^{(2)J\wat p} \,.
\ea \ee
 We note that the curvature
\be 
\Phi_{IJ} = [\nabla_I, \nabla_J] \,,
\ee
where $\nabla_I = \partial_I + [E_I, \cdot\,]$, satisfies the equation
\be  \label{FieldStrengthEoM}
\mathcal{D}^2_{\theta}\Phi_{IJ} + [\varphi_{\wat a}, [\varphi^{\wat a}, \Phi_{IJ}]] = 2\left( [\Upsilon_{I\wat a}, \Upsilon^{\wat a}_J] + [\Upsilon_I^{\theta}, \Upsilon_J^{\theta}]\right) .
\ee
It can be used to solve the equations of motion by
\be \label{GaugeFieldSol}
\ba
\varphi &=  8i r \Phi_{IJ} \lambda_{\wat p}^{(1)I} \lambda^{(1) J \wat p} \cr 
\bar\varphi &=  -8i r \Phi_{IJ} \lambda_{\wat p}^{(2)I} \lambda^{(2) J \wat p} \cr 
A_{\mu} &= E_I \partial_{\mu} X^I + 8i \Phi_{IJ} \lambda_{\wat p}^{(1)I} \gamma_\mu \lambda^{(2) J \wat p} \,.
\ea
\ee
Inserting this back in the action the terms with $\varphi, \bar\varphi$ results in 
\be
S_{\varphi,\bar\varphi} = \frac{16}{r\ell} \int d\theta d^4x \gf \, \text{Tr}\left( \mathcal{D}_{\theta}\Phi_{IJ}\mathcal{D}_{\theta}\Phi_{KL} + [\Phi_{IJ}, \varphi^{\wat a}][\Phi_{KL}, \varphi_{\wat a}]\right) \lambda^{(1)I\wat p}\lambda^{(1)J}_{\wat p}\lambda^{(2)K\wat q}\lambda^{(2)L}_{\wat q} \,.
\label{quartic1}
\ee
The terms we obtain by integrating out $A_\mu$ will be grouped into three types of terms. The first type are such that $X^I$ appear quadratically
\be 
\ba
S_{A_{\mu}, {\rm type \, 1}} &= -\frac{1}{4r\ell} \int d\theta d^4x \gf \, \text{Tr}\Big( \mathcal{D}_{\theta}E_I\mathcal{D}_{\theta}E_J - 2\partial_IA_{\theta}\mathcal{D}_{\theta}E_J + 2 \partial_I\varphi^{\wat a}[E_J, \varphi_{\wat a}] \cr 
&\qquad \qquad + [E_I, \varphi^{\wat a}][E_J, \varphi_{\wat a}]\Big) \partial_{\mu}X^I \partial^{\mu}X^J\,.
\ea
\ee
These terms combine with terms in the scalar action \eqref{SigmaScalars} to give the usual sigma-model kinetic term
\be 
S_{{\rm scalars}} + S_{A_{\mu, {\rm type \, 1}}} = \frac{1}{4r\ell} \int d^4x \gf \, G_{IJ} \partial_{\mu}X^{I}\partial^{\mu}X^J\,.
\ee
Terms of  type 2 are linear in $X^I$ and covariantise the kinetic terms of the spinor
\be \ba 
S_{A_\mu, {\rm  type \, 2}} &= - \frac{4i}{r\ell} \int d\theta d^4x \gf\, \text{Tr} \left(2\Upsilon^{\wat a}_{I}[E_J, \Upsilon_{K \wat a}]+ 2\Upsilon_I^{\theta}[E_J, \Upsilon_{K}^{\theta}]\right) \lambda^{(1)I\wat p}\gamma^{\mu}\lambda^{(2)K}_{\wat p}\partial_{\mu}X^J\,.
\ea \ee
The terms involving the connection $E_I$ are promoted to covariant derivatives $\nabla_I$ when combined with the terms in the spinor action \eqref{SigmaSpinors}. Using the identities 
\be \ba 
\nabla_I\Upsilon_J^{\wat a} &= \Gamma_{IJ}^{K}\Upsilon_K^{\wat a} + \half[\Phi_{IJ}, \varphi^{\wat a}]  \\
\nabla_I\Upsilon_J^{\theta}&= \Gamma_{IJ}^{K}\Upsilon_K^{\theta} -\half\mathcal{D}_{\theta}\Phi_{IJ}  \,,
\ea \ee
where 
\be 
\Gamma_{IJ,K} =-\int d\theta\, \text{Tr}\left( \Upsilon^{\wat a}_{K}\nabla_{(I}\Upsilon_{J)\wat a} + \Upsilon^{\theta}_{K}\nabla_{(I}\Upsilon_{J)}^{\theta} \right),
\ee
the kinetic term in the spinor action is covariantised. 
Lastly, the terms of type 3  give rise to the quartic fermion interaction. Using \eqref{FieldStrengthEoM} these terms simplify to
\be
\ba 
S_{A_{\mu , {\rm type \, 3}}}&= -\frac{16}{r\ell}\int d^4x d\theta \gf \,\text{Tr} \left( \mathcal{D}_{\theta}\Phi_{IJ}\mathcal{D}_{\theta}\Phi_{KL} + [\Phi_{IJ}, \varphi^{\wat a}][\Phi_{KL}, \varphi_{\wat a}]\right) \cr 
&\qquad \qquad \qquad \times  \lambda^{(1)I\wat p}\gamma^{\mu}\lambda^{(2)J}_{\wat p}  \lambda^{(1)K\wat q}\gamma_{\mu}\lambda^{(2)L}_{\wat q}  \,.
\ea \ee
Using various identities, including Fierz-type identities,
\be
\ba
 (\lambda^{(1)\wat p[I}\lambda^{(1)J]}_{\wat p}) (\lambda^{(2)\wat q[K}\lambda^{(2)L]}_{\wat q}) 
 &= 2(\lambda^{(1)\wat p[I}\lambda^{(1)J]\wat q}) (\lambda^{(2)[K}_{\wat p}\lambda^{(2)L]}_{\wat q}) \cr 
 \omega^{\wat a}{}_I{}^{K} \nabla_{[K}\Upsilon^\theta_{J]}
 & = \nabla_{[I} \Upsilon^{\wat a}_{J]} \cr 
\nabla_{[I} \Upsilon^{\wat a}_{J]}\lambda^{(i)J}_{\wat p} 
&= i \nabla_{[I}\Upsilon^\theta_{J]} (\sigma^{\wat a})_{\wat p}^{\wat q} \lambda^{(i)J}_{\wat q} \cr 
 \nabla_{[I} \Upsilon^{\wat a}_{J]}\nabla_{[K} \Upsilon_{L]\wat a} \lambda^{(i)J}_{\wat p}\lambda^{(i)L}_{\wat q} 
 &= 3 \nabla_{[I} \Upsilon^{\theta}_{J]}\nabla_{[K} \Upsilon^{\theta}_{L]} \lambda^{(i)[J}_{\wat p}\lambda^{(i)L]}_{\wat q} \,,
\ea
\ee
 it can be shown that this quartic fermion interaction combines with the term \eqref{quartic1} to make the Riemann tensor of the target space appear
\be 
S_{A_{\mu}, {\rm type \, 3}} + S_{\varphi,\bar\varphi} = - \frac{32}{r\ell}\int d^4x \gf\, R_{IJKL} (\lambda^{(1)I\wat p}\lambda^{(1)J}_{\wat p}) (\lambda^{(2)K\wat q}\lambda^{(2)L}_{\wat q}) \,,
\ee
where the Riemann tensor is given by
\be \ba 
R_{IJKL} 
&= -\int d\theta \Tr \lp 2 \nabla_{[I}\Upsilon^{\wat a}_{J]} \nabla_{[K}\Upsilon_{L] \wat a} + \nabla_{[I}\Upsilon^{\wat a}_{K]} \nabla_{[J}\Upsilon_{L]\wat a}- \nabla_{[I}\Upsilon^{\wat a}_{L]} \nabla_{[J}\Upsilon_{K]\wat a}   \right. \\
& \qquad \qquad  \left. + 2 \nabla_{[I}\Upsilon^{\theta}_{J]} \nabla_{[K}\Upsilon^{\theta}_{L]} + \nabla_{[I}\Upsilon^{\theta}_{K]} \nabla_{[J}\Upsilon^{\theta}_{L]}- \nabla_{[I}\Upsilon^{\theta}_{L]} \nabla_{[J}\Upsilon^{\theta}_{K]} \rp \\
&= -\frac{1}{4}\int d\theta \,\Tr \Big(2\mathcal{D}_{\theta}\Phi_{IJ}\mathcal{D}_{\theta}\Phi_{KL} + 2[\Phi_{IJ}, \varphi^{\wat a}][\Phi_{KL}, \varphi_{\wat a}]  \\
& \phantom{\frac{1}{4}\int d\theta \,\Tr \qquad }   + \mathcal{D}_{\theta}\Phi_{IK}\mathcal{D}_{\theta}\Phi_{JL} + [\Phi_{IK}, \varphi^{\wat a}][\Phi_{JL}, \varphi_{\wat a}] \\
& \phantom{\frac{1}{4}\int d\theta \,\Tr \qquad }   - \mathcal{D}_{\theta}\Phi_{IL}\mathcal{D}_{\theta}\Phi_{JK} - [\Phi_{IL}, \varphi^{\wat a}][\Phi_{JK}, \varphi_{\wat a}]  \big) \,. 
\ea \ee
Combining all the terms we obtain the final sigma-model \eqref{4dTwistedLambdas}.


\section{Sigma-model for Hyper-K\"ahler $M_4$ from 5d SYM}
\label{sec:TopSigM4}

In this appendix we provide a comprehensive discussion of the topological twist of the 5d SYM on an interval with Nahm pole boundary conditions, and its dimensional reduction to 4d for $M_4$ a Hyper-K\"ahler manifold. This results in the same 4d topological sigma-model as we obtained in section \ref{sec:TopSigmaHK}, by twisting the 4d sigma-model on flat $M_4$. 


\subsection{Topological Twist}
\label{subsec:TopoTwist5d}

Let us first consider the topological twist 1 of section \ref{ssec:TwistsM4} applied to the 5d SYM theory. From now on we switch to Euclidean signature \footnote{For this twist we change from Lorentzian to Euclidean signature. In what follows $\gamma_0$ as defined in appendix \ref{app:ConvSpinors} is replaced with $\gamma_{0'} = i\gamma_0$, where the prime will be omitted.}. The twisted 5d theory was already considered in \cite{Witten:2011zz, Anderson:2012ck}. 

Twist 1 of the 6d $N=(0,2)$ theory identifies $\su(2)_\ell \subset \su(2)_\ell \oplus \su(2)_r$ of the 4d Lorentz algebra with the $\mathfrak{su}(2)_R \subset \su(2)_R \oplus\so(2)_R \subset \mathfrak{sp}(4)_R$. 
Under dimensional reduction to 5d the symmetries after the twist are
\be
\mathfrak{sp}(4)_R \oplus \mathfrak{so}(5)_L \quad \rightarrow \quad \mathfrak{g}_{\rm twist} \ = \ \mathfrak{su}(2)_{\rm twist} \oplus \mathfrak{su}(2)_r \oplus \mathfrak{u}(1)_R \,.
\ee
The fields of the 5d theory become forms in the twisted theory, according to their transformations with respect to the $\mathfrak{g}_{\rm twist} $, as summarized in the following table:
\be
\begin{array}{ccc}
\hbox{Field} & \hbox{$\mathfrak{g}_{\rm twist}$ Representation} & \hbox{Twisted Field}\cr \hline
A_\mu & ({\bf 2}, {\bf 2})_{0}  & A_\mu \cr
\varphi & ({\bf 1}, {\bf 1})_{2} & \varphi \cr
\bar\varphi & ({\bf 1}, {\bf 1})_{-2} & \bar\varphi \cr
\varphi^{\wat{a}} & ({\bf 3}, {\bf 1})_{0} & B_{\mu\nu} \cr 
\rho^{(1)}_+ & ({\bf 2}, {\bf 2})_{1} & \psi^{(1)}_{\mu} \cr 
\rho^{(2)}_+ & ({\bf 2}, {\bf 2})_{-1} & \psi^{(2)}_{\mu} \cr 
\rho^{(1)}_- & ({\bf 1}, {\bf 1})_{1}  \oplus  ({\bf 3}, {\bf 1})_{1}  & (\eta^{(1)} , \chi^{(1)}_{\mu\nu}) \cr 
\rho^{(2)}_- & ({\bf 1}, {\bf 1})_{-1}  \oplus  ({\bf 3}, {\bf 1})_{-1}  & (\eta^{(2)} , \chi^{(2)}_{\mu\nu}) \cr 
\end{array}
\ee
The fields $A_\mu, \varphi, \bar\varphi$ do not carry $\su(2)_R$ charge and are thus unaffected. The scalars $\varphi^{\wat a}$ transform as a triplet of $\su(2)_R$. In the twisted theory they become a triplet $\varphi^a$ of $\su(2)_{\rm twist}$, defining a self-dual two-form $B_{\mu\nu}$ on $M_4$:
\be\label{Bdef}
B_{\mu\nu} = - (j^{\wat a}){}_{\mu\nu} \varphi^{\wat a} \,,
\ee
where the three local self-dual two-forms $j^{\wat a}$ transforming as a triplet of $\su(2)_{\rm twist}$. They can be defined in a local frame $e^A_\mu$ as
$(j^a){}_{\mu\nu} = e^A_\mu e^B_\nu (j^a){}^{A}{}_{B}$, $a=1,2,3$, with 
\be
(j^a){}_{0b} = - \delta^a_b \,, \quad (j^a){}_{bc} = - \epsilon^a{}_{bc} \,, \quad a,b,c = 1,2,3\,.
\ee
In this local frame we have
\be
B_{0a} = \varphi^{a}, \quad B_{ab} = \epsilon_{abc} \varphi^{c}, \quad a,b,c = 1,2,3 \,.
\ee
The self-dual tensors $j^a$ are used to map the vector index $a$ of $\so(3)$ to the self-dual two-form index $[AB]^+$. 
The tensors $ (j^a){}^{\mu}{}_{\nu}$ define an almost quaternionic structure, since they satisfy
\begin{align}
(j^a){}^{\mu}{}_{\rho} (j^b){}^{\rho}{}_{\nu} &= - \delta^{ab} \delta^\mu_\nu + \epsilon^{ab}{}_c (j^c){}^{\mu}{}_\nu \,.
\end{align}

The spinor fields transform as doublets of $\su(2)_R$. They become scalar, self-dual two-forms and one-form fields on $M_4$ as indicated in the table. The explicit decomposition, is obtained using the Killing spinor associated to the scalar supercharge in the twisted theory. This Killing spinor can be found as follows. 
The spinor $\epsilon_{\wat m}$ generating the preserved supersymmetry is a constant spinor and is invariant under the  twisted Lorentz algebra $\su(2)_{\rm twist} \oplus \su(2)_r$. 
As explained in section \ref{subsec:5dGenf} and in appendix \ref{app:SpinorDecomp} $\epsilon_{\wat m}$ decomposes  under $\mathfrak{sp}(4)_R \to \su(2)_R \oplus \mathfrak{u}(1)_R$ into two spinors doublets of $\su(2)_R$: $\epsilon_{\wat m} \to \epsilon^{(1)}_{\wat p} , \epsilon^{(2)}_{\wat p}$, satisfying the projections \eqref{KSrelation5d2}
\begin{align}
\epsilon^{(1)}_{\wat p} - \gamma^5 \epsilon^{(1)}_{\wat p} = 0 \,, \quad  \epsilon^{(2)}_{\wat p} + \gamma^5 \epsilon^{(2)}_{\wat p} = 0 \,.
\label{KSrelation5d2bis}
\end{align}
As explained in section \ref{subsec:TopoTwist}, $\epsilon^{(2)}_{\wat p}$ has one scalar component under $\su(2)_{\rm twist} \oplus \su(2)_r$ selected out by the projections
\be 
\ba
& (\gamma_{0a} \delta^{\wat q}_{\wat p} +  i (\sigma_{\wat a})^{\wat q}_{\wat p})\epsilon^{(2)}_{\wat q} = 0 \,, \quad a\simeq \wat a = 1,2,3 \,,
\ea
\ee
where the indices $a$ and $\wat a$ gets identified in the twisted theory. The spinor $\epsilon^{(2) \wat p}$ parametrizing the preserved supercharge is then decomposed as
\be 
\epsilon^{(2) \wat p}  = u \, \tilde{\epsilon}^{\wat p} \,,
\ee
where $u$ is complex Grassmann-odd parameter and $\tilde{\epsilon}^{\wat p}$ is a Grassmann-even spinor with unit normalisation.
The decomposition of the spinors into the twisted fields is then given by
\be 
\ba 
\rho^{(1)}_{+\wat p} &=\gamma^{\mu}\psi^{(1)}_{\mu}\tilde{\epsilon}_{\wat p} \\
\rho^{(2)}_{+\wat p} &=\gamma^{\mu}\psi^{(2)}_{\mu}\tilde{\epsilon}_{\wat p} \\
\rho^{(1)}_{-\wat p} &=\left(\eta^{(1)} + \frac{1}{4} \gamma^{\mu \nu} \chi^{(1)}_{\mu \nu}\right)\tilde{\epsilon}_{\wat p} \\
\rho^{(2)}_{-\wat p} &=\left(\eta^{(2)} + \frac{1}{4} \gamma^{\mu \nu} \chi^{(2)}_{\mu \nu}\right)\tilde{\epsilon}_{\wat p} \,.
\ea 
\ee


\subsection{Twisted 5d Action}
\label{app:5dTwistedAction}

We rewrite now the action in terms of the twisted fields and provide the preserved supersymmetry transformations. The bosonic part of this action has appeared in \cite{Witten:2011zz}, and related considerations regarding the supersymmetric versions of the twisted model can be found in \cite{Anderson:2012ck}. 

The action in \eqref{RescaledAction} in terms of the twisted fields takes the form
\be
\label{5dTwistedAction}
\ba
S_F
&= - \frac{r}{8\ell}\int d\theta d^4 x  
\gf \, \text{Tr}\left( F_{\mu\nu} F^{\mu\nu}+ \frac{2}{r^2}(\partial_\mu A_\theta-\partial_\theta A_\mu+[A_\mu,A_\theta])^2\right) \cr 
S_{\rm scalars}
&= - \frac{1}{4r\ell}\int d\theta d^4 x  \gf \, \text{Tr}
\left(\frac{1}{4}D^{\mu}B_{\rho \sigma} D_{\mu}B^{\rho \sigma}+\frac{1}{4r^2}D_{\theta}B_{\rho \sigma} D_{\theta}B^{\rho \sigma}\right. \cr 
& \qquad\qquad\qquad\qquad\qquad\qquad \left. +D^{\mu}\varphi D_{\mu}
\bar{\varphi}+\frac{1}{r^2}D_{\theta}\varphi D_{\theta}
\bar{\varphi}\right)\cr 
S_\rho 
&= \frac{2 i}{r\ell} \int d\theta d^4 x \gf\, \text{Tr}
\biggl[ \eta^{(2)}D_{\mu}\psi^{(1) \mu} - \psi^{(1)}_{\mu} D_{\nu}\chi^{(2) \mu \nu}  +
\eta^{(1)}D_{\mu}\psi^{(2) \mu} -\psi^{(2)}_{\mu} D_{\nu} \chi^{(1) \mu \nu} \cr
&\qquad\qquad\qquad   \left.+ \frac{1}{r}\left(
\psi^{(1)}_{\mu}D_{\theta}\psi^{(2)\mu}-\eta^{(1)}D_{\theta}\eta^{(2)} - \frac{1}{4}\chi^{(1)}_{\mu \nu} D_{\theta} \chi^{(2) \mu\nu} \right)\right]\cr 
S_{\rm Yukawa}
&= - {i\over r^2 \ell}\int d\theta d^4 x  \gf \, \text{Tr} \left(- \half B_{\mu \nu}\left[\eta^{(2)},\chi^{(1)\mu \nu}\right] + \half B_{\mu \nu}\left[\eta^{(1)}, \chi^{(2)\mu \nu}\right] \right.\cr 
& \qquad\qquad  \left. - \half B_{\mu \nu}\left[\chi^{(2)\mu \tau}, \chi^{(1)}{}^{\nu}{}_{\tau}\right]- 2B_{\mu \nu} \left[\psi^{(2)\mu}, \psi^{(1) \nu}\right] 
\right.\cr 
& \qquad\qquad  \left.
+\bar\varphi\left[\eta^{(1)}, \eta^{(1)}\right] +\frac{1}{4}\bar\varphi \left[\chi_{\mu \nu}^{(1)}, \chi^{(1)\mu \nu} \right] + \bar \varphi \left[ \psi^{(1)}_{\mu}, \psi^{(1)\mu} \right] \right. \cr 
&\qquad\qquad  \left.- \varphi\left[\eta^{(2)}, \eta^{(2)}\right] -\frac{1}{4}\varphi \left[\chi_{\mu \nu}^{(2)}, \chi^{(2)\mu \nu} \right] - \varphi \left[ \psi^{(2)}_{\mu}, \psi^{(2)\mu} \right]  \right) \\
S_{\rm quartic}
&= - {1\over 16r^3\ell}
\int d\theta d^4 x  \gf \, \text{Tr} \left( \frac{1}{4} \left[B_{\mu\rho}, B_{\nu}{}^{\rho}\right]\left[B^{\mu}{}_{\sigma},B^{\nu \sigma}\right] 
+[B_{\mu\nu},\varphi][B^{\mu \nu},\bar{\varphi}]-[\varphi,\bar{\varphi}][\varphi,\bar{\varphi}]\right) \\
S_{\text{bdry}} &= \frac{1}{16r^3\ell} \int d\theta d^4x \gf \tr\left(\partial_{\theta}B_{\mu \nu}[B^{\mu \rho}, B^{\nu}{}_{\rho}]\right)\,.
\ea
\ee
The supersymmetry transformations of this 5d topologically twisted SYM theory are
\be \ba 
\delta A_{\mu} &= -\frac{u}{r}\psi^{(1)}_{\mu} &\quad \delta A_{\theta} &= u\eta^{(1)} \\
\delta B_{\mu \nu} &= u\chi^{(1)}_{\mu \nu}  \\
\delta \varphi & = 0 &\quad \delta \bar\varphi &= 2u \eta^{(2)} \\
\delta \psi^{(1)}_{\mu} &= -\frac{iu}{4}D_{\mu}\varphi  &\quad \delta \psi^{(2)}_{\mu} &= -\frac{iu}{4}F_{\mu \theta} -\frac{iu}{4}D^{\nu} B_{\nu \mu} \\
\delta \eta^{(1)} &= \frac{iu}{4r} D_{\theta} \varphi   &\quad \delta \eta^{(2)} &= -\frac{iu}{8r}[\varphi, \bar\varphi]\\
\delta \chi^{(1)}_{\mu \nu} &= -\frac{iu}{4r}[\varphi, B_{\mu \nu}]  &\quad \delta \chi^{(2)}_{\mu \nu} &= \frac{iur}{2}F^+_{\mu \nu} + \frac{iu}{4r}D_{\theta}B_{\mu \nu} -\frac{iu}{8r}[B_{\mu \tau}, B_{\nu}{}^{\tau}]\,,
\ea \ee
where  the self-dual part of the gauge field is defined as 
\be
F^+ = {1\over 2}(1+*)F\,.
\ee
To define the twisted action for curved $M_4$, in addition to covariantising the derivatives, the curvature terms 
\be \label{CurvatureTerms}
\mathcal{R} B_{\mu \nu} B^{\mu \nu} \text{ and } \mathcal{R}_{\mu \nu \rho \sigma} B^{\mu \nu}B^{\rho \sigma}\,,
\ee
must be added to the action in order to preserve supersymmetry. These terms can be repackaged with the kinetic term for $B_{\mu \nu}$ changing the action for the scalars to 
\be\label{NewScalarTwistedSYM}
\ba
&S_{\rm scalars}\cr 
&= - \frac{1}{4r\ell}\int d\theta d^4 x  \sqrt{|g|} \, \text{Tr}
\left(\mathcal{D}^{\mu}B_{\mu \rho} \mathcal{D}^{\nu}B_{\nu}{}^{\rho} - \half F_{\mu \nu} B^{\mu}{}_{\sigma}B^{\nu \sigma} + \frac{1}{4r^2}D_{\theta}B_{\rho \sigma} D_{\theta}B^{\rho \sigma}+\mathcal{D}^{\mu'}\varphi\mathcal{D}_{\mu'}
\bar{\varphi}\right)\,,
\ea\ee
where $\mathcal{D}$ is defined to be covariant with respect to the curvature connection on $M_4$ and the gauge connection. The 5d twisted action on curved $M_4$ can be written in the form
\be 
S_{5d} = Q V + S_{5d,{\rm top}}\,,
\ee 
where the $Q$-exact and topological terms are given by
\be  \ba 
V &= - \frac{1}{r\ell} \int d\theta d^4x \sqrt{|g|} \,\text{Tr}\left[ \chi^{(2) \mu \nu} \left(P_{\mu \nu} -i(r F_{\mu \nu} +\frac{1}{2r}(D_{\theta}B_{\mu \nu} - \half[B_{\mu \tau}, B_{\nu}{}^{\tau}] ))\right) \right. \\
&\left.\qquad \qquad + 2\psi^{(2) \mu}\left(2P_{\mu} + i(F_{\mu \theta} + D^{\nu}B_{\nu \mu}) \right)  + i \psi^{(1) \mu}D_{\mu} \bar{\varphi}-\frac{i}{2r}\eta^{(2)}[\varphi, \bar{\varphi}] - \frac{i}{r}\eta^{(1)}D_{\theta} \bar{\varphi} \right. \\
&\left. \qquad \qquad +\frac{i}{4r}\chi^{(1) \mu \nu}[\bar{\varphi},B_{\mu \nu}] \right] \\
S_{5d,{\rm top}} &= \frac{r}{ 4 \ell} \int_{M_4 \times I} \,\text{Tr}\,  F\wedge * F-\frac{1}{2r\ell}\left[\int_{M_4}\text{Tr} \,F\wedge B\right]^{\theta = \pi}_{\theta = 0}\,,
\ea \ee
where $P_{\mu \nu}$ and $P_{\mu}$ are auxiliary fields. The supersymmetry transformations are 
\be \ba 
Q A_{\mu} &= -\frac{1}{r}\psi^{(1)}_{\mu} \ &\quad Q A_{\theta} &= \eta^{(1)}  &\quad Q B_{\mu \nu} &= \chi^{(1)}_{\mu \nu}  \cr 
Q \varphi & = 0  &\quad Q \bar\varphi &= 2 \eta^{(2)} \\
QP_{\mu} &= \frac{i}{4r} [\psi^{(2)}_{\mu}, \varphi]  &\quad QP_{\mu \nu}  &= \frac{i}{4r}[\chi^{(2)}_{\mu \nu}, \varphi] \\ 
Q \eta^{(1)} &= \frac{i}{4r} D_{\theta} \varphi  &\quad
Q \psi^{(1)}_{\mu} &= -\frac{i}{4}D_{\mu}\varphi &\quad 
Q \chi^{(1)}_{\mu \nu} &= -\frac{iu}{4r}[\varphi, B_{\mu \nu}] \\
Q \eta^{(2)} &= -\frac{i}{8r}[\varphi, \bar\varphi]  &\quad Q \psi^{(2)}_{\mu} &= P_{\mu} 
 \,, &\quad Q\chi^{(2)}_{\mu \nu} &= P_{\mu \nu}\,.
\ea \ee
The auxiliary fields are integrated out by 
\be \ba 
P_{\mu} &=-\frac{i}{4}(F_{\mu \theta} + D^{\nu}B_{\nu \mu}) \\
P_{\mu \nu} &= \frac{ir}{2}F^+_{\mu \nu} + \frac{i}{4r}\left(D_{\theta}B_{\mu \nu} - \half[B_{\mu \tau}, B_{\nu}{}^{\tau}]\right) \,.
\ea \ee
We can now proceed with the dimensional reduction to four-dimensions. 


\subsection{Triholomorphic Sigma-model with Hyper-K\"ahler $M_4$}

We now reduce the twisted 5d SYM theory to 4d on Hyper-K\"ahler $M_4$.
We proceed similar to the analysis in section \ref{sec:4dUntwistedRedux} and in appendix \ref{app:4dFlat}, and expand all fields in powers of $r$ and demand that the leading order terms in $\frac 1r$ in the action \eqref{5dTwistedAction} vanish. This sets $\varphi = \bar\varphi =O(r)$ and leads to Nahm's equations for the self-dual two-forms
\be\label{NahmEqnBfield}
D_\theta B_{\mu\nu} - \frac 12 [B_{\mu\rho}, B_{\nu}{}^{\rho}] = 0 \,,
\ee
with $\varrho=[k]$ Nahm pole boundary condition. Locally this is the same situation as in the untwisted theory, but not globally. In the untwisted theory the scalars $\varphi^{\wat a}$ were scalar fields on $\bbR_4$ and the solutions to the Nahm's equations are described by a map $\bbR^4 \to \cM_{k}$. In the twisted theory $B$ belongs to the bundle $\Omega^{2,+}(M_4)$ and the global solutions to \eqref{NahmEqnBfield} are generically more involved. However this complication does not happen when the bundle of self-dual two-forms $\Omega^{2,+}(M_4)$ is trivial, namely when $B$ transforms as a scalar. In this case one can regard the components $B_{\mu\nu}$ as scalars on $M_4$ and the solutions to \eqref{NahmEqnBfield} are again given in terms of a map
\begin{equation}
X : M_4 \rightarrow \cM_{k}  \,,
\end{equation}
where $\cM_k$ is the moduli space of solutions to Nahm's equations with $\varrho$ Nahm pole boundary conditions. As before we define  coordinates $X = \{ X^{I} \}$ on $\cM_k$.
The case when $\Omega^{2,+}(M_4)$ is trivial corresponds to $M_4$ having reduced holonomy $SU(2)_r \subset SU(2)_\ell \times SU(2)_r$, which is the definition of a Hyper-K\"ahler manifold. 

The zero modes around a solution $B_{\mu\nu}(X^{I})$ can be expressed as 
\begin{equation}
\ba
\delta B_{\mu\nu} &= \Upsilon_{I , \mu\nu} \delta X^I \cr
\delta A_{\theta} &= \Upsilon^{\theta}_{I} \delta X^I   \,,
\ea
\ee
where the expansion is in terms of the cotangent vectors $\Upsilon$, which satisfy
\be
\ba
\Upsilon_{I ,\mu\nu} &=  \p_I B_{\mu\nu} + [E_I , B_{\mu\nu}] \cr
  \Upsilon^{\theta}_{I} &= \p_I A_\theta - \p_\theta E_I - [A_\theta , E_I ] \,,
\ea
\ee
with $E_I$ defining a gauge connection on $\mathcal{M}_{N}$.
We will choose the convenient `gauge fixing condition'
\begin{equation}
 D_{\theta} \Upsilon^\theta_I - \frac 14 [\Upsilon_{I , \mu\nu}, B^{\mu\nu}]  = 0 \,.
 \label{UpsilonGauge}
\end{equation}
The equations obeyed by the cotangent vectors $\Upsilon^{\mu \nu}_I$, $\Upsilon^{\theta}_I$ are
\begin{align}
D_\theta \Upsilon_{I , \mu\nu} + [\Upsilon^\theta_{I}, B_{\mu\nu}] - \frac 12 \lp  [\Upsilon_{I ,\mu\rho}, B_{\nu}{}^{\rho}] -  [\Upsilon_{I ,\nu\rho}, B_{\mu}{}^{\rho}] \rp = 0 \,.
\label{UpsilonEqn}
\end{align}
A natural metric on  $\mathcal{M}_{N}$ can be defined as
\be\label{eqn:metricmodspace2}
G_{I J}= - \int d\theta \, \text{Tr} \lp  \frac 14 \Upsilon_{I}^{\mu\nu} \Upsilon_{J, \mu\nu } {+} \Upsilon_{I}^{\theta} \Upsilon_{J}^{\theta} \rp \,.
\ee
Similarly we can write down an expression for the three symplectic forms $\omega^a{}_{IJ}$ (see e.g. \cite{Gaiotto:2008sa}), repackaged into $\omega_{\mu\nu, IJ} = - (j^a)_{\mu\nu} \omega^a{}_{IJ}$, as 
\begin{equation}\label{SympForm2}
\omega_{\mu\nu, IJ} = -\int d\theta \, \text{Tr} \lp  \frac 12 \Upsilon_{I , \mu\rho} \Upsilon_{J}{}^{\rho}{}_{\nu}- \frac 12 \Upsilon_{I, \nu\rho} \Upsilon_{J}{}^{\rho}{}_{\mu}- \Upsilon_{I,\mu\nu} \Upsilon^{\theta}_J + \Upsilon^{\theta}_I\Upsilon_{J,\mu\nu} \rp \,.
\end{equation}
These provide the Hyper-K\"ahler structure of the moduli space $\mathcal{M}_k$. The quaternionic relations on the three complex structures $\omega^a{}^I{}_J$ becomes
\begin{align}
\omega_{\mu\rho, I}{}^J \omega_{\nu}{}^\rho{}_{J}{}^{K}&=  2 \omega_{\mu\nu,I}{}^{K} - 3 g_{\mu\nu}\delta_I^K \,.
\end{align}
Using the orthogonality of the $\Upsilon^{\mu \nu}_I$, $\Upsilon^{\theta}_I$ modes we derive the relations 
\begin{equation}
\ba
\omega_{\mu\nu, I}{}^J \Upsilon^{\theta}_{J} &= - \Upsilon_{I, \mu\nu} \\
\omega_{\mu\rho, I}{}^J \Upsilon^{\nu\rho}_{J} &= 2\Upsilon_{I, \mu}{}^{\nu} + 3 \delta^\nu_\mu \Upsilon^{\theta}_I \,.
\ea
\label{UpsilonRelations}
\end{equation}


\noindent At order $r^{-2}$ in the 5d action we find terms involving fermions. They vanish upon imposing 
\be 
\eta^{(2)} = O(r) , \quad \psi_{\mu}^{(1)} = O(r), \quad \chi^{(2)}_{\mu \nu} = O(r) \,.
\ee
The 4d action arises with overall coupling $\frac{1}{4r\ell}$ and at this order in $r$ the above fermions appear as Lagrange multipliers and can be integrated out to give the constraints
\be
\ba
 D_\theta \chi^{(1)}_{\mu\nu} {+} [\eta^{(1)}, B^{\mu\nu}] - \frac 12 \lp  [\chi^{(1)}_{\mu\rho}, B_{\nu}{}^{\rho}] -  [\chi^{(1)}_{\nu\rho}, B_{\mu}{}^{\rho}]  \rp &= 0 \cr 
D_{\theta} \eta^{(1)} - \frac 14 [\chi^{(1)}_{\mu\nu}, B^{\mu\nu}]  &= 0 \cr 
 D_\theta \psi^{(2)}_\mu  -  [\psi^{(2)}_\nu, B_\mu{}^{\nu}] &= 0 \,.
\ea
\ee
These equations are solved using the basis of the contangent bundle, which obey \eqref{UpsilonEqn} and \eqref{UpsilonGauge}, with the following relations
\be\label{FermDecomp}
\ba
\chi^{(1)}_{\mu \nu} &= \Upsilon_{I \, \mu \nu} \lambda^{I} + \Upsilon^{\theta}_I \zeta_{\mu \nu}^I + \Upsilon_{I\,\sigma [\mu}\zeta^{I \, \sigma}{}_{\nu]} \cr 
\eta^{(1)} &= \Upsilon^\theta_I \lambda^{I} -\frac{1}{4}\Upsilon_{I \, \mu \nu} \zeta^{I\, \mu \nu}\cr 
\psi^{(2)}_{\mu} &= \Upsilon_{I \,  \mu}{}^{\nu} \kappa^{I}_\nu - \Upsilon^\theta_I  \kappa^I_\mu \,,
\ea
\ee
where the fields $\lambda^I, \kappa^I_{\mu}$ and $\zeta^I_{\mu \nu}$ are Grassmann-odd scalars, vectors and self-dual two-forms on $M_4$, respectively.
The identities \eqref{UpsilonRelations} imply that the fermionic fields obey the constraints
\be
\ba
\omega_{\mu \nu}{}^I{}_J \lambda^J &=  \xi^I_{\mu \nu} \cr 
\omega_{\mu \sigma}{}^I{}_J \xi^{J}{}_\nu{}^\sigma &=  2 \xi^I_{\mu \nu} -3 \delta_{\mu \nu} \lambda^I  \cr 
\omega_{\mu \nu}{}^I{}_J \kappa^{J \nu} &= -3 \kappa^{I}_\mu \,.
\ea
\label{KappaConstr}
\ee
or more generally
\be 
\omega_{\mu \nu}{}^I{}_J \kappa_{\sigma}^J = g_{\mu \sigma} \kappa_{\nu}^I - g_{\nu \sigma} \kappa_{\mu}^I + \epsilon_{\mu \nu \sigma}{}^{ \rho} \kappa^{I}_{ \rho}\,.
\ee 
This decomposition satisfies the fermion equtaions of motion, which can be seen by using the identity
\be 
\Omega_{\rho\mu} \tilde{\Omega}^\rho{}_{\nu} = \frac{1}{4}\Omega_{\rho\sigma}\tilde{\Omega}^{\rho\sigma} g_{\mu\nu} + \Omega_{\rho[\mu} \tilde{\Omega}^\rho{}_{\nu]}\,,
\ee
where $\Omega_{\mu\nu}, \tilde{\Omega}_{\mu\nu}$ are self-dual two-forms.

\subsection{Dimensional Reduction to 4d Sigma-Model}
\label{app:TwistedDetails}

 After reduction to four dimensions the bosonic fields of the theory will be the collective coordinates $X^I$ describing a map $M_4 \rightarrow \mathcal{M}_{k}$ and the  fermionic fields will be the scalars $\lambda^I$, one-forms $\kappa^I$ and self-dual two-forms $\zeta_{\mu \nu}^I$, which are valued in the pull-back of the tangent bundle to $\mathcal{M}_{k}$
\be
\ba
& \lambda \in \Gamma(X^* T\mathcal{M}_k) \\
& \kappa \in \Gamma(X^* T\mathcal{M}_k \otimes \Omega^1) \\
& \zeta \in \Gamma(X^* T\mathcal{M}_k \otimes \Omega^2) \,.
\ea
\ee
The bosonic and fermionic zero modes lead to a four-dimensional effective action with overall coupling constant $\frac{1}{r \ell}$ for the fields $X^I$, $\lambda^I$, $\kappa^I_\mu$, $\zeta_{\mu \nu}^I$, $A_\mu$ and the scalars $\varphi$, $\bar\varphi$. 

As mentioned previously the kinetic term for $A_\mu$, namely $F_{\mu\nu}^2$ is of order $r$ and drops from the action in the small $r$ limit.  The gauge field $A_\mu$ becomes an auxiliary field and can be integrated out using its equation of motion, and likewise for the scalars $\varphi$ and $\bar\varphi$. Their equations of motion are
\be \ba
D_{\theta}^2 \varphi + \frac{1}{4}[B_{\mu \nu},[B^{\mu \nu},\varphi]] =& 4i r \left([\eta^{(1)}, \eta^{(1)}] + \frac{1}{4}[\chi^{(1)}_{\mu \nu}, \chi^{(1) \mu \nu}]\right) \cr 
D_{\theta}^2 \bar\varphi + \frac{1}{4}[B_{\mu \nu},[B^{\mu \nu},\bar\varphi]] =& -4i r\left([\psi^{(1)}_{\mu}, \psi^{(1)\mu}] \right) \cr 
D_{\theta}^2 A_{\mu} + \frac 14 \left[ B_{\nu\rho}, \left[ B^{\nu\rho}, A_{\mu}\right]\right] =& \left[A_{\theta}, \partial_{I} A_{\theta}\right]\partial_\mu X^I + \frac 14 \left[B_{\nu\rho}, \partial_{I} B^{\nu\rho}\right]\partial_\mu X^I \cr 
&+ 4i ([\eta^{(1)},\psi^{(2)}_{\mu}] -[\chi^{(1)}_{\nu \mu}, \psi^{(2) \nu}])\,.
\ea \ee
The spinor bilinears can be further simplified by applying the expansion for the spinors \eqref{FermDecomp}
\be \label{TwistedReduxGF}
\ba
\phantom{1}[\eta^{(1)}, \eta^{(1)}] + \frac{1}{4}[\chi^{(1)}_{\mu \nu}, \chi^{(1) \mu \nu}] &=4([\Upsilon_I^\theta, \Upsilon_J^\theta]  +\frac 14 [\Upsilon_{I \mu\nu}, \Upsilon_J^{\mu\nu }])( \lambda^I \lambda^J + \frac{1}{4}\zeta^I_{\sigma \rho} \zeta^{J \sigma \rho}) \\
\phantom{1}[\psi^{(1)}_{\mu}, \psi^{(1)\mu}] &= - 4( [\Upsilon_I^\theta, \Upsilon_J^\theta]  +\frac 14 [\Upsilon_{I \sigma\rho}, \Upsilon_J^{\sigma \rho}] ) \kappa^{I}_{ \mu} \kappa^{J \mu} \\
\phantom{1}[\eta^{(1)},\psi^{(2)}_{\mu}]-[\chi^{(1)}_{\nu \mu}, \psi^{(2) \nu}] &=  - 4( [\Upsilon_I^\theta, \Upsilon_J^\theta]  +\frac 14 [\Upsilon_{I \nu\rho}, \Upsilon_J^{\nu \rho}] )\lambda^I \kappa^J_\mu\,.
\ea
\ee
To solve these equations we note that the curvature
\be 
\Phi_{IJ} = [\nabla_I, \nabla_J] \,,
\ee
where $\nabla_I = \partial_I + [E_I, \cdot\,]$, satisfies the equation
\be  \label{FieldStrengthEoM2}
D^2_{\theta}\Phi_{IJ} + \frac 14 [ B_{\nu\rho}, [ B^{\nu\rho}, \Phi_{IJ}]] =  \frac 12 [\Upsilon_{I \nu\rho}, \Upsilon_J^{\nu\rho}]  + 2 [\Upsilon_I^\theta, \Upsilon_J^\theta]  \,.
\ee
Combining the information above the solutions are 
\be \label{GaugeFieldSol2}
\ba 
\varphi &= 8i r\Phi_{IJ} \lambda^I\lambda^J + 2i r\Phi_{IJ} \zeta^I_{\mu \nu} \zeta^{J \mu \nu} \\
\bar \varphi &= -8i r \Phi_{IJ} \kappa_{\mu}^I \kappa^{\mu J} \\
A_{\mu} &= E_I \partial_{\mu} X^I - 8i \Phi_{IJ}(  \lambda^I \kappa^J_{\mu} - \zeta^I_{\nu \mu} \kappa^{J \nu})  \,.
\ea \ee

Replacing the fermionic and bosonic zero modes in the action one obtains
\be
\ba
S_{\rm scalars} &= -\frac{1}{4r\ell}\int d\theta d^4x \gf \, \left[\text{Tr} \left( \partial_IA_{\theta}\partial_JA_{\theta} + \frac{1}{4}\partial_IB_{\rho \sigma}\partial_J B^{\rho \sigma} \right)\partial_{\mu} X^I \partial^{\mu} X^J \right]\\
S_{\rm fermions} &= {+} \frac{2i}{r\ell}\int d^4x \gf \, \left[\left(G_{IJ}g^{\mu \nu} -\omega^{\mu \nu}_{IJ}\right) (\lambda^I\partial_{\mu}\kappa_{\nu}^J -\xi^I{}_{\mu}{}^\sigma\partial_{\sigma}\kappa^J_{\nu})\right. \\
&\phantom{=\,}\left. {-} (\delta_I^K g^{\sigma \nu} - \omega^{\sigma \nu}{}_I{}^K)\text{Tr}\lp \frac{1}{4} \Upsilon_{K\, \rho \tau}\partial_J\Upsilon^{\rho \tau}_L + \Upsilon_K^\theta \partial_J \Upsilon_L^\theta \rp \partial_{\mu}X^J(\delta_{\sigma}^{\mu} \lambda^I\kappa_{\nu}^L - \xi^{I}{}_{\sigma}{}^{\mu}\kappa^L_{\nu})\right] \,.
\ea \ee
Substituting in the solution for the gauge field \eqref{GaugeFieldSol2} we obtain three different types of terms, which we address in turn. Terms of type \, 1 are proportional to $\partial_{\mu}X^I \partial_{\nu}X^J$ and combine with the terms in the scalar action to give
\be 
S_{{\rm scalars}} + S_{A_{\mu}, {\rm type \, 1}}
=  \frac{1}{4r\ell}\int d^4 x \gf\,G_{IJ}g^{\mu \nu} \partial_{\mu}X^I\partial_{\nu}X^J\,.
 \ee
Terms of type \, 2 combine with terms from the  action of the fermions to give  
\be \ba 
S_{A_{\mu}, {\rm type \, 2}} &= -\frac{2i}{r\ell}\int d\theta d^4x  \gf \, (\delta_I^K g^{\sigma \nu} - \omega^{\sigma \nu}{}_I{}^K)\text{Tr}\lp \frac{1}{4} \Upsilon_{K\, \rho \tau}\nabla_J\Upsilon^{\rho \tau}_L + \Upsilon_K^\theta \nabla_J \Upsilon_L^\theta \rp \partial_{\mu}X^J \cr 
&\qquad \qquad \qquad \times (\delta_{\sigma}^{\mu} \lambda^I\kappa_{\nu}^L - \xi^{I}{}_{\sigma}{}^{\mu}\kappa^L_{\nu}) \,.
\ea \ee 
Using the identities
\be \ba 
\nabla_I\Upsilon_J^{\mu \nu} &= \Gamma_{IJ}^{K}\Upsilon_K^{\mu \nu} + \frac{1}{2}[\Phi_{IJ}, B^{\mu \nu}]  \\
\nabla_I\Upsilon_J^{\theta}&= \Gamma_{IJ}^{K}\Upsilon_K^{\theta} -\half D_{\theta}\Phi_{IJ}  \,,
\ea \ee
where 
\be 
\Gamma_{IJ,K} = - \int d\theta\, \text{Tr}\left(\frac{1}{4} \Upsilon^{\mu \nu}_{K}\nabla_{(I}\Upsilon_{J)\mu \nu} + \Upsilon^{\theta}_{K}\nabla_{(I}\Upsilon_{J)}^{\theta} \right)\,,
\ee
these terms simplify to
\be \ba
S_{A_{\mu}, {\rm type \, 2}} &= \frac{2i}{r\ell}\int d\theta d^4x  \gf\, (G_{IJ} g^{\sigma \nu} - \omega^{\sigma \nu}{}_{IJ})\Gamma^J_{KL}\partial_{\mu}X^K (\delta_{\sigma}^{\mu} \lambda^I\kappa_{\nu}^L - \xi^{I}{}_{\sigma}{}^{\mu}\kappa^L_{\nu}) \,.
\ea \ee
and covariantise the kinetic terms for the fermions. Lastly the terms of type three contribute towards quartic fermion interactions. These take the form 
\be\label{QuarticFromGF}
\ba 
S_{A_{\mu}, {\rm type \, 3}} 
&= \frac{16}{r\ell}\int d\theta d^4x  \gf \text{Tr} \left(D_{\theta}\Phi_{IK}D_\theta\Phi_{JL} 
+ \frac{1}{4}[\Phi_{IK}, B_{ \mu \nu}][\Phi_{JL}, B^{\mu \nu}]\right)\cr 
&\qquad \qquad\qquad \qquad  \times
 (\lambda^I \lambda^J  \kappa^K_{\tau} \kappa^{L \tau} + \frac{1}{4} \zeta^I_{\rho \sigma} \zeta^{J \rho \sigma} \kappa^K_{\tau} \kappa^{L \tau}) \cr 
&=\frac{8}{r\ell}\int d\theta d^4x  \gf \,
\text{Tr} \Big(D_{\theta}\Phi_{IK}D_\theta\Phi_{JL} + \frac{1}{4}[\Phi_{IK}, B_{\mu \nu}][\Phi_{JL}, B^{\mu \nu}]  \cr 
&\qquad \qquad \qquad \qquad \qquad  -D_{\theta}\Phi_{IL}D_\theta\Phi_{JK} + \frac{1}{4}[\Phi_{IL}, B_{\mu \nu}][\Phi_{JK}, B^{\mu \nu} ]\Big)
\cr 
& \qquad \qquad\qquad \qquad  \times \left(\lambda^I \lambda^J  \kappa^K_{\tau} \kappa^{L \tau} + \frac{1}{4} \zeta^I_{\rho \sigma} \zeta^{J \rho \sigma} \kappa^K_{\tau} \kappa^{L \tau}\right)\,,
\ea \ee
where we have made use of the identity
\be \label{OmegaNableId}
\omega^{\mu \nu}{}_M{}^I \nabla_{[I}\Upsilon^{\theta}_{J]} = -  \omega^{\mu \nu}{}_J{}^I \nabla_{[I}\Upsilon^{\theta}_{M]}\,,
\ee 
and the analogous relation for $\Upsilon_I^{\mu \nu}$, and antisymmetrized in $KL$ indices. To obtain a quartic fermion interaction involving the Riemann tensor of the target we need to combine the terms in \eqref{QuarticFromGF} with the term which arises from integrating out $\varphi$ and $\bar\varphi$ 
\be  \label{QuarticFromScalars}
\ba 
S_{\varphi/\bar \varphi} &= \frac{16}{r\ell}\int d\theta d^4x  \gf \,
\text{Tr} \left(D_{\theta}\Phi_{IJ}D_\theta\Phi_{KL} + \frac{1}{4}[\Phi_{IJ}, B_{\mu \nu}][\Phi_{KL}, B^{\mu \nu}]\right) \\
& \qquad \qquad  \times(\lambda^I \lambda^J  \kappa^K_{\tau} \kappa^{L \tau} + \frac{1}{4} \zeta^I_{\rho \sigma} \zeta^{J \rho \sigma} \kappa^K_{\tau} \kappa^{L \tau}) \,.
\ea \ee
 Combining \eqref{QuarticFromScalars} and \eqref{QuarticFromGF}, as well as the fact that the Riemann tensor on the target is given by
\be \ba 
R_{IJKL} 
&= -\int d\theta \tr \Big( {1\over 2} \nabla_{[I}\Upsilon^{\mu \nu}_{J]} \nabla_{[K}\Upsilon_{L] \mu \nu} 
+ {1\over 4}\nabla_{[I}\Upsilon^{\mu \nu}_{K]} \nabla_{[J}\Upsilon_{L]\mu \nu}
-{1\over4} \nabla_{[I}\Upsilon^{\mu \nu}_{L]} \nabla_{[J}\Upsilon_{K]\mu \nu} \\
&\qquad \qquad\qquad    + 2 \nabla_{[I}\Upsilon^{\theta}_{J]} \nabla_{[K}\Upsilon^{\theta}_{L]} + \nabla_{[I}\Upsilon^{\theta}_{K]} \nabla_{[J}\Upsilon^{\theta}_{L]}- \nabla_{[I}\Upsilon^{\theta}_{L]} \nabla_{[J}\Upsilon^{\theta}_{K]} \big) \,,
\ea \ee
we obtain the four fermi interaction 
\be 
S_{{\rm fermi}^4} = - \frac{32}{r\ell} \int d^4x  \gf \,
R_{IJKL} \left(\lambda^I \lambda^J  \kappa^K_{\tau} \kappa^{L \tau} + \frac{1}{4} \zeta^I_{\rho \sigma} \zeta^{J \rho \sigma} \kappa^K_{\tau} \kappa^{L \tau}\right) \,.
\ee
The final action upon combining all the above terms is
\be \ba 
S &=\frac{1}{r\ell}\int d^4x \gf \,\left[\frac{1}{4}G_{IJ}g^{\mu \nu} \partial_{\mu}X^I \partial^{\nu}X^J + 2i \left(G_{IJ}g^{\mu \nu}- \omega^{\mu \nu}_{IJ}\right) (\lambda^I D_{\mu}\kappa_{\nu}^J - \zeta^I{}_{\mu}{}^{\sigma}\mathcal{D}_{\sigma}\kappa_{\nu}^J) \right. \cr 
& \qquad \qquad\qquad \qquad \qquad \left.-32 R_{IJKL}  
\left(\lambda^I \lambda^J  \kappa^K_{\tau} \kappa^{L \tau} + \frac{1}{4} \zeta^I_{\rho \sigma} \zeta^{J \rho \sigma} \kappa^K_{\tau} \kappa^{L \tau}\right)\right]\,,
\ea \ee
where 
\be
D_{\mu}\kappa^I_{\nu} = \partial_\mu \kappa^I_{\nu} + \Gamma^I_{JK} \partial_{\mu}X^J \kappa_{\nu}^K\,. 
\ee
The action can be further simplified by using relations between the complex structures $\omega_{\mu \nu}{}^I{}_J$ and the fermions \eqref{KappaConstr} to eliminate the self-dual two-form $\zeta^I_{\mu \nu}$. 
In addition we know that the target space $\mathcal{M}_{k}$ is Hyper-K\"ahler, which means that the three complex structures $\omega_{\mu \nu}{}^I{}_{J}$ define covariantly constant on $\mathcal{M}_{k}$
\be
D_I \omega_{\mu \nu}{}^J{}_{K} = 0\,.
\ee
This in turn implies the relations with the Riemann tensor on $\mathcal{M}_{k}$
\be \label{RiemannOmegaRel}
R_{IJK}{}^M \omega_{\mu \nu, M L} = R_{IJL}{}^M \omega_{\mu \nu, M K}\,,
\ee
and other relations obtained using the standard symmetries of the Riemann tensor.
With \eqref{KappaConstr} and \eqref{RiemannOmegaRel}, and after rescaling $\lambda \rightarrow \frac{1}{4}\lambda^I$ and $\kappa_\mu \rightarrow {i\over 16} \kappa_\mu$, the action simplifies to
\be\label{TopoAction}
 \ba 
S_{HK} &= \frac{1}{4 r\ell}\int d^4x \gf \,  \Big(   G_{IJ}g^{\mu \nu} \partial_{\mu}X^I \partial_{\nu}X^J -  2 G_{IJ}  g^{\mu \nu}\kappa_{\mu}^I  D_{\nu}\lambda^J + \frac{1}{8} g^{\mu\nu}  R_{IJKL} \kappa_{\mu}^I\kappa_{\nu}^J\lambda^K \lambda^L  \Big) \,.
\ea \ee
The constraint on the fermions $\kappa^I_\mu$ can be re-expressed as
\begin{align}
\kappa^I_\mu + \frac 13 (j^a)_\mu{}^\nu \kappa^J_\nu \omega^a{}_{J}{}^I = 0 \,,
\end{align}
The supersymmetry transformations are
\be \ba 
\delta X^I &=  u\lambda^I \\
\delta \lambda^I &= 0 \\
\delta \kappa_{\mu}^I &= u\left(\partial_{\mu}X^I - (j^a)_{\mu}{}^\nu \partial_{\nu} X^J \omega^{a}{}_J{}^I \right) -  u   \Gamma^I_{JK}\lambda^J\kappa_{\mu}^K \,.
\ea \ee
This dimensional reduction of the 5d topologically twisted SYM theory, thus gives precisely the same action we obtained in (\ref{4dHKSigmaModel}), by topologically twisting the 4d sigma-model for Hyper-K\"ahler $M_4$. 



\providecommand{\href}[2]{#2}\begingroup\raggedright\endgroup


\end{document}